# High Spatial-Resolution Fast Neutron Detectors for Imaging and Spectrometry

**Thesis submitted in partial fulfillment
of the requirements for the degree of
"DOCTOR OF PHILOSOPHY"**

**by**

**Ilan　　　　Mor**

**Submitted to the Senate of Ben-Gurion University
of the Negev**

**January 2013**

**Beer-Sheva**

# High Spatial-Resolution Fast Neutron Detectors for Imaging and Spectrometry

**Thesis submitted in partial fulfillment
of the requirements for the degree of
"DOCTOR OF PHILOSOPHY"**

**by**

**Ilan Mor**

**Submitted to the Senate of Ben-Gurion University
of the Negev**

**Approved by the advisor**

**Approved by the Dean of the Kreitman School of Advanced Graduate Studies**

**January 2013**

**Beer-Sheva**

This work was carried out under the supervision of

Dr. David Vartsky
Soreq Nuclear Research Center

Dr. Itzik Orion
In the Department: Nuclear Engineering
Faculty:  Engineering

Prof. Amnon Moalem
In the Department:  Physics
Faculty:  Natural Sciences

# Acknowledgments

Many people have participated in this work. First and fore most, I would like to express my gratitude to my primary dissertation advisor and head of this project, Dr. David Vartsky from Soreq Nuclear Research Center. For his patient and open minded guidance throughout which made me an equal partner and left me scope to express my independence while making mistakes and learning from them. In my perspective, Dr. Vartsky has been a role-model for a true physicist. I have benefited immensely, both professionally and personally, from his vast knowledge and expertise.

I also want to thank my secondary thesis advisors, Prof. Amnon Moalem and Dr. Itzik Orion from the Ben-Gurion University. Both were always available to help with any academic issues that arose on occasion and provided valuable comments to this dissertation which increased readability and reduced ambiguity.

As an important member of this research group, Dr. Mark B. Goldberg contributed from his extensive knowledge and expertise during numerous discussions with David and myself at various stages of development and testing. I found those discussions most educative. In addition, this dissertation would not be as eloquent and well structured as it is without Dr. Goldberg's assistance.

Special thanks are due to my lab colleague Michal Brandis. Michal participated and provided valuable contribution during beam-time experiments at PTB. In addition, Michal supported my preparations for beam-time experiment with the capillary detector at PTB by performing preliminary simulations of the capillary array which aided in deciding on irradiation geometry.

Image acquisition and data collection were controlled via a special computer program written by Doron Bar, who also participated and contributed to all beam-time experiments at PTB. For these I am grateful. I would also like to acknowledge the contribution of Dr. Gennadi Feldman to the image registration topic.

The detailed engineering design of TRION was done by David Katz with the supervision of Eli Sayag, from the drawing-board to the workshop, with the active supervision and participation of Shaul Levy and many of the workers there.

Our lab technicians, Shaul Levy and Zion Vagish, found creative solutions for the continuous technical development of our system.

The electronic wiring of the system was done with the kind assistance of Yair Cohen.

I would like to give heartfelt special thanks to Dr. Volker Dangendorf from PTB, Germany. Dr. Dangendorf played a key role in this work and was involved in all stages of the development. I have benefited greatly from his broad knowledge, experience and creative ideas. Dr. Dangendorf has played a major role in the success of this project.

Moreover, as a post-Doc. at Dr. Dangendorf's lab at PTB, I was granted full support for the final stage of the lengthy writing process of this dissertation. His efforts on my behalf are immensely appreciated.

Many thanks are due to the PTB group headed by Dr. Dangendorf. I want to thank the group members: Kai Tittelmeier, Benjamin Bromberger, Mathias Weierganz and the accelerator team for their tremendous support and involvement in this project.

Last but certainly not least, I would like to thank my dearest wife and beloved son, Liat and Ofek, who coped with a part time husband/dad during these long months of writing. Liati, your support during this time was invaluable. Thank you for supporting me through this demanding marathon.

# CONTENTS

## List of Figures

# List of Tables

# Abstract

The present work seeks to provide efficient, large-area fast-neutron detectors for combined sub-mm spatial imaging and energy spectrometry, capable of operation in mixed, high-intensity neutron-gamma fields. The combination of high spatial-resolution fast-neutron imaging with energy spectrometry is <u>novel and unique</u>.

Within the framework of this work two detection systems based on optical readout were developed:

a. Integrative optical detector

A 2$^{nd}$ generation of Time-Resolved Integrative Optical Neutron (TRION) detector (TRION Gen. 2) was developed. It is based on an integrative (as opposed to event counting) optical technique, which permits fast-neutron energy-resolved imaging via time-gated optical readout. This mode of operation allows loss-free operation at very high neutron-flux intensities. The TRION neutron imaging system can be regarded as a stroboscopic photography of neutrons arriving at the detector on a few-ns time scale. As this spectroscopic capability is based on the Time-of-Flight (TOF) technique, it has to be operated in conjunction with a pulsed neutron source, such as an ion accelerator producing 1-2 ns wide beam pulses at MHz repetition rates. TRION Gen. 2 is capable of capturing 4 simultaneous TOF frames within a single accelerator pulse and accumulating them over all pulses contained within a finite acquisition time. It exhibits a 4-fold increase in the utilization efficiency of the broad neutron spectrum created following each accelerator pulse. This presents a significant step towards a future real-time operational system. In addition, it has demonstrated spatial resolution comparable to that of TRION Gen.1, as well as reduced intensifier thermal noise and improved temporal resolution.
The detector principle of operation, simulations and experimental results are described.

b. Fibrous optical detector

A fast neutron imaging detector based on micrometric glass capillaries loaded with high-refractive-index liquid scintillator has been developed. Neutron energy spectrometry is based on event-by-event detection and reconstruction of neutron energy from the measurement of the recoil proton track projection length and the amount of light produced in the track. In addition, the detector can provide fast-neutron imaging with position resolution of tens of microns. The detector principle of operation, simulations and experimental results obtained with a small detector prototype are described. Track-imaging of individual recoil protons from incident neutrons in the range of 2-14 MeV are demonstrated as well as preliminary results of detector spectroscopic capabilities.

Keywords: Fast neutron resonance radiography; Time-of-Flight; Fast neutron imaging; Energy-resolved imaging; Neutron spectrometry; Capillary array; Liquid scintillator

# I. Introduction

The present work seeks to provide efficient, large-area fast-neutron detectors for combined sub-mm spatial imaging and energy spectrometry, capable of operation in high-intensity mixed high-intensity neutron-gamma fields.

These detectors can be used for energy-selective radiography, as performed, for example, in Fast-Neutron Resonance Radiography (FNRR), for the detection of contraband by identification of the presence of light elements such as carbon, nitrogen and oxygen within the inspected object.

The combination of high-spatial resolution fast-neutron imaging with energy spectrometry presented here is <u>novel and unique</u>.

Within the framework of this work two types of detection systems based on optical readout were developed:

a. Integrative optical detector

The Time-Resolved Integrative Optical Neutron (TRION) detector is the continued development of my M.Sc. work (Mor, 2006). It is based on an integrative (as opposed to event counting) optical technique, which permits fast-neutron energy-resolved imaging based on time-gated optical readout. This mode of operation permits loss-free operation at very high neutron-flux intensities.

The TRION neutron imaging system can be regarded as a stroboscopic photography of neutrons arriving at the detector on a few-ns time scale. As this spectroscopic capability is based on the Time-of-Flight (TOF) technique, it needs to be integrated with a pulsed neutron source, such as an ion accelerator producing 1-2 ns wide beam pulses.

The TRION detection system will lead toward the next generation systems for automatic detection of small quantities of standard and improvised explosives as well as special nuclear materials concealed in cargo and passenger vehicles.

b. Fibrous optical detector

This detector is based on event-by-event detection and extraction of various parameters related to its response to an impinging fast neutron, which are combined in a fashion that permits neutron spectroscopy. These parameters are: length of the recoil proton track-projection within the detector and the amount of energy deposited by the recoil proton. The detector is based on micrometric glass capillary array loaded with high-refractive index liquid scintillator. Capillaries with diameter of about 11 microns diameter should allow measurement and reconstruction of the recoil-proton track with sufficient accuracy. The expected spatial-resolution defined by this detector is of the order of tens of microns. This technique is <u>not</u> based on TOF spectroscopy and is designed to work with an un-pulsed neutron sources. This fact should render the detector suitable for application with any fast-neutron source.

**1. Review of fast neutron spectroscopic methods**

Accurate measurement of a neutron energy-spectrum is not a trivial task since the techniques common in charged particle spectrometry are not applicable for neutrons. Nevertheless, many fast-neutron spectrometric techniques have been developed over the years, such as: Time-of-Flight (TOF), $He^3$ detectors, hydrogenous gas detectors based on proton recoil, recoil-proton telescope detector and other hydrogenous detectors based on proton track reconstruction and capture gated detectors.

Out-of the above mentioned spectrometric techniques, the common techniques employed for neutrons between 0.8-14 MeV are:
- Proton recoil detectors such as: recoil-proton telescope, hydrogenous gas, liquid and plastic detectors
- $He^3$ gas proportional counter detector
- Capture gated detectors.
- Time of flight technique

The following will summarize physical principles underlying these techniques.

1.1 Proton recoil detectors
In this group of detectors the incident neutron interacts with hydrogenous medium in a gaseous, liquid or solid form. The neutron is detected by measuring the properties (energy and/or angle) of the recoil proton resulting from the elastic scattering interaction. This category of detectors can be subdivided into detectors in which the measured protons arrive at a set recoil angle, (e.g. recoil proton telescope); and to hydrogenous detectors (gas, liquid and plastic) which allow usage of mathematical unfolding to calculate the neutron spectrum.

The following will provide a short summary of these two detector groups.

1.1.1 Proton recoil telescope detectors

These detectors are based on narrow choice of the recoil-proton scattering angle (Baba et al., 1999; Knoll, 2000; Mori et al., 1999; Schuhmacher et al., 1999).

The energy gained by a recoil nucleus $E_R$ in terms of its own angle of recoil θ (in the lab reference system) from an incident neutron (with non-relativistic kinetic energy) is:

$$E_R = \frac{4A}{(1+A)^2}(cos^2 \theta) E_n \tag{1}$$

Where A is the atomic mass of the recoil nucleus and $E_n$ is the incident neutron energy. From Eq. 1 we can see that, for fixed *A*, the energy imparted to the recoil nucleus is uniquely determined by the scattering angle *θ*.

When the recoil nucleus is hydrogen, Eq. 1 is reduced to:

$$E_R = E_n \cos^2(\theta_R) \tag{2}$$

Proton recoil telescopes can be applied only to situations in which the incoming direction has been defined by collimation or other means (Knoll, 2000). Neutrons impinge upon a very thin hydrogenous slab, composed usually of an organic polymer of thickness less than the range of the lowest energy recoil-proton which is to be measured.

In Figure 1, the angle $\theta_R$, at which recoil protons are observed, is defined by locating the detector in vacuum at a certain distance and angle relative to the thin slab, in order to prevent the protons from losing energy on their way to the detector.

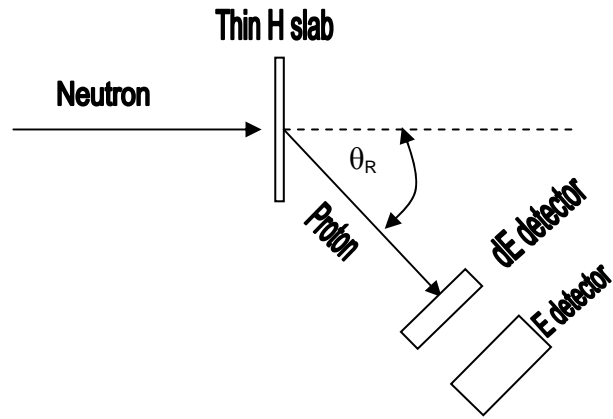

**Fig. 1 Schematic drawing of recoil proton telescope**

Most of these detectors are positioned at a small angle (different from 0) in order to avoid background resulting from interaction of the incident neutrons with the detectors.

Although a single detector allows measurement of the proton energy (thus permitting determination of the neutron energy using Eq. 2), in most cases, several detectors are used in order to reduce background resulting from competing reactions and other spurious events (Baba et al., 1999; Schuhmacher et al., 1999).

The configuration shown in Figure 1 is very common; a very thin detector (dE) is positioned in front of a thicker detector (E) which fully stops the recoil protons. When operating these detectors in coincidence mode, only particles emitted along (or close to) the thin hydrogenous slab axis are registered. If the response of both detectors is linear and known, then an appropriately-weighted sum of their signals will scale with the total energy of the proton.

The obvious drawback of this type of detector is its extremely low detection efficiency (typically, 1 count per $10^5$ neutrons (Knoll, 2000). It results from two principal causes which unavoidably degrade the energy resolution:

a. Thickness of the hydrogenous slab – needs to be as thin as possible in-order to minimize proton energy-losses before they leave the slab. Typical thicknesses used (0.5-2 mm) lead to a probability of $10^{-3}$- $10^{-4}$ that a neutron will be scattered inside the slab.

b. The solid-angle subtended by the proton detectors must be small, in-order to prevent broadening of the peak due to the response function (see Eq. 2).

The advantage of this detector is that the proton detection efficiency can be calculated reliably and accurately, since complications resulting from neutron multiple scattering and wall effects are significantly reduced.

1.1.2 Hydrogenous detectors and the unfolding technique

In order to overcome the problem of telescope-detector low efficiency, hydrogenous detectors (gas, liquid and plastic) which measure neutrons scattered at all angles were developed. Therefore, the response function (rectangular shaped, approximately) of these detectors to mono-energetic neutrons incorporates all recoil proton energies.

The field of spectrometry using detector response unfolding has been widely explored in the past and also in the present. The following will describe briefly the general idea at the basis of the unfolding technique.

The detector pulse-height distribution when counting $Z_i$ counts in channel i can be described by Eq. 3 (Bartlett et al., 2003; Matzke, 2003):

$$Z_i = \int R_i(E) \cdot \Phi(E) dE \quad (i=1,...,M) \quad (3)$$

Where $R_i(E)$ represents the detector response function at channel i to mono-energetic neutrons of energy E, $\Phi(E)$ is the mono-energetic neutron flux at energy E.

The reading $Z_i$ in Eq. 3 is a linear function of the spectral flux.

In order to perform numerical calculations, Eq. 3 has to be transformed into a matrix of discrete linear equations:

$$Z = R \cdot \Phi \qquad (4)$$

Finding the flux vector is called 'Solution of the inverse problem of spectrometry', or:

$$\Phi = R^{-1} \cdot Z \qquad (5)$$

Eq. 3 represents a model of an ideal situation, however in reality, the real measured readings ($Z_0$) contain an element of noise $\varepsilon$. $Z_0$ can be written as the sum of readings in an ideal situation with the noise component:

$$Z_0 = Z + \varepsilon \qquad (6)$$

where $\varepsilon$ represents fluctuations resulting from statistical and systematic uncertainty.

Some of the mathematical techniques used for the solution of the inverse problem in a situation which involves noise are: The maximal entropy principle , Bayesian parameter estimation (Reginatto et al., 2002), iterative solution technique for linearized least-squares (Koohi-Fayegh et al., 2001), differential unfolding technique (Koohi-Fayegh et al., 2001) and so on.

Several algorithms were developed in order to solve the inverse problem for particle spectrometry. These algorithms were incorporated into different computer programs which tested successfully. A technique should be chosen according to the problem needed to be solved. Below are several names of programs which perform spectral unfolding: FORIST (Koohi-Fayegh et al., 2001), FERDOR (Koohi-Fayegh et al., 2001), MAXED (Reginatto et al., 2002), FLYSPEC (Koohi-Fayegh et al., 2001), RADAK (Reginatto et al., 2002; Reginatto, 2010).

The unfolding techniques require an exact knowledge of the detector response function for the reconstruction of the neutron spectra. This requirement is by no means a trivial one.

1.1.3 H and D gas proportional counters

Hydrogenous gas proportional counters can be used to measure fast-neutrons through the recoil process (Verbinski and Giovannini, 1974). They are useful in the energy range 50 keV to several MeV as wall effects limit the usefulness of these counters for neutrons of energies above about 5 MeV (Ing et al., 1997; Rosenstock et al., 1997).

The proportional counter is a type of gas-filled detector that relies on the phenomenon of gas multiplication to amplify the charge represented by the original ion pairs created within the gas by incident radiation.

In these neutron-sensitive gas proportional counters the fill gas is usually hydrogen, a-hydrogen containing gas such as methane or some other low-Z gas such as helium. In the case of hydrogen, the expected proton-recoil spectrum for mono-energetic incident neutrons should be a simple rectangular shape extending from zero to the full incident neutron energy. This is due to the fact that the scattering process is isotropic in the center-of-mass coordinate system. However, complicating effects that distort this simple response function often arise and render the task of unfolding measured pulse-height spectra much more complicated. The most important of these complicating effects is the effect of the finite size of the active volume of the chamber which leads to proton tracks being truncated either in the walls or the end of the tube.

Although hydrogen is in many ways the ideal target for neutron scattering, its use as a fill gas in proportional counters is limited by the low density and relatively low stopping power for recoil protons. Methane is a more common choice, but the presence of carbon adds complications to the response function. It has been estimated (Knoll, 2000) that carbon will create about 75 % as much ionization in the gas as the equivalent energy proton. Because the maximum energy of a carbon ion is 28 % of the incident neutron energy, one would expect that all carbon-recoil pulses to lie below about 21 % of the maximum proton recoil pulse amplitude.

Because the detection medium is a gas of relatively low density, recoil proportional counters have a markedly lower counting efficiency (~1 %) than typical organic scintillators. Nevertheless, the low density does make for lower probability of multiple-scattering interactions.

Recoil proportional counters are generally more sensitive to technical problems than organic scintillators. Purity of the gas fill is of utmost importance and microscopic air leaks will ultimately lead to detector failure. The fact that these detectors are not large compared to the range of recoil nuclei in a gas means that the correction for wall-and-end effects, usually quite small in scintillators, become an important consideration in determining the response of these detectors. Another consequence of the low-density gas detection medium is the behavior of the detector in the presence of gamma-rays. As the size of the recoil proportional counters is comparable to the range of the recoil nuclei, it is likely that neutron-induced events will deposit all their energy within the detector whereas gamma rays will deposit only part of their energy due to the longer range of gamma-ray-produced-electrons.

An additional drawback is the nature of gamma-ray interactions in the detector. In the gas counter, neutrons must interact in the fill gas but gamma-rays may interact either in the gas or, more likely, in the much denser walls and other construction materials of the counter, leading to secondary electrons that can escape into the gas volume. Provided the neutron energy is fairly low, proton recoil track will tend to be rather short and consequently confined to a limited range in radius. Gamma-ray induced fast electrons will almost always pass completely through the gas and, on average, involve a much greater range of radii in the tube. As a result, neutron-induced pulses will tend to have shorter rise times than those induced by gamma rays, and pulse-shape discrimination methods can serve to differentiate between the two (Hawkes, 2007; Knoll, 2000; Verbinski and Giovannini, 1974).

As sources of mono-energetic neutrons are not commonly available, the energy calibration of a proton-recoil proportional counter is not a straightforward process.

One possible method is to incorporate a small amount of $^3$He into the fill gas, so that when irradiated with thermal neutrons, proton and triton pairs sharing 764 keV kinetic energy (see Figure 2, section 1.2) are generated internally. The corresponding peak in the recorded spectrum then provides an energy marker. Errors in such mode of calibration may occur if one of the particles created is not a proton.

1.1.4 Liquid and solid hydrogenous scintillator spectrometers

The most common method employing proton recoil in the detection of fast neutrons in the energy range 0.1-20 MeV is via the use of hydrogen-containing scintillators. The properties of scintillator detectors are discussed in detail in textbooks and reviews (Bromley, 1979). A relative recent review of scintillation detectors for neutrons was presented by Klein and Brooks (Klein and Brooks, 2006).

These detectors can be in the form of organic crystals, liquids and plastics and are useful for spectrometry, either by TOF or by the unfolding of pulse height spectra. Mono-energetic fast neutrons incident on the scintillator give rise to recoil protons whose energy distribution should be approximately rectangular, ranging from zero to the full neutron energy. As the range of the recoil protons is usually small compared to the dimensions of the scintillator, their full energy is deposited in the scintillator and the expected pulse height distribution is also approximately rectangular.

If a set of such measurements are performed for the same scintillator using several different incident neutron energies one obtains a set of response functions which can be used as a response matrix for the unfolding procedure. One can also obtain these matrices from Monte-Carlo simulations. For high neutron energies above 20 MeV contributions of nuclear reactions on C that produce protons, alphas and neutrons, complicate the procedure (Novotny et al., 1997).

In order to measure neutron spectra in mixed gamma-neutron fields the unfolding procedure must be performed on a clean proton energy spectrum without electron interference. This can be achieved by the technique of pulse shape discrimination (PSD) a property possessed by certain organic crystals and liquid scintillators.

1.1.4.1 Solid hydrogenous scintillator spectrometers

Much of the early work in developing proton recoil scintillators was done using crystals of anthracene or stilbene (Knoll, 2000). Anthracene has the largest light output of any organic scintillator. However, attention gradually shifted to stilbene due to its superior gamma-ray rejection characteristics. Nevertheless, both crystals are expensive and difficult to obtain in large sizes (greater than a few centimeters) and are also subject to damage from thermal and mechanical shock. A further disadvantage stems from the directional variation of the light output from such crystals which depends on the orientation of the path of the charged particles with respect to the crystal axis. This effect greatly complicates the job of unfolding an observed pulse height spectrum in order to derive the incident fast neutron energy spectrum.

It is far more common to use plastic and liquid scintillators for fast neutron applications. These materials are relatively inexpensive, can be tailored to a wide variety of sizes and shapes and are totally non-directional in their response. Plastics can be obtained in large sizes including rods, bars and sheets or in small sizes such as fiber scintillators (cross-sectional dimensions of several hundreds of microns) and are relatively simple to machine. The attenuation length for the scintillation light in typical plastics can be several meters or more, an important consideration when scintillators of large size are required.

1.1.4.2 Liquid hydrogenous scintillator spectrometers

This category of scintillators is generally produced by dissolving an organic scintillator in an appropriate solvent. Liquid scintillators can consist of simply these two components, or a third constituent may be added as a wavelength shifter, such that the emission spectrum better matches the spectral response of common photomultiplier tubes, image-intensifiers and so on.

In many such liquids, the presence of dissolved oxygen can serve as a strong quenching agent resulting in substantially reduced fluorescence efficiency.

Due to their lack of ordered structure, liquid scintillators are expected to be more resistant to radiation damage effects than crystalline or plastic scintillators.

For certain applications, liquid scintillators are sometimes preferred because of their pulse shape discrimination capability, allowing suppression of gamma-ray events.

1.2 Detectors based on the $^3$He(n,p)$^3$H reaction

These are gas proportional counters filled with $^3$He gas. The $^3$He(n,p) reaction has been widely applied for fast neutron detection and spectroscopy (Verbinski and Giovannini, 1974).

The pulse-height spectrum from a detector based on the $^3$He reaction should show three distinct features. Neglecting the wall effect, the full energy of the reaction products is always totally absorbed within the detector. These features are:
  a) Full energy peak corresponding to all (n,p) events induced directly by the incident neutrons. This peak occurs at an energy equal to the neutron energy plus the Q-value of the reaction (764 keV).
  b) A pulse height continuum resulting from elastic scattering of the neutron and a partial transfer of its energy to a recoiling He nucleus (up to 75% of the neutron energy for $^3$He).
  c) An epithermal peak corresponding to the detection of incident neutrons which have slowed down to thermal velocities by moderation in external materials. All such reactions deposit energy equal to the Q-value, or 764 keV.

Wall effects arise whenever the dimensions of the detector are not large compared to the ranges of secondary particles produced in these reactions.

Several other types of detectors are based on the $^3$He(n,p) reaction: proportional counter, ionization chamber, scintillator and semiconductor-sandwich.

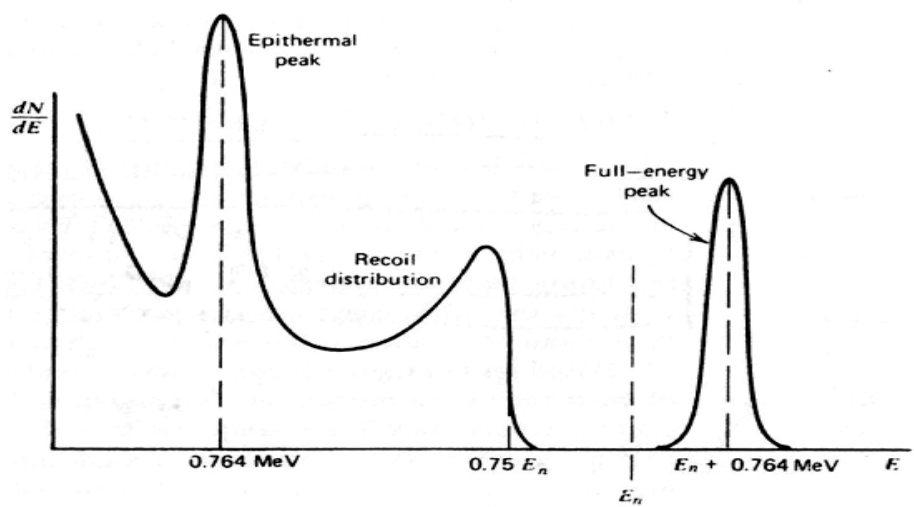

**Fig. 2 Differential energy spectrum of charged particles expected from fast neutrons incident on $^3$He (Knoll, 2000)**

During recent years, there has been an increasing shortage of $^3$He causing sharp prices increases of available material. In addition, applications which required large-area detectors were brought to a halt.

1.3 Capture gated $^{10}$B or $^6$Li enriched detectors

Capture-gated neutron spectrometry (Fisher et al., 2011; Hyeonseo, 2010) is based on the coincidence measurement of the scattered signal and the neutron capture signal. As illustrated in Figure 3, an incident fast neutron entering a scintillation medium undergoes a series of scatterings with hydrogen (creating a recoil proton) or carbon nuclei in the scintillator. The number of collisions depends on the size and density of the scintillator. After one or two collisions most of the incident neutrons escape from the detector volume. If the detector is sufficiently large, some neutrons will lose their entire energy by successive collisions, thus become thermalized and captured within the detector volume.

The use of organic scintillator provides high density of protons that will efficiently moderate the neutron; a neutron with energy of several MeV loses 90% of its energy in this manner during the first 10 ns in the scintillator (Fisher et al., 2011).

This delayed-coincidence signature indicates that the fast neutron dissipated essentially all of its energy within the scintillator and provides a very strong rejection of uncorrelated backgrounds such as escaped events. Within approximately 10-50 μs, depending upon the size and geometry of the detector, the thermalized neutron may be captured by an isotope of large neutron capture cross-section and thus providing an unambiguous signature of the neutron capture reaction.

One must consider several factors when selecting which capture isotope is appropriate for an application. Typical elements are gadolinium, boron, and lithium. Gadolinium is commercially available and has a very large neutron capture cross section, but it produces a high energy gamma-ray that may be difficult to detect with small volume detectors. With neutron capture on $^{10}$B, the $^{7}$Li ejectile can be formed in the ground or first excited states, leading to two reaction branches (Fisher et al., 2011):

$$^{10}B + n \rightarrow {}^{7}Li^{*} (0.84 MeV) + \alpha (1.47 MeV) \quad (1)$$
$$^{7}Li^{*} \rightarrow {}^{7}Li_{g.s.} + \gamma (0.477 MeV)$$
$$^{10}B + n \rightarrow {}^{7}Li_{g.s.} (1.01 MeV) + \alpha (1.78 MeV) \quad (2)$$

The branching ratios for the first reaction is 93.7% and for the second 6.3%. The cross section is also large (in the 1000 b range), and the alpha particle makes efficient thermal detection possible with a small-volume detector. $^{10}$B detectors are expensive, which tends to make scaling to larger-volume detectors cost-prohibitive. A good alternative to $^{10}$B is $^{6}$Li (Fisher et al., 2011):

$$^{6}Li + n \rightarrow t (2.05 MeV) + \alpha (2.73 MeV) \quad (3)$$

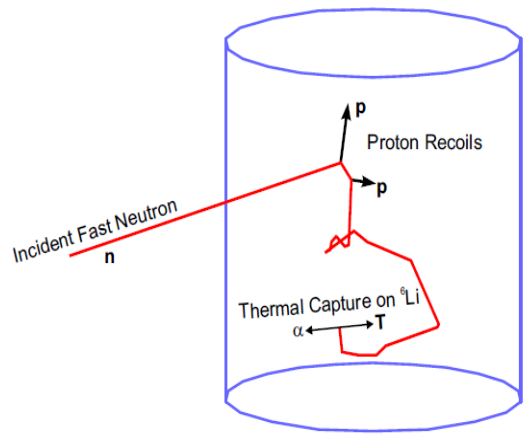

Fig. 3 An illustration of the principle of capture-gated detection. A fast neutron impinges on the detector rapidly gives up its energy (Fisher et al., 2011)

The use of enriched $^{6}$Li as the dopant has the advantage of a large Q-value and the production of two energetic charged-particles. Due to scintillation quenching, the light yield of the 2 MeV triton is nearly a factor of 10 higher than that of the 1.5 MeV alpha from neutron capture on $^{10}$B. The neutron capture signature from $^{6}$Li is well-separated from the noise and most background sources. The concern about energy leaving the scintillator does not exist, contrary to dopants that produce gamma rays. The capture cross section is high, but is less than for Gd and $^{10}$B.

1.4 Time of Flight (TOF) method

At non-relativistic velocities $v_n$, TOF (see Eq. 7 below) is the time required for a neutron to traverse a flight path of length x between source and detector:

$$TOF = \frac{x}{v_n} = x\sqrt{\frac{m_n}{2E_n}} \quad (7)$$

where $m_n$ and $E_n$ are the neutron mass and energy, respectively.

The arrival time of these neutrons at the detector depends upon their energy. Hence, by measuring the neutron arrival time and knowing their creation time (the time of the accelerator pulse), their energy can be determined.

The energetic resolution of this technique depends upon the pulse-width (1-2 ns), detector thickness, source-to-detector distance and times related to the response of the detector to the impinging neutron and to signal-processing electronic components.

The TOF technique allows good separation between gamma-rays and neutrons emitted from the source, since the TOF of gamma-rays is significantly shorter than that of the fastest neutron.

In a typical single-event-counting TOF measurement, the accelerator pulse acts as a start trigger for the time measurement by a Time-to-Amplitude Converter (TAC). Neutron detection signal from the detector acts as a stop for the time measurement by the TAC. The latter generates a signal of amplitude proportional to the elapsed time between the start and stop pulses mentioned above. In fact, the "start" pulses are periodic and have much higher repetition rate than the "stop" detector pulses; hence, it is common to reverse the sources of the "start" and "stop" pulses. The neutron detector pulse is used to start the TAC, while the accelerator burst pulse is delayed and becomes a "stop" pulse. With this arrangement every useful "start" pulse is accompanied by a "stop" pulse and the TAC operates at much lower rate set by detected neutrons, with a consequence of negligible TAC dead time.

Neutron arriving at the detector after the stop trigger will not be counted. Thus, in this mode of operation, only a single event will be counted for each accelerator beam-pulse.

If the probability to detect more than one neutron for each accelerator pulse is high, then the TAC will give preference to measuring faster neutrons which arrive sooner. This bias in favor of the faster neutrons will cause distortion in the energy spectrum. Reducing the detection probability for each accelerator pulse (by reducing beam intensity) to ca. 0.01, allows reduction of the spectral distortion to negligible levels. This requirement prohibits utilization of the event-counting-mode at high count rates.

Recent progress in the field of digital-signal processing has permitted counting more than one event per beam pulse. Such systems are capable of working at count-rates in the MHz range.

The weak point of this technique is the requirement of a start time for the neutron, which is unavailable in many applications, such as in neutron fields at a nuclear power plant.

Summary:

Table 1 presents the characteristics of neutron spectrometers discussed in previous sections, based on Brooks and Klein (Brooks and Klein, 2002).

**Table 1 Characteristics of neutron spectrometers (Brooks and Klein, 2002)**

| Method | Energy range (MeV) | Resolution % | Detection efficiency | High resolution imaging possible? |
|---|---|---|---|---|
| **Recoil telescope** | 1-250 | 4% at 60 MeV | 0.05% | No |
| **Organic scintillator** | 2-150 | 4% at 8 MeV | 20% | Yes with optical readout |
| **He$^3$ gas proportional counter detector** | 0.05-10 | 2% at 1 MeV | 0.3% | Yes with multiwire gas cells |
| **Capture gated detectors** | 1-20 | 50% at 5 MeV | 1% | No |
| **TOF method** | 1-15 | 5% at 2.5 MeV | Few percent with organic scintillator | Yes with optical readout |

One can observe that, to achieve high efficiency, good resolution imaging and good spectroscopy the most favorable method is TOF with organic plastic scintillator. Indeed, one of the detectors developed in this work –TRION is based precisely on this approach.

## II. Background

Detection of explosives concealed in air-cargo or passenger-baggage presents a considerable challenge which has not been fully met by currently-deployed X-ray inspection systems. These systems provide <u>only limited information</u> about specific cargo contents, such as their shape and density, and their performance capabilities <u>rely heavily on human operator skill</u>. Furthermore, only very limited differentiation amongst elements in the low atomic-number (Z) range can be achieved. The sole X-ray based explosives detection technique capable of automatic detection utilizes coherent x-ray scattering (Strecker et al., 1994). However, its penetration through items that are substantially more massive than carry-on baggage is severely limited by the low energy of the relevant X-rays (<100 keV).

For this and other related reasons, there has been notable interest in the use of neutrons for non-destructive detection of explosives and illicit drugs in the last twenty years. Up-to-date overviews of fast neutron contraband detection techniques can be found in (Buffler and Tickner, 2010; Buffler, 2004; Mor, 2006).

In this context, Fast-Neutron Resonance Radiography (FNRR) is one of the most promising methods for **<u>fully-automatic</u>** detection and identification of explosives concealed in passenger luggage and air-cargo.

The reasons are as follows: **a)** Neutron transmission depends only weakly on absorber Z (with the exception of hydrogen), allowing neutrons to penetrate voluminous objects and high-Z materials. In addition, **b)** neutrons probe the nuclear properties of the absorber and exhibit highly characteristic structure in the neutron-energy dependence of interaction cross-sections with different isotopes, particularly for low-Z substances. The knowledge of the total cross-section allows the measured transmission spectra to be unfolded, and the areal densities of elements present in the interrogated object to be determined. The method can thus reveal the presence and abundance of C, O, N and neighboring elements in relatively small sub-volumes of the interrogated object, thereby allowing discrimination between explosives and benign material. The latter is a crucial pre-requisite for ensuring that false alarm rates in the inspection system are kept at tolerably low levels.

Since the FNRR method provides the operational basis of the imaging system presented in this work, it will be described here in detail below.

## 2. Fast Neutron Resonance Radiography (FNRR) methods

FNRR is based on the fact that the absorption of a beam of neutrons varies with neutron energy for a given element. Typically, the total cross-section of each element displays a series of narrow peaks (resonances) and valleys as function of neutron energy.

In this section the different approaches to FNRR will be reviewed and the degree of fulfillment of required conditions will be examined for each variant of the method.

The techniques are divided into two categories: un-pulsed and pulsed FNRR.
In both the above applications, the precise knowledge of the neutron energy is a pre-requisite for resonance radiography. There are 3 approaches to achieve this goal:

- **Variable quasi-monoenergetic beam FNRR** (Chen and Lanza, 2002). In this approach, continuous neutron energy variability is achieved by viewing the neutron source from different angles.
- **Switchable quasi-monoenergetic beam FNRR** (Hamm et al., 1998; Tapper, 2000). In this approach, toggling between two discrete neutron energies is performed.
- **Pulsed broad-energy beam FNRR** (Dangendorf et al., 2004; Miller, 1997; Vartsky et al., 2005). In this approach, a pulsed (ns) broad-energy fast neutron beam is used in combination with time-of-flight (TOF) neutron spectroscopy, thus giving rise to Pulsed Fast Neutron Transmission Spectroscopy (PFNTS).

The first two approaches can utilize un-pulsed (DC or CW) neutron beams, whereas the third approach requires a ns-pulsed neutron beam.

2.1 Un-pulsed FNRR
2.1.1 Variable quasi-monoenergetic beam FNRR
This method was developed by the MIT-LLNL collaboration (Chen and Lanza, 2002; Raas et al., 2005). In this approach the neutrons are produced by the $D(d,n)^3He$ reaction with 2.3-2.5 MeV deuterons on a 4 atm. gas target. Due to the kinematics of the reaction the neutron energy varies continuously with emission angle as shown in Figure 4a. Thus the neutron energy is about 5.6 MeV and 2.0 MeV for emission angles of $0^o$ and $130^o$, respectively. For a small angular aperture at a given angle, the neutrons produced in this reaction are effectively mono-energetic, however, due to deuteron beam energy loss of about 600 keV in the gas target, the neutron spectrum has a kinematic spread which is angle dependent. Figure 4b shows the energy spread vs. neutron energy. It can be observed that the energy spread varies from about **10 keV to about 630 keV** over the energy range 2 - 5.6 MeV. The size of the gas target is about **1 cm diameter by 1 cm length** and the neutron flux at 1m distance from the target is about **$10^4$ n/s/cm$^2$** (Lanza and Demining, 2007).

The magnitude of the flux decreases rapidly with emission angle and at 90°, it drops down to about 10% of the 0° value (Marion and Fowler, 1963). Resonant radiography is performed by positioning the inspected object and detector at **10** selected angles that correspond to energies of specific resonances in C, O and N. Due to finite detector dimensions there is an energy variation across the detector and the angular aperture subtended by the detector must therefore be less than 10°. This restriction limits the dimensions of the inspected object.

The neutron detector employed in (Chen and Lanza, 2002; Raas et al., 2005) consists of a plastic scintillator slab viewed by a cooled, low-noise CCD camera. Since it was incapable of discriminating between neutron events and gamma-rays emitted from the source, the elemental contrast was significantly degraded and the experimentally-derived attenuation coefficients deviated substantially from expected data (5 – 50 %) (Blackburn et al., 2007).

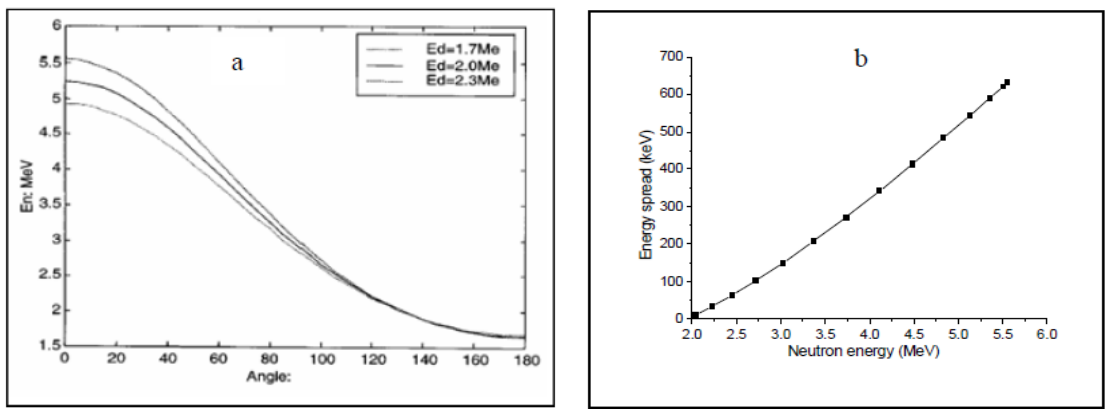

**Fig. 4 a) Neutron energy variation vs. emission angle and b) Energy spread vs. neutron energy (Chen and Lanza, 2002)**

Thus, a new LLNL source design was implemented using tungsten as the dominant source material (IAEA, 2006), in order to reduce the gamma-ray yield.

Recently, a different detection approach was proposed (Blackburn et al., 2007; Lanza and Demining, 2007), using 1-dimensional arrays of discrete liquid or plastic scintillators operating in counting mode with pulse shape discrimination for separation of neutrons from gamma-rays. Nevertheless, the problem caused by variation of neutron energy across the array apparently still persists. In addition, the spatial resolution dictated by detector elements is rather low, not better than 2 cm.

2.1.2 Switchable quasi-monoenergetic beam FNRR

This variant of the FNRR method (Hamm et al., 1998; Tapper, 2000) was proposed by the De-Beers company, South Africa, for detection of diamonds in kimberlite rock (a dense rock formation in South Africa in which it is well known that diamonds may be found). Due to commercial reasons, not much information about this system is available. A significant amount of development work on the neutron source, detectors and system simulation was performed in collaboration with the group of the University of Witwatersrand, SA (Ambrosi and Watterson, 2004; Guzek et al., 1999; Watterson and Ambrosi, 2003). In this application only the detection of carbon is of interest, therefore only two neutron energies, that correspond to the broad cross-section fluctuations around 6.8 MeV and 7.8 MeV have been considered.

The neutron beam is produced by switching the deuteron energy between 4 and 5 MeV using a 4 MeV RFQ accelerator followed by a second accelerating stage, which can either transmit the deuteron beam as is, or boost its energy to 5 MeV. Neutrons are produced by interaction of the deuteron beam with deuterium a gas target. Similarly to neutron beams produced in the previous system, the spread of the neutron energy is about **600 keV** for the windowed gas target (Ambrosi and Watterson, 2004). This large energy spread is of no consequence, since the dip and peak in the carbon cross-section at 6.8 and 7.8 MeV are both broad anyway.

The detection system consists of a few-centimeters thick scintillating fiber detector viewed either by a CCD camera or coupled to an amorphous Si plate (Tapper, 2000), as schematically illustrated in Figure 5.

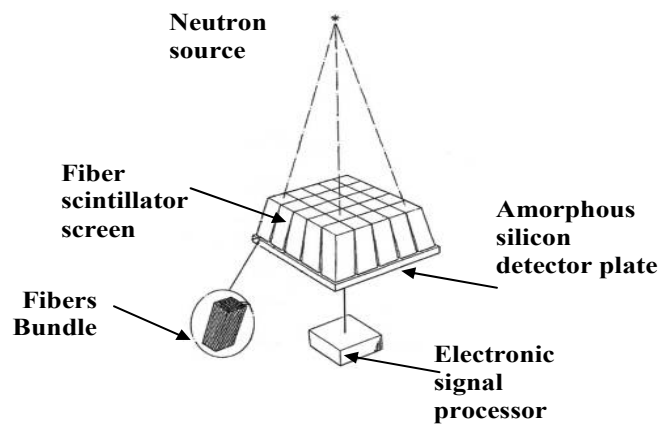

**Fig. 5 Fast neutron detection apparatus built by De-Beers (Tapper, 2000)**

2.2 Pulsed broad beam FNRR (PFNTS)

The Pulsed Fast Neutron Transmission Spectroscopy (PFNTS) method (Overley, 1985) is a variant of Fast-Neutron Resonance Radiography that requires an intense "white" spectrum of ns wide neutron bunches. These are usually produced via the $^9$Be(d,n) (see Figure 6 of (Micklich, 1995) or $^3$H(t,n)He and $^3$H(t,2n) reactions. In this method, a broad energy-spectrum fast-neutron beam (0.8 – 10 MeV) transmitted through an object is modified according to the resonant features present in the cross-sections of its constituent elements. Using the time-of-flight (TOF) technique, the spectrum of the transmitted neutrons is measured by a position-sensitive neutron detector and the attenuation at particular neutron energies, that corresponds to specific cross-section structures ("resonances") of light elements, such as C, N and O, is determined (Ambrosi and Watterson, 2004; Guzek et al., 1999; Watterson and Ambrosi, 2003).

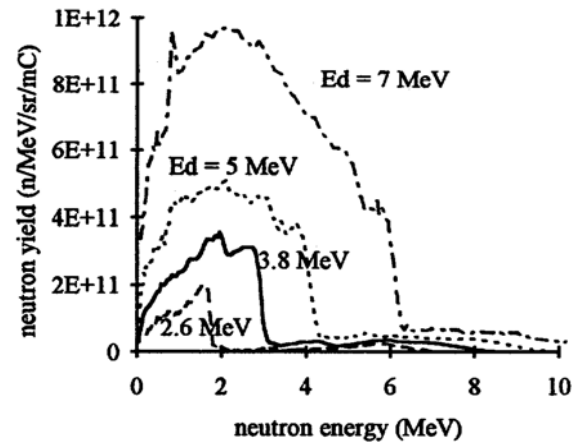

**Fig. 6 Neutron yield and energy spectrum at different deuteron energies for the Be(d,n) reaction (Micklich, 1995)**

The knowledge of the total cross-section allows the measured transmission spectra to be unfolded, and the areal densities of elements present in the interrogated object to be determined. The method reveals the presence of C, O, N, and other elements in the interrogated object, in a quantitative fashion.

Since the PFNTS method provides the operational basis of the imaging system presented in this work, it will be described here in detail.

2.2.1 The University of Oregon PFNTS system

Overley et al. (Overley, 1985, 1987) at the University of Oregon were the first to demonstrate that knowledge of the total cross-sections for H, C, N, O, and other elements of interest allows the measured transmission spectra to be unfolded, thus providing information on areal densities of elements present in the interrogated object. The combined effective contribution from other elements (which are of no interest in the present context) was taken together as that of a fictitious element X and assigned an energy-independent cross-section of 3 barns (Overley et al., 1997).

For every voxel in the inspected object, the energy-dependent transmitted neutrons are measured and used to determine relative amounts of the above mentioned elements.

Overley's group developed a TOF spectrometer composed of a 16-detector linear horizontal array with $3\times 3$ cm$^2$ pixels (Overley et al., 1997). The neutron spectrometer consisted of a 10 cm-diameter, 2.5 cm-thick liquid scintillator coupled to a fast photomultiplier (Overley, 1985)

. Timing signals derived from the deuteron beam pulse and from the photomultiplier were routed to a time-to-amplitude converter (TAC). Resulting pulses were digitized by an analog-to-digital converter (ADC) and stored by an online computer as 1024-channel neutron flight-time spectra. Proton recoil events of energies less than 0.3 MeV were rejected at the ADC. Overall time resolution was about 2 ns (Overley et al., 1997).

In 1996, a series of blind tests was conducted by the Federal Aviation Administration (FAA) at the University of Oregon (National Research Council (U.S.)., 1999; Overley et al., 1997), in order to evaluate the effectiveness of the PFNTS technique for explosive detection in real luggage. The tests involved 134 different luggage items and 8 different nitrogen-based explosives, which were interrogated using neutron time-of-flight (TOF) spectrometer configured as described above.

As a result of these tests, Overley et al. (National Research Council (U.S.)., 1999; Overley et al., 1997) reported detection probability of 93.3% (for all explosive types, after algorithm adjustments), 4% false alarm probability (for all explosive types after algorithm adjustments). The undetected

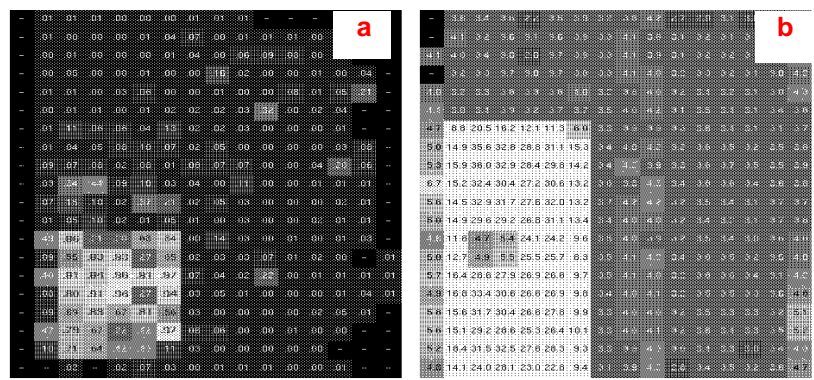

Fig. 7 Gray-scale maps obtained from one suitcase from Oregon's blind tests. a) Shows explosive likelihood (b-values), b) contains Z values (Lefevre et al., 1998)

explosives weighed 450 grams or less and were less than 1.2-cm in thickness (Overley et

al., 1997).

Figure 7 shows two gray-scale maps obtained from a suitcase in one of the Oregon blind test experiments (Lefevre et al., 1998). The first (a) shows explosive likelihood (b-values), where black is innocuous and white is explosive. A sheet of explosive is indicated in the lower left corner. The second grey-scale map (b) contains Z values. The large white rectangular area has Z values up to 40.

After several years, the University of Oregon discontinued its work on PFNTS.

2.2.2 The Tensor Technology Inc. system

During the 1990's, Tensor Technology also developed a TOF fast neutron spectrometer, assembled for identification of contraband in sealed containers.

The system was based on a two-dimensional 99-element array with pixel size 4×4×3cm$^3$ (National Research Council (U.S.)., 1999; Van Staagen et al., 1997; Tensor Technology, 2004), as seen in Figure 8. At that time, it was stated that reduction of pixel size would entail an increase in the quantity of electronics to an unmanageable level (Tensor Technology, 2003).

TOF measurements were performed using a Time-to-Digital Converter (TDC) (Van Staagen et al., 1997) which digitized the TOF measurements without recourse to any analog circuitry. The TDC method directly counts the number of clock cycles between the start and the stop signals with a fast clock, and then determines the position of the stop pulse within the clock pulse, using an interpolator. This method resulted in faster conversion times

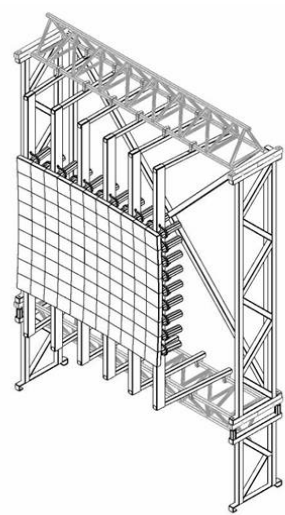

(compared to those obtainable with a time-to-analog converter (TAC) followed by an analog-to-digital converter (ADC)), as well as a significantly reduced the electronics complexity.

In 1997, 150 suitcases were scanned at Tensor Technology as part of an FAA blind test (National Research Council (U.S.)., 1999), aimed at evaluating explosives detection using the TOF spectrometer described above. The setup constructed by Tensor Technology (National

**Fig. 8** Two dimensional matrix of 99 individual scintillation detectors, each coupled to a light guide, photomultiplier and electronics. Pixel size 4×4×3 cm$^3$ (National Research Council (U.S.)., 1999; Van Staagen et al., 1997; Tensor Technology, 2004)

Research Council (U.S.)., 1999; Van Staagen et al., 1997; Tensor Technology, 2004) is comprised of a matrix of individual scintillation detectors positioned as a 2-dim array, as shown in Figure 8.

Detection and false alarm probabilities probability reported in these tests (National Research Council (U.S.)., 1999) were 88% and 24% respectively, for all explosive sizes and thicknesses studied. Detection probability for thin explosives (1.2 cm or less thick and weighing 450 gr or less, based on Overley (Overley et al., 1997), was 40% (National Research Council (U.S.)., 1999). Figure 9 illustrates neural net values obtained during Tensor's blind testing for a slurry sample in a suitcase (National Research Council (U.S.)., 1999).

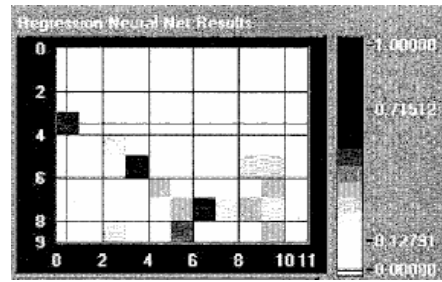

**Fig. 9 Neural net values obtained during Tensor blind testing (National Research Council (U.S.)., 1999)**

2.2.3 Summary of Oregon & Tensor PFNTS tests

It has been claimed by these groups that both series of blind tests provided good detection levels for bulk explosives (of thickness greater than 1.2 cm and weighing more than 450 gr (Overley et al., 1997). Thus the PFNTS method appears to provide a very robust set of measurements for explosives detection, in the sense that they lead to unequivocal quantitative assignment of C, N & O areal densities. Table 2 provides a summary of detection performance levels obtained in these tests.

**Table 2  Summary of detection performance level**

| Group | Explosive category | Detection probability $P_d$ | False alarm probability $P_f$ |
|---|---|---|---|
| Oregon University | All explosives | 93.3% | 4% |
| Tensor Tech. | All explosives | 88% | 24% |

However, the pixel size determined by these detectors (few centimeters) posed an intrinsic limitation on the position resolution, which did not permit reliable detection of objects that are smaller and thinner (<1cm thick).

Thus, a National Academy of Science Panel (NAS) advised in 1999 (National Research

Council (U.S.)., 1999) against building an operational airport prototype, since no compact, suitable neutron source was available at that time, nor was the detector spatial resolution adequate for reliable detection of sheet explosives (thinner than approx. 1.2 cm and weighing less than 450 gr (Overley et al., 1997). Nevertheless, it recommended developing existing and new technologies in the field of neutron sources and neutron detectors.

In order to respond to the above recommendation of the NAS panel in 1999 a next-generation PFNTS detector should be capable of providing mm-size spatial resolution with good TOF spectroscopy per pixel and the capability to work at high neutron fluxes. The following section describes a detector that attempts to meet the above requirements.

2.2.4 Soreq NRC TRION generation- I

TRION is a novel, fast-neutron imaging device based on time-gated optical readout. The concept was first proposed by Dangendorf et al. (Dangendorf et al., 2002) of PTB and subsequently developed at Soreq NRC and PTB. The first generation detector (Mor, 2006; Mor et al., 2009), developed and evaluated in several beam experiments, was capable of imaging only a single TOF-frame per beam burst. The basic design of the single-frame TRION detector is shown in Figure 10.

The detector is designed to detect fast-neutron pulses produced, for example, in the $^9$Be(d,n) reaction using a pulsed deuteron beam (~1-2 ns bursts, 1-2 MHz repetition rate).

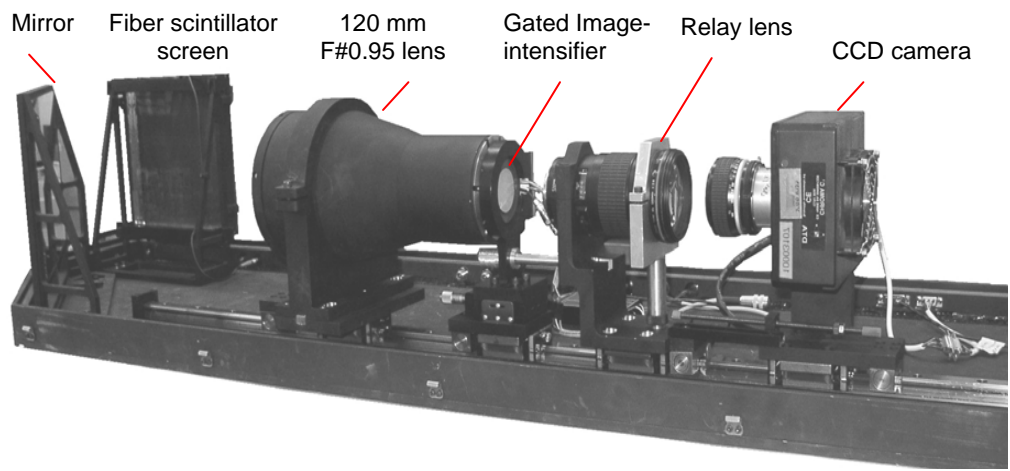

Fig. 10 TRION Gen. 1

After a specific Time-of-Flight (TOF) that depends on their energy, the fast-neutrons impinge on the plastic fiber scintillator, causing the emission of light from the screen via knock-on protons. The light is reflected by a front-coated bending mirror (98% reflectivity) positioned at an angle of 45° relative to the neutron beam direction, towards a large aperture 120 mm F#0.95 lens and subsequently focused on the image-intensifier. The latter not only amplifies the light intensity but, more importantly, acts as an electronic shutter that is opened for a gate period of Δt (as short as 8 ns) at a fixed, pre-selected TOF relative to each beam burst. Repetition rate for the beam pulses was up to 2 MHz and images were integrated by cooled CCD camera over many beam bursts, with acquisition times ranging from tens to several hundreds of seconds. Figure 11 provides a schematic illustration of TRION's imaging concept.

The system components are mounted on linear guides that can move freely along a precision rail. All system components are mounted in a light-tight enclosure.

With TRION Gen.1, TOF multi-frame imaging is performed sequentially, i.e., for each TOF frame, the intensifier is triggered to capture just one specific energy interval. This procedure is time consuming and inefficient in its use of the broad-energy neutron spectrum. Thus, in order to progress towards a real-time operational system, it is necessary to acquire images for several TOF frames (or corresponding pre-selected energy regions) simultaneously. This can be achieved by employing several ns-triggered intensified CCD cameras, such that each camera acquires a transmission image corresponding to a different energy region.

Although stroboscopic time-resolved optical imaging techniques have previously been used to determine various physical properties, no time-resolved neutron imaging has yet been performed.

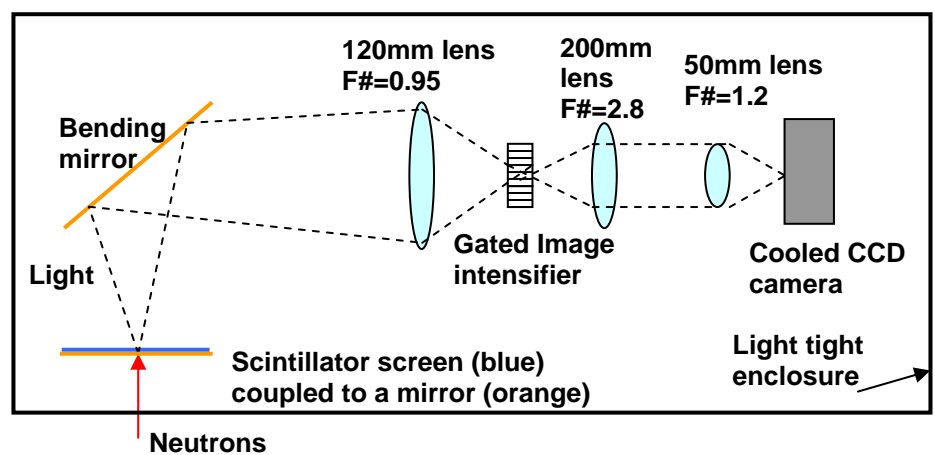

**Fig. 11 Schematic description of TRION concept (top view perspective)**

Figures 12 a & b show the fast-neutron radiograph and a photographic image of a phantom, respectively. The TRION detector was positioned 12 m away from the neutron source while the phantoms were ca. 30 cm from the detector. For the neutron image, the full spectral distribution of the neutron beam was utilized, excluding the gamma peak by setting an appropriate gate on the TOF spectrum.

The phantom consists of a plastic toy gun, a trumpet mouthpiece, two vials containing water and acetone mixtures (numerical markings on the vials indicate water volume

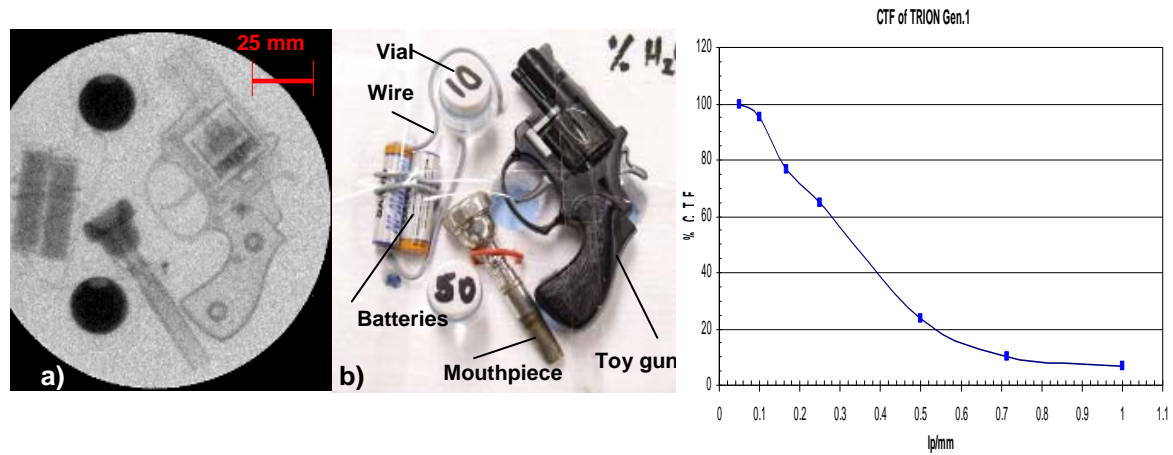

**Fig. 12 a) Radiography image of a phantom containing a plastic toy gun, a trumpet mouthpiece, vials containing water and acetone mixtures, batteries wrapped with electrical wire; b) A photographic image of the actual phantom; c) CTF curve of TRION Gen. 1**

percentage) and two regular AA batteries wrapped with electrical wire. As can be seen, image resolution is good enough to allow visual inspection, as with X- and γ-ray images. But, unlike the latter, which would only exhibit strong contrast for high-Z materials (such as the brass mouth-piece), the fast neutron image displays high contrast for low-Z materials as well. Figure 12c shows the Contrast-Transfer Function (CTF) curve for TRION Gen.1, displaying quantitatively image contrast for different spatial frequencies. As can be seen, image quality is suitable for visual inspection.

Fig. 13 shows the fast-neutron flux transmission vs. neutron TOF for a 10 cm block of graphite and 22 cm thick liquid nitrogen absorber. The characteristic spectral features of each element can be used for identifying and determining its elemental content.

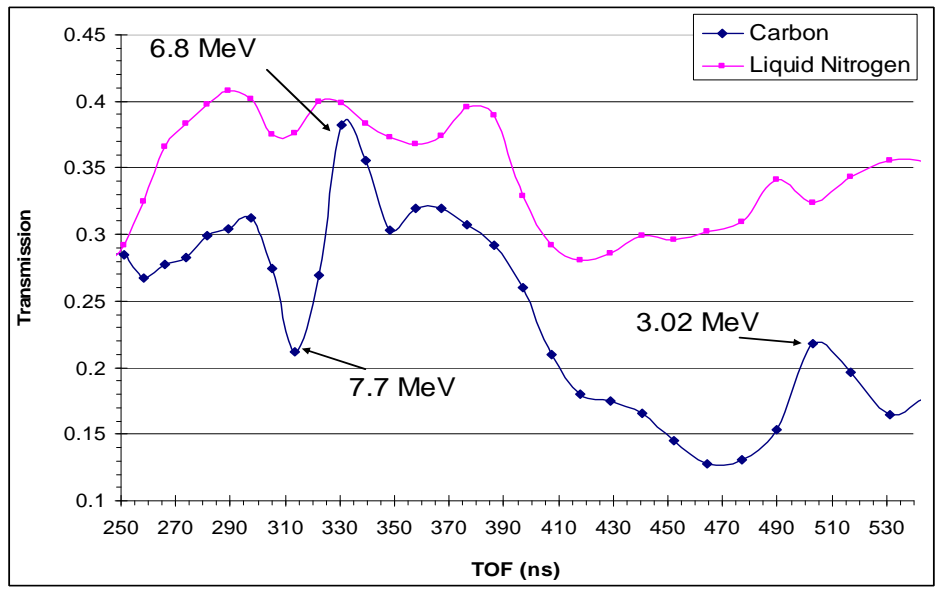

**Fig. 13 TOF spectra of d-Be fast neutrons transmitted through a block of graphite and liquid nitrogen, obtained with TRION Gen.1**

Fig. 14 shows an object consisting of melamine ($C_3H_6N_6$ which simulates here a commercial nitrogen-rich explosive), several carbon rods and a steel wrench, that was consecutively imaged at various neutron energies.

The top row of Fig. 14 shows 6 neutron images, marked with the corresponding TOF windows. With no previous information, they all look quite identical; however, with appropriate least-squares fit solutions, the net C and N distribution of the objects can be derived (Fig. 18, lower right). Also, as expected, the steel wrench disappears completely in this process.

In order to allow extraction of elemental information from a series of time-resolved radiographs, as seen in fig. 13, the neutron energies should be chosen such that images are taken where a significant difference in the cross-section for a single element exists. With nitrogenous material, for example, some of the energies of importance are: 3.6, 4.2, 4.9 and 7.7 MeV.

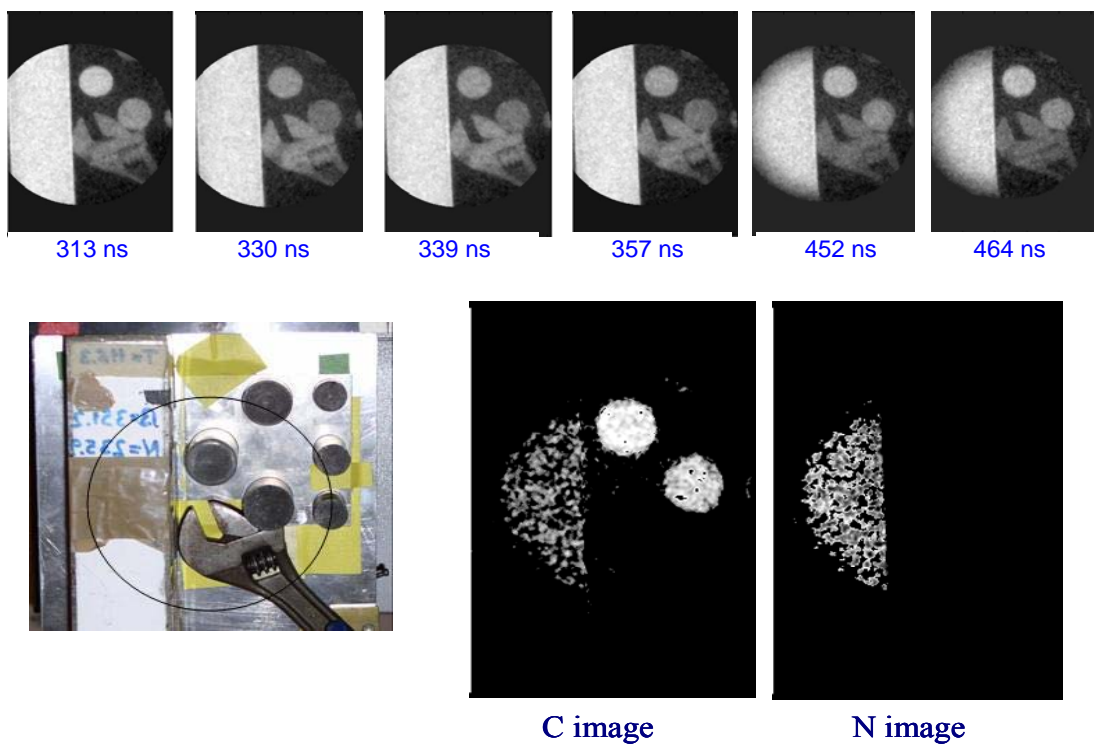

**Fig. 14 Example of resonance imaging using TOF for energy selection** The bottom left image shows a photo of the sample. The circle marks the region which was imaged by the neutron detector (ca. 10 cm diameter). The top row of shows 6 neutron images taken at different energies (indecated is the corresponding TOF below each image)

A more comprehensive description of TRION Gen. 1 can be found in (Mor, 2006; Mor et al., 2009).

## III. Spectroscopic neutron detectors developed in this work – part 1

As mentioned in the introduction, two spectrometric detectors were developed as part of this work: an integrative optical detector (TRION) and a fibrous optical detector.

Chapter 3 will present the 2$^{nd}$ generation of the integrative optical detector (TRION), its properties and improvements in comparison to the first generation of TRION.

## 3. TRION - Time Resolved Integrative Optical Neutron detector, 2$^{nd}$ generation

With TRION Gen.1, TOF multi-frame imaging is performed sequentially, i.e., for each TOF frame, the intensifier is triggered to capture just one specific energy interval. This procedure is time consuming and inefficient in its use of the broad-energy neutron spectrum. Thus, in order to progress towards a real-time operational system, it is necessary to acquire images for several TOF frames (or corresponding pre-selected energy regions) simultaneously. This can be achieved by employing several ns-triggered intensified CCD cameras, such that each camera acquires a transmission image corresponding to a different energy interval. This was accomplished with TRION Gen. 2 which contains 4 independently gated optical channels.

As illustrated by Figure 15, the front section of TRION Gen.2, comprising the fiber scintillator screen, mirror and F#0.95 collecting lens remained the same as in TRION Gen.1. However, the large-area gated image-intensifier was replaced by an ungated intensifier with a fast phosphor employed as an Optical Pre-Amplifier (OPA).

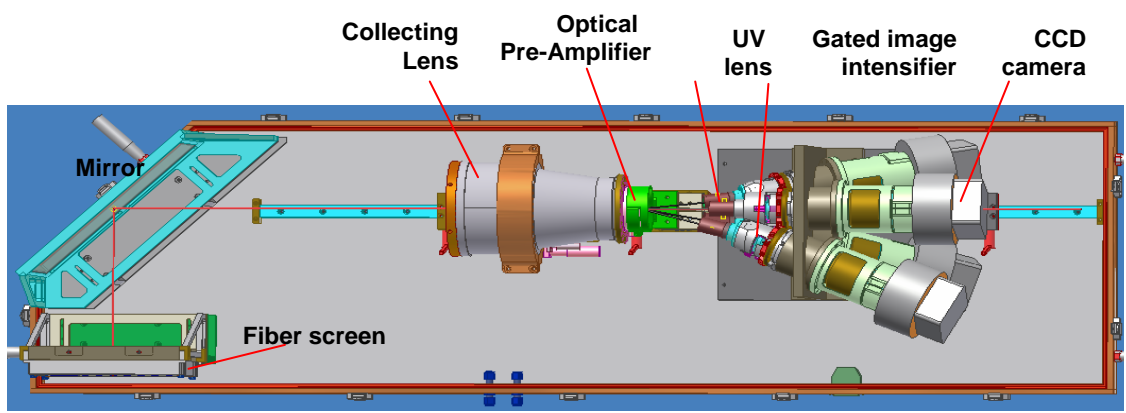

**Fig. 15 Schematic illustration of TRION Gen. 2**

As the light image from the OPA is emitted in the near-UV range, custom-designed UV lenses were employed to relay this light to gated image-intensifiers, incorporated in each of the 4 optical channels.

Figure 16 shows TRION Gen. 2, which comprises 4 optical channels.

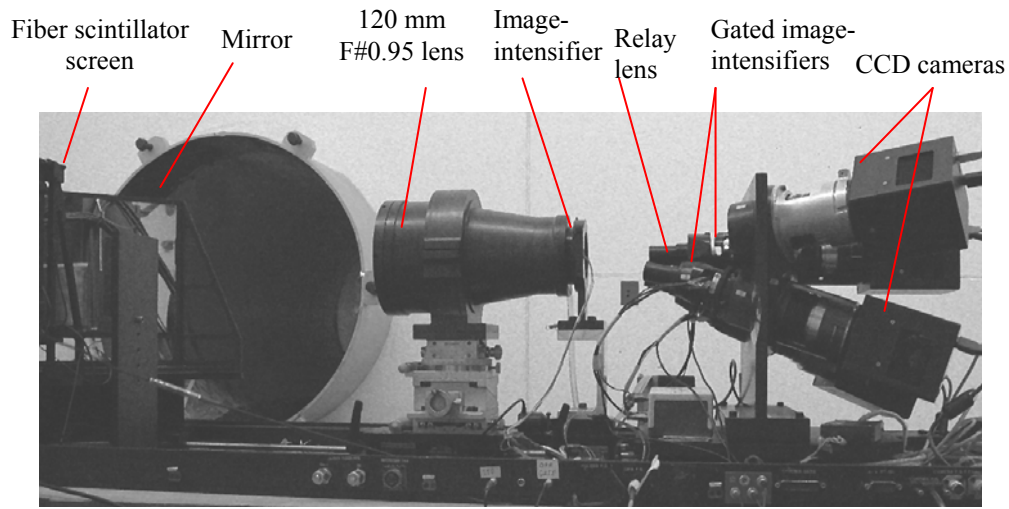

**Fig. 16 Actual view of TRION Gen. 2**

Figure 17 shows the rear section of TRION, containing the fully-equipped 4 optical channels.

The assembly contains a central optical channel aligned on the principal optical axis, and 3 other channels that view the phosphor screen of the OPA at various azimuthal orientations around the optical axis. Each optical channel contains a gated, small (18 mm diameter) image-intensifier (manufactured by Photonis-DEP (PHOTONIS, 2011) that captures the image at the OPA phosphor screen after a preselected neutron TOF and relays it, after further amplification, to a cooled CCD camera (ML0261E), manufactured by FLI (Finger Lakes Instrumentation, 2011).

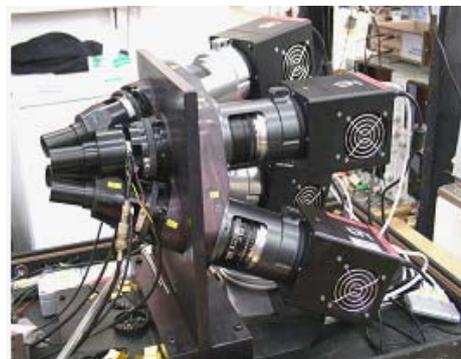

**Fig. 17 Side view of TRION's rear section containing all 4 optical channels**

3.1 TRION Gen. 2 components

TRION Gen. 2 incorporates improvement in three key aspects:
- Acquisition speed – Simultaneous acquisition of multiple energy windows, achieved by replacement of a single optical channel by four channels

- Reduction of thermal noise - Achieved by cooling the image intensifier
- Improving temporal resolution – Achieved by employing a different gating electronics and smaller diameter gated intensifiers

The following provides a more detailed description of each of TRION components and their relevant contribution to the above improvements.

3.1.1 Scintillator screen

A plastic fiber scintillator screen, shown in Figure 19, was used in this work as neutron detector. The most common fast-neutron detection scheme is based on conversion of the neutron into a proton, via neutron elastic-scattering by hydrogen present in the detector medium.

The fiber screen, BCF 99-55 (based on BCF-12 fiber), was manufactured by Saint-Gobain (formerly Bicron), USA (Saint-Gobain, 2011). Its light emission peaks at 435 nm (see Figure 18) with a decay time of 3.2 ns (Saint-Gobain, 2011). The screen surface area is 200×200 mm$^2$ and is composed of 30 mm long scintillating fibers, each consisting of a 0.5×0.5 mm$^2$ polystyrene core (refractive index-$n_1$=1.60), a 20 μm thick Poly-Methyl-MethAcrylate (PMMA) cladding (refractive index-$n_2$=1.49), and a 16 μm thick, TiO$_2$ doped, white polyurethane paint coating, which acts as an Extra Mural Absorber (EMA) to prevent light cross-talk.

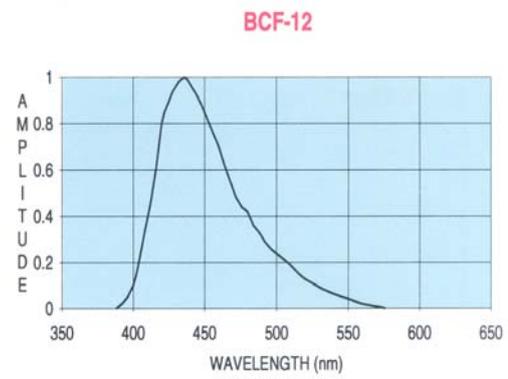

**Fig. 18 BCF 12 emission spectrum (Saint-Gobain, 2011)**

The fibers were first assembled in 10×10 mm$^2$ square bundles, which were then glued together to form the entire screen. The face of the fiber screen facing the incident neutron beam is covered by a backing mirror (of 98% reflectivity, made by Praezisions Glas & Optik GmbH (Praezisions Glas & Optik GmbH (PGO), 2011) with the reflective side facing the fiber scintillator. The backing mirror is not to be confused with the mirror on the optical readout side (see Figures. 15, 16).

This permits collecting also the scintillation photons that were originally emitted in the backward direction, i.e., away from the lens.

Figure 19a shows the scintillating fiber screen while figure 19b shows a magnified view of one of its sections. As can be seen, the screen is not flawlessly constructed, displaying voids and misaligned fiber rows, both of which contribute to the non-uniform light output. The latter is visible in Fig. 19c, which shows a full-transmission neutron radiography image (henceforth denoted "flat" image).

The local net sensitive area-fraction was evaluated for different regions of the screen and was found to range between 59 % and 72 %. The mean net sensitive area is 63 %.

The light output of a typical plastic scintillator based on polyvinyl-toluene (BC-400 or EJ-200) is about **10,000 photons** per MeV for a minimum ionizing particle. For a bare polystyrene based fiber-scintillator (BCF-12), the manufacturer quotes a light yield of **7000-8000 photons** per MeV for a minimum ionizing particle (Saint-Gobain, 2011).

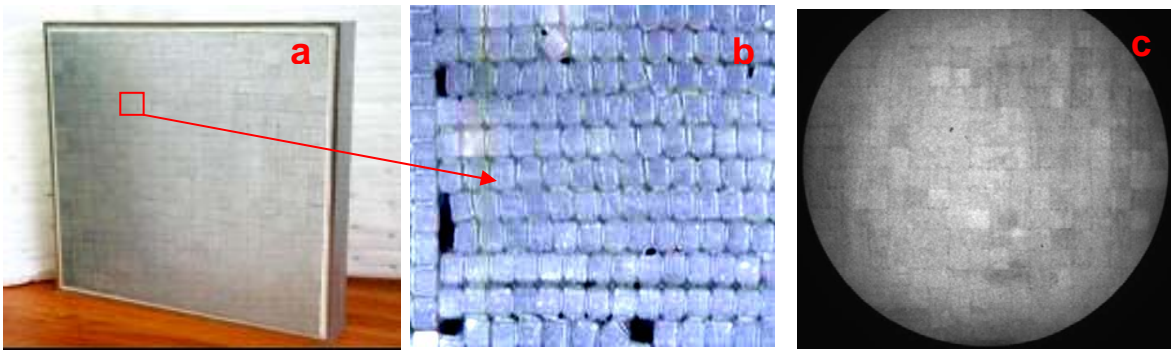

**Fig. 19 a) Fiber scintillator screen used in TRION, b) Enlarged photograph of a section from the screen, c) Flat image of the fiber screen**

The lower emission value compares unfavorably with inorganic scintillators such as NaI(Tl) or CsI(Tl), which emit 38000-54000 photons per MeV (Saint-Gobain, 2011) but have much longer decay times. In our fiber screen the light output is further reduced by the use of EMA (Saint-Gobain, 2011), as well as by dead-layers and imperfections (broken fibres, incomplete or destroyed cladding) which reduce the effective scintillator screen area.

The emission of light from the fiber was simulated using the ZEMAX optical design program (ZEMAX, 2011). A source was placed at three positions (both extremities and center of fiber) along the central axis of a scintillating fiber. The simulation indicates that the light is emitted into a shape shown in Figure 20 and the light trapping efficiency along the fiber is 4.5 % in each direction.

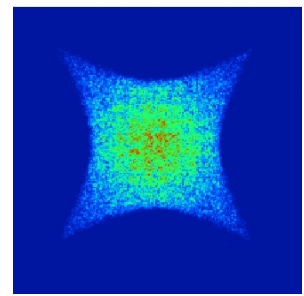

**Fig. 20 Shape of the light beam emitted from the fiber. Hottest color (red) indicates highest intensity**

3.1.2 Bending mirror

In order to protect the electro-optical components (CCD and image-intensifier) from radiation damage, they must be positioned out of the direct neutron-beam. To this end, a bending mirror is installed at an orientation of 45° facing the scintillator screen in order to deflect the scintillation light toward the F#=0.95 collecting lens (see Figure 15).

This mirror (3.3 mm thick), also manufactured by PGO (Praezisions Glas & Optik GmbH (PGO), 2011), is composed of a borosilicate substrate, front-coated by a multi-layer dielectric thin-film. It exhibits reflectivity of >99% at wavelengths up to 700 nm (see Figure 21), mechanical robustness and high temperature stability.

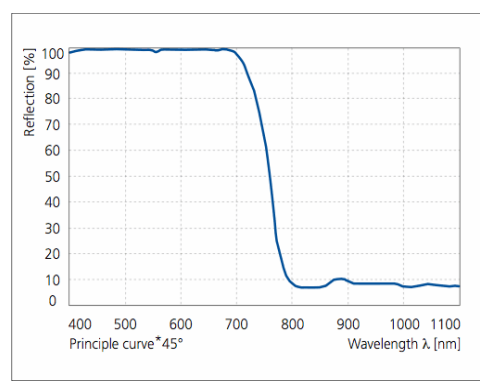

**Fig. 21 Reflectivity curve as function of wavelength (Praezisions Glas & Optik GmbH (PGO), 2011)**

3.1.3 Large-aperture collecting lens

The amount of scintillation-light photons created in a single fiber of the screen per detected 7.5 MeV neutron is rather small, i.e., 5424 photons on average (Mor et al., 2009). It is thus important to collect this light on an image intensifier as efficiently as possible. This is done with the aid of a large-aperture lens, 120 mm focal-length with a relative aperture F# = 0.95.

Since such a collecting lens was not commercially available, it had to be custom-designed and manufactured for this application. A schematic view of this lens is shown in Figure 22.

The lens consists of 5 elements, 3 positive and 2 negative. Its effective focal length is 120 mm, the relative aperture F# is 0.95 and its entrance pupil diameter is 126 mm.

The lens elements were manufactured by A. Optical Components Ltd., Azur, Israel, and the lens was assembled at Soreq NRC. Figures 23 show engineering illustrations of the collecting lens with the image-intensifier coupled at its rear-end (right hand side in the Figure).

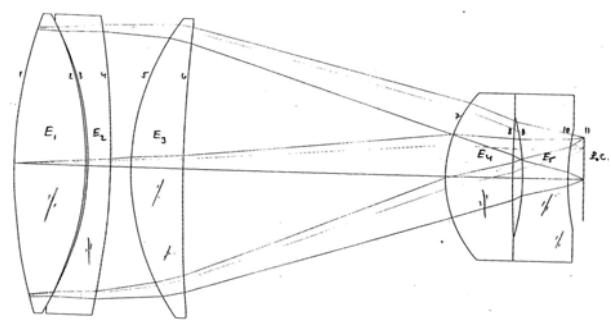

**Fig. 22 Schematic drawing of the collecting lens**

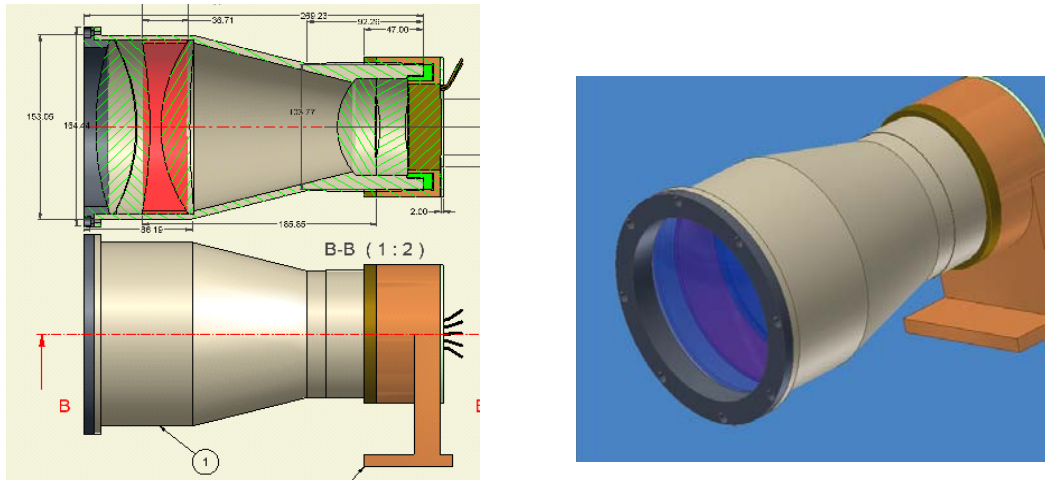

**Fig. 23 left) Side view of the F#=0.95 lens. Top: cross-section view of the collecting lens and image-intensifier coupling. Bottom: outside view of the lens and intensifier holder (brown), right) General view of the lens**

The fraction of light collected by the lens (positioned at distance of 750 mm and exhibiting ~90% light transmission) out of total light emitted from the fiber was also determined by the ZEMAX simulation mentioned in section 3.1.1 and found to be **1.3%.**

3.1.4 Optical Amplifier (OPA)

To achieve high photo-detection yield a special, large area (40 mm diameter) image intensifier, employed as an Optical Pre-Amplifier (OPA) was proposed and jointly manufactured by two companies (Photek Ltd, U.K, (Photek Ltd., 2011) and El-Mul, Israel (El-Mul, 2010). It is of crucial importance that the glow curve of the optical preamplifier be as fast as that of the scintillator screen (of the order of 2 ns FWHM) to preserve the good TOF resolution of the imaging system. While all components of a Multi-Channel

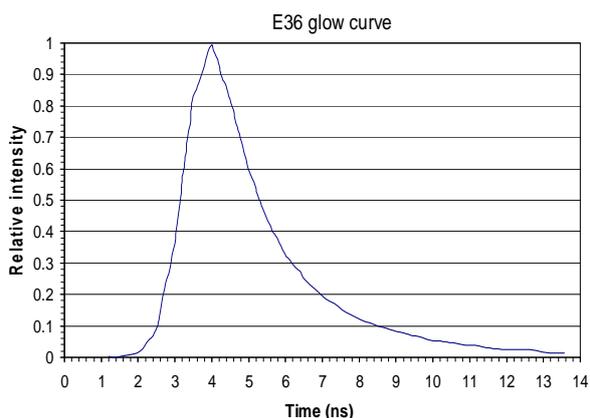

**Fig. 24 Glow curve of the ultra-fast phosphor E36 manufactured by El-Mul (Dangendorf et al., 2009)**

Plate-based (MCP), proximity-focus image-intensifier are very fast (sub ns), the sole critical aspect is the decay time of the phosphor screen. All standard, present-day image-intensifier phosphors exhibit glow curves with primary decay times (not including various long after-glow components) ranging from 100-nanoseconds up to tens of milliseconds (Photek Ltd., 2011). However, El-Mul Technologies Ltd. commercially offers a fast phosphor screen referred to as E36 for open multi-channel plate systems used in TOF electron microscopy. Figure 24 (Dangendorf et al., 2009) shows the E36 glow curve. The light decay constant of the phosphor screen is 2.4 ns. The disadvantage of the E36 is its relatively low light output.

The specs of the OPA are (Photek Ltd., 2008):
Diameter: 40 mm; photocathode type- Bialkali; QE @ 420 nm is 18 %; quartz entrance window; phosphor screen type (El-Mul) E36; electron gain- $5 \times 10^4$ e/e; phosphor emission range is 370 – 410 nm with maximum at 394 nm; borosilicate output window.

3.1.4.1 Improvement of signal to noise ratio

Due to the relatively low intensity of the PTB neutron beam, long exposure times were required. Consequently, the images acquired contained a non-negligible fraction of thermal noise from the photocathode. In order to reduce this noise the photocathode window had to be cooled. The cooling was achieved by blowing dry cold air on the photocathode window. Figure 25a depicts a schematic drawing of the plastic cooling ring containing 10 air nozzles, while Figure 25b shows the OPA and the large-aperture lens with the cooling ring mounted on its rear end. During operation, the OPA is in close proximity to the ring. Cooling tests showed that the thermal noise was reduced by a factor of 24 to insignificant levels, yielding a signal to noise ratio of 200 (Mor et al., 2009).

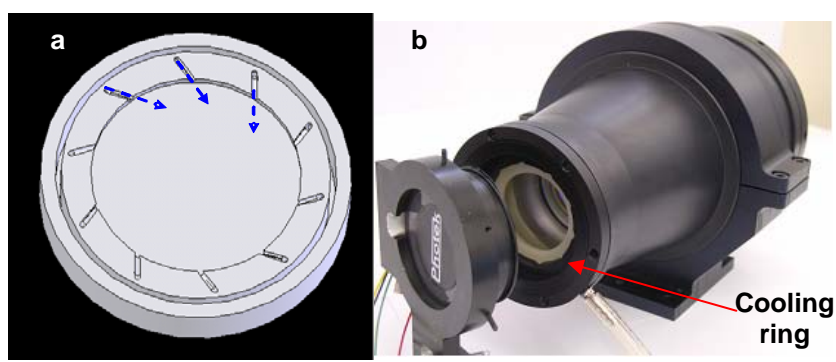

**Fig. 25 a) A schematic illustration of the cooling ring with the direction of air flow marked by blue arrows.  b) The OPA (seen on left) and the large-aperture lens. The cooling ring is mounted at the rear end of the lens**

3.1.5 UV relay lens

As the OPA phosphor screen emits in the near-UV range (394 nm), a UV lens for each of the 4 optical channels was custom-designed and manufactured. This lens is highly efficient (about 85% transmission) at 394 nm; it has focal length of 38 mm and F#3.8. Figure 26 shows a schematic diagram of such a UV lens, which consists of 3 elements mounted within a special housing. The light enters from the right-side towards the left-side of Figure 26.

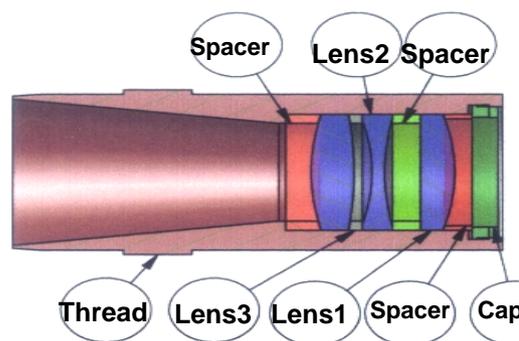

**Fig. 26 Custom-designed UV lens**

3.1.6 Gated image-intensifiers

For each optical channel, the image from the fast phosphor of the OPA is focused by the UV lens on the photocathode of a gated image-intensifier (I-I). The gated intensifier is an 18 mm diameter, type XX1450AAF produced by Photonis-DEP, Holland (PHOTONIS, 2011). To serve as electrical contact, a highly transmissive low-resistance mesh is positioned between the photocathode (S20-UV) and the entrance window of the I-I. This, in principle, allows gating with pulses as short as 1 ns. With our gating electronics (see section 3.1.6.3, below) and this intensifier we have achieved a gating speed of 4.6 ns at a repetition rate of 2 MHz. The specs of the gated intensifiers are:

Cathode type- S20-UV, Quantum Efficiency (QE) at a wavelength of $\lambda$ = 394 nm is 15 %, phosphor screen type is P43 with fiber optic output window and the photon gain at $\lambda$ = 400 nm is $4 \times 10^3$ emitted photons per incident photon. The tube is equipped with a single MCP.

3.1.6.1 Defocusing correction

Each of the optical channels views the OPA output screen from a different orientation relative to the principal optical axis. Only the central channel is positioned on this axis, while the optical axes of the other channels are positioned at an angle of 17.5° relative to it.

In the central optical channel, oriented along the principal optical axis, the image planes of the OPA phosphor, the UV lens and the photocathode of the 18 mm I-I are all parallel (see Figure 27a). In contrast, for the other channels, the image plane of the OPA and those of the UV lenses are tilted at 17.5º to each other. In such a configuration, certain parts of the viewed object (the phosphor of the OPA) will be outside the focal plane of the UV lens, resulting in a partially-defocused and distorted image. This problem can be alleviated (to a large extent) by tilting the image plane (photocathode of gated I-I) using the Scheimpflug principle (Merklinger, 1993). This principle states that the image plane is rendered sharp when the three planes (object, lens and image) intersect along one line. Hence, if the angle between the object plane and the lens plane is α, the image plane should be tilted relatively to the lens plane by an angle β, such that:

$$\tan(\beta) = m \times \tan(\alpha) \qquad (8)$$

Where m is lens magnification. In our case α is 17.5º and m=0.45, so the image intensifier should be positioned at 7.8º relative to the lens plane, as seen in Figure 27b.

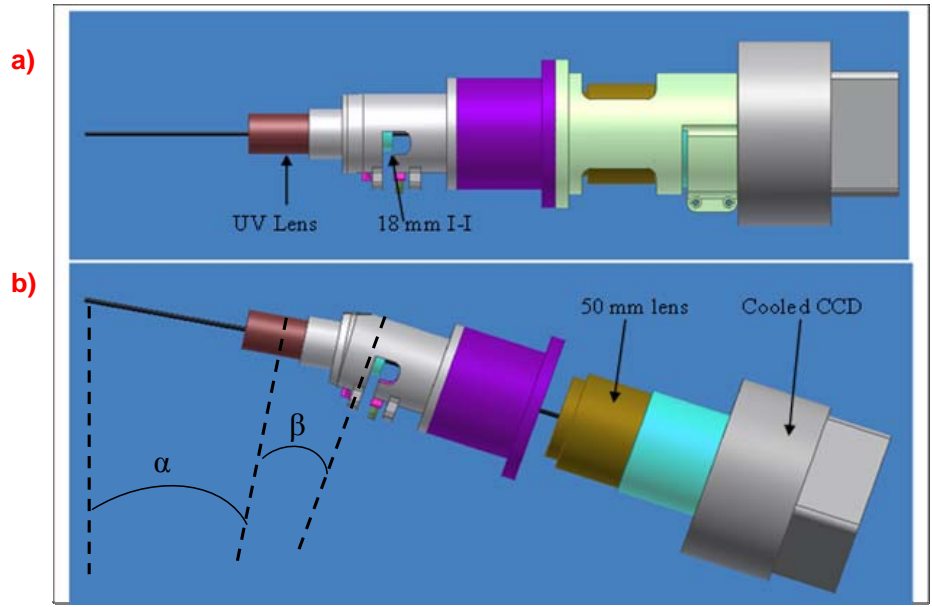

Fig. 27 a) Central and b) off-axis optical channel

3.1.6.2 Off-axis images correction (Image registration)

Image registration is the process of aligning two or more images of the same scene. Typically, one image, denoted the base image or reference image, is considered the reference to which the other images, denoted input images, are compared. The object of image registration is to bring the input image into alignment with the base image by applying a spatial transformation to the input image.

A spatial transformation maps locations in one image to new locations in another image.

The key to the image registration process, is determining the parameters of the spatial transformation that are needed to bring the images into alignment.

The MATLAB (MathWorks, 2011) Image Processing Toolbox software provides tools to support point mapping to determine the parameters of the transformation required to bring an image into alignment with another image. In point mapping, one picks points in a pair of images that identify the same feature or landmark in the images. Then, a spatial mapping is inferred from the positions of these fiducial points.

In order to facilitate the process of point mapping a transparent mask with a grid pattern printed on it, as seen in Figure 28 was imaged by each of the 4 optical channels. The mask was attached to a luminescent screen which acted as a light source, positioned instead of the scintillator screen. For point mapping, the image from the central optical channel was used as the reference image for all the other 3 channels, resulting in a spatial map for each channel to be used later for automatic alignment of experimental images.

Figure 28 shows images of a grid mask taken by the different optical channels. Figure 28(0) shows an image taken by the central channel located on the principal optical axis and Figure 28(1-3) show the different off-axis views. The fact that the off-axis optical channels view the OPA screen at an angle of $17.5^o$ relative to the principal optical axis results in a distorted image. As described above, with an alignment program that incorporates MATLAB (MathWorks, 2011) built-in image-registration functions, the off-axis images 1-3 were aligned with the on-axis image 0. Figure 28(1a – 3a) show the off-axis images after alignment. Pixel-to-pixel correspondence among images acquired at different energies is a pre-requisite for deriving elemental distributions via pixel-by-pixel arithmetic.

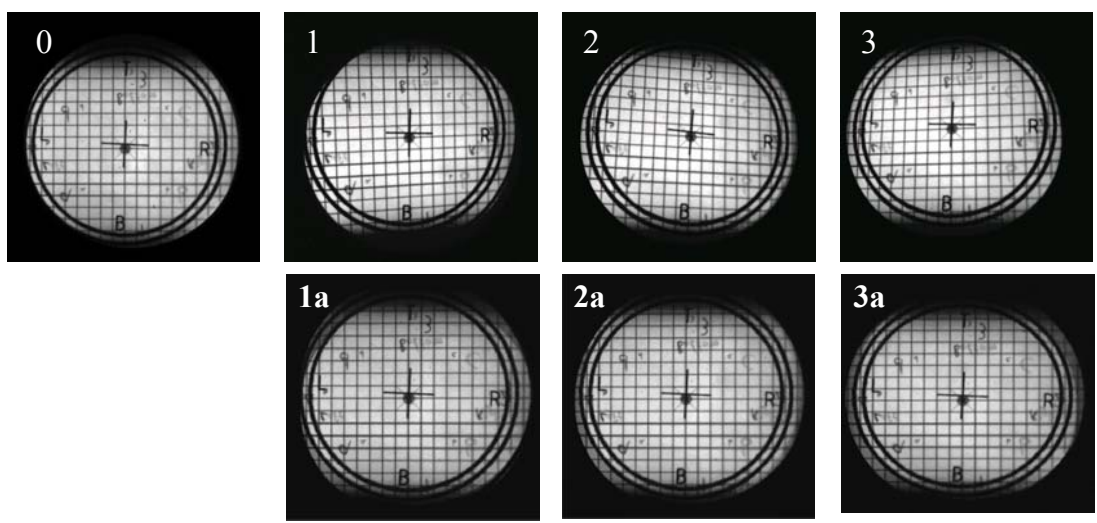

**Fig. 28 0) Central camera on optical axis; 1-3) Off-axis views before correction; 1a-3a) Off-axis images after correction**

3.1.6.3 Gating of the intensifiers

The gate pulsers for the four intensifiers were developed at PTB-Braunschweig, Germany. They are based on an IXYS (IXYS RF, 2010) fast transistor switch and a self-matched clipping-line, as schematically shown on Figure 29.

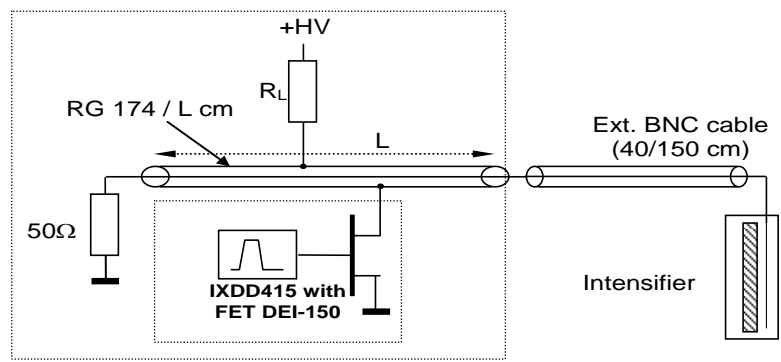

**Fig. 29 IXYS pulser with self-terminating clipping line**

The advantage of this concept is its proper (impedance-matched) termination at the end of the clipping line. It permits positioning the pulsers at some distance from the image-intensifier. This is important for densely-packed multiple-intensifier configurations, for which it is preferable to keep the electronics outside the optical enclosure.

However, the use of this pulser presented several disadvantages compared to the programmable gate unit used for TRION Gen.1 (Mor et al., 2009):

- Due to the use of the clipping line, the pulse width Δt is not programmable but fixed by cable length L (Δt = L/20 [ns], L in cm)
- At high pulsing frequencies the baseline drifts slightly. This calls for rather small values of $R_L$, which results in power consumption of about 50 W per gate module at the maximum gating speed of 2 MHz.

During the evaluation of TRION Gen.1, in which ROENTDEK pulsers (RoentDek, 2011) were used, gate-widths no shorter than 10 ns could be realized. With the new pulsers we have been able to achieve gate-widths as short as 4.2 ns (FWHM) and repetition rates of 2 MHz. The gating of the I-I requires switching the photocathode potential relatively to MCP-IN between +30V to -170V, as illustrated in Figure 30.

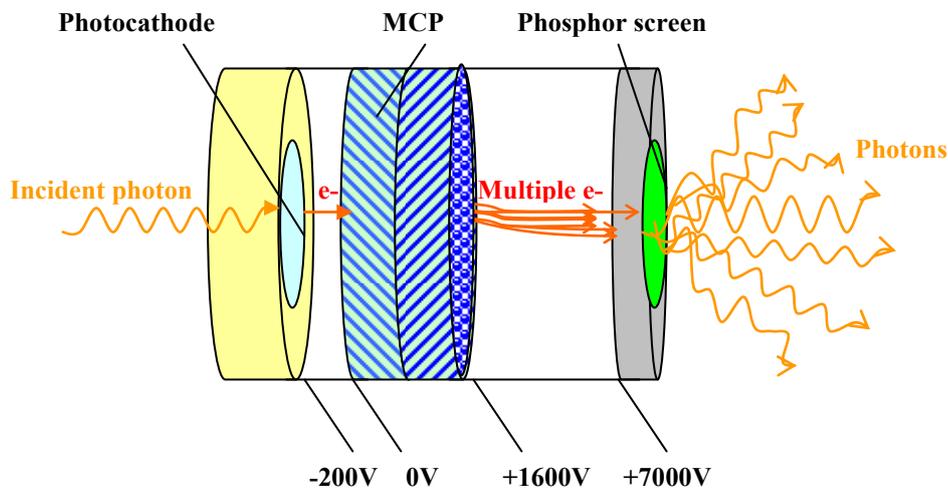

**Fig. 30 Schematic description of an image intensifier (not to scale). An incident light photon striking the photocathode causes the release of an electron via the photoelectric effect. This electron is accelerated towards the MCP, where it is multiplied and accelerated again toward the phosphor screen. These electrons strike the phosphor screen, causing the emission of fluorescence photons**

3.1.7 CCD cameras

The CCD camera used in TRION Gen.2, ML0261E, was manufactured by Finger Lake Instrumentation (FLI) (Finger Lakes Instrumentation, 2011), USA. The CCD sensor is a KAF 0261E, 512×512 pixels, pixel size 20×20 $\mu m^2$. All four cameras were controlled simultaneously via the computer using a custom-made program with integrated FLI drivers.

3.2 Properties of TRION Gen. 2

The following provides an analysis of the parameters which affect TRION's performance.

3.2.1 Spatial resolution

As demonstrated in my M.Sc. work (Mor, 2006) and in a recent publication (Mor et al., 2011) each of TRION's components (scintillating screen, optical system, electro-optical components) influences the spatial resolution obtained with the detector. Image quality was quantitatively assessed in terms of the Contrast Transfer Function (CTF). CTF provides a measure of how well the detection system reproduces the real contrast of the radiographed object at different spatial frequencies.

The CTF was determined using a bar-patterned steel or tungsten mask containing a series of slits with increasing spatial frequency, as illustrated in Figure 31.

CTF is defined as the following ratio and is measured as function of spatial-frequency, commonly described in units of line-pairs/mm (lp/mm):

$$CTF\,(\%) = \frac{[(C_{dk} - C_{bt})/(C_{dk} + C_{bt})]_{obs}}{[(C_{dk} - C_{bt})/(C_{dk} + C_{bt})]_{exp}} \times 100 \qquad (9)$$

Where $C_{dk}$ and $C_{bt}$ are the grey-level values of dark and bright regions of the picture, respectively, "*obs*" and "*exp*" refer to observed and expected values, respectively.

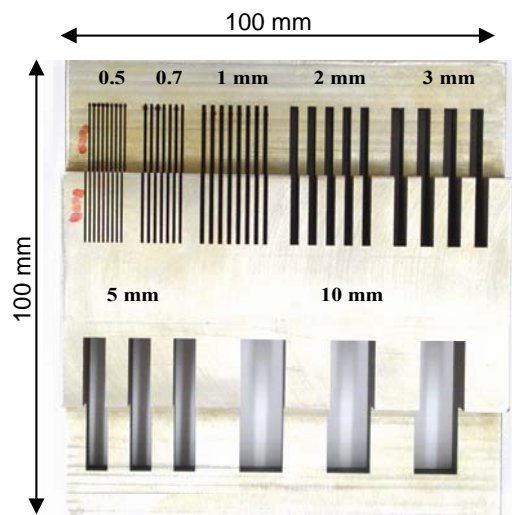

Fig. 31 Mask for CTF calculation. The numbers indicate the width of each line in the corresponding series below

Figure 32 provides a CTF comparison between TRION Gen. 1 and 2. As can be seen, TRION Gen. 2 presents comparable CTF to the one obtained with TRION Gen. 1. However, the CTF of TRION Gen. 2 extends only down to 0.5 lp/mm (1 mm wide lines, see Figure 31). This is to be expected as TRION Gen. 2 differs from the first generation by a smaller number of CCD pixels (512×512 vs. 1536×1024 pixels) and larger pixel dimensions (20×20 vs. 9×9 μm$^2$) as well as an additional image-intensifier added to the optical chain, further adding to the loss of contrast caused by the preceding OPA.

As 1 mm spatial resolution is adequate for visual inspection and the fact that reduction of the number of CCD pixels reduces CCD camera cost, it was decided to perform the above mentioned changes.

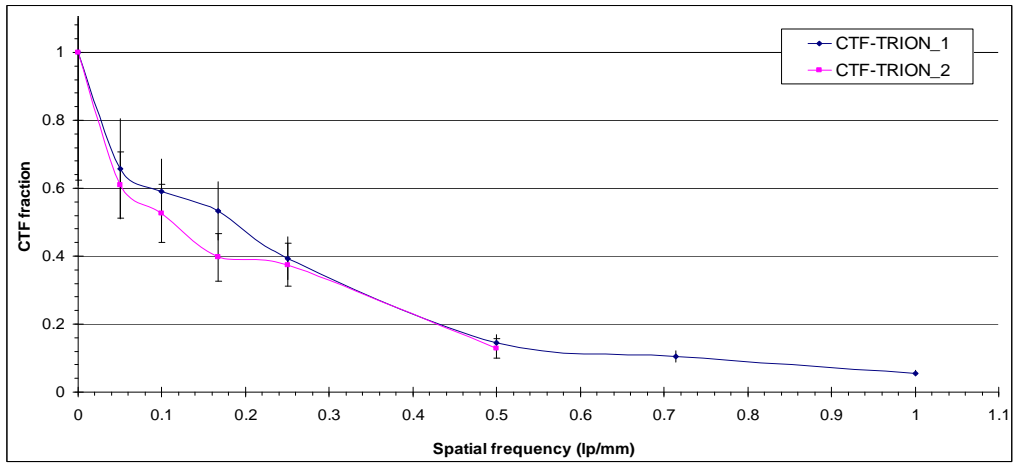

**Fig. 32 CTF comparison between the two generations of TRION**

Table 3 provides a summary of the key differences between the two generations of TRION in terms of components which affect spatial-resolution.

**Table 3 key differences between the two generations of TRION in terms of components which affect spatial-resolution**

|  | # CCD pixels | CCD Pixel size (µm$^2$) | Electro-optical chain |
|---|---|---|---|
| **TRION Gen. 1** | 1536 × 1024 | 9×9 | Singe image-intensifier |
| **TRION Gen. 2** | 512 × 512 | 20×20 | Two image-intensifiers |

3.2.1.1 Processes in the scintillating fiber screen

The spatial-resolution of a scintillating fiber screen is affected by:

- Range of the recoil proton generated by the incident neutron
- The fiber cross-sectional dimensions
- Multiple scattering of neutrons within the screen
- Light cross-talk between fibers

The fiber diameter determines the inherent image granularity, provided that the granularity introduced by the following system elements such as the image-intensifier and the CCD is of much smaller dimensions. The recoil proton generated by an incident fast neutron in the scintillator may cover a distance of several hundred microns (for example: a 3.75 MeV proton from a 7.5 MeV n,p collision) will travel 203 µm (Janni, 1982) before coming to rest within the scintillator and reach an average radial distance from the interaction point of ca. 140 µm. Thus, the best achievable spatial-resolution is of the order of the average radial proton range within the scintillator for a given neutron energy. For neutron at energies of up to 14 MeV, a 500-700 μm fiber presents a suitable compromise between required spatial-resolution and the significantly increasing costs of fibers of smaller diameters.

Multiple neutron scattering, charged particle and light cross-talk between fibers will reduce the contrast. The cross-talk of light in our screens is prevented by coating the fibers with EMA; however, the neutron scattering within the screen cannot be suppressed. In order to determine the effects of scattered neutrons and proton cross-talk, I have calculated by Monte-Carlo the processes which take place in the fiber screen. A detailed calculation of the Point-Spread Function (PSF) in the scintillating screen has been performed using the GEANT 3.21 code (CERN, 2003). The simulated setup consisted of a 200×200×30 mm$^3$ fiber screen, irradiated at 5 different neutron energies (2, 4, 7.5, 10 and 14 MeV) by a mono-energetic, infinitesimally thin pencil-beam of neutrons impinging upon the axis of the central fiber.

Figure 29 (left) shows a schematic configuration of 9 fibers located in the center of the screen and a magnified view (right) of the central fiber (0.5×0.5 mm$^2$ polystyrene core, 20 μm thick PolyMethyl-MethAcrylate (PMMA) cladding and 16 μm thick EMA paint).

The simulation calculates the energy deposited by protons and amount of light created in the fibers, taking into account the non-linear light yield as function of proton energy. For tracking and tallying purposes, light created by knock-on protons created in the central fiber by incident neutrons is termed the "primary signal". Light created outside the central fiber by protons escaping it is defined as a "secondary signal" and signal created by neutrons scattered within the screen is defined as a "tertiary signal".

Only the primary signal carries the correct radiographic information. The other signals will cause deterioration in spatial resolution and loss of contrast.

For a given neutron energy, the secondary signal depends on the proton range relative to fiber dimensions and the tertiary signal depends on neutron beam dimensions and screen geometry. Since the tertiary protons are created by neutrons scattered within the screen, their contribution will vary with position across the screen.

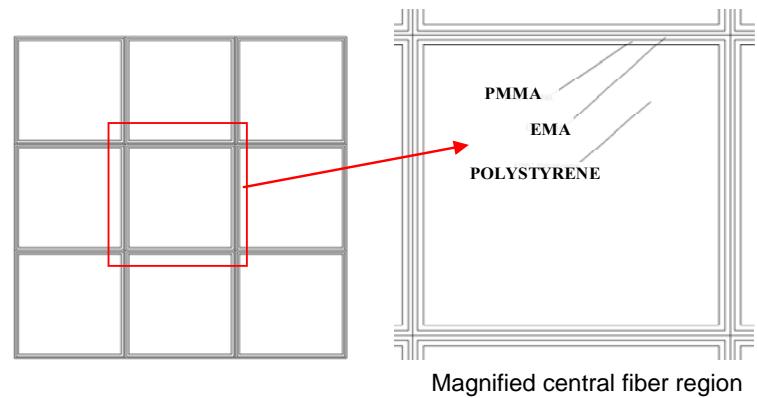

Fig. 33 Schematic of 9-fiber array (left) and magnified view of central fiber (right)

Figure 34a shows the PSF obtained with 7.5 MeV neutrons and Figure 34b the profile of primary, secondary and tertiary signal contributions to the PSF. As can be observed, the primary and secondary profiles are quite narrow; The Full Width Half Maximum

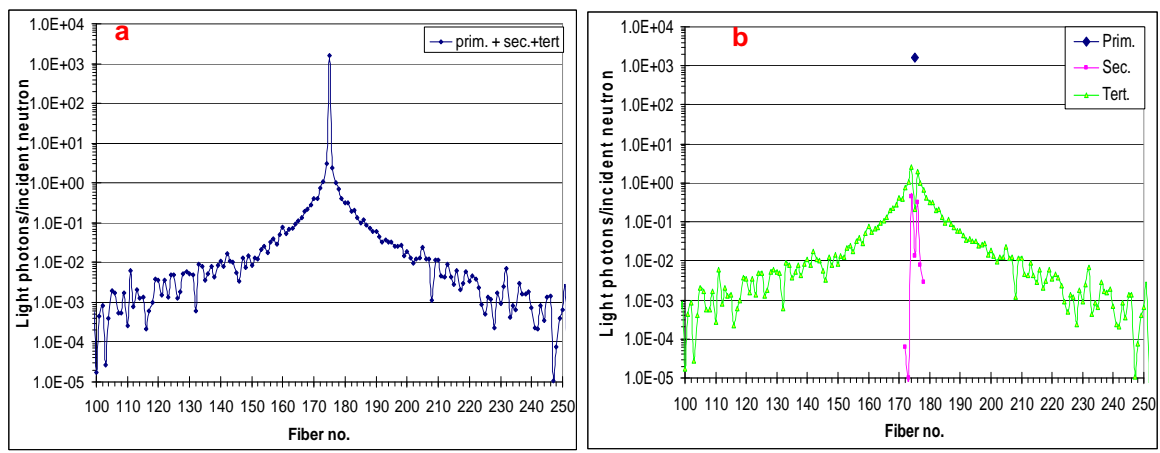

Fig. 34 a) neutron induced total PSF and b) contribution of each component to the total PSF

(FWHM) of the total PSF (figure 34a) is 1 fiber, which corresponds to 0.572 mm (including cladding and EMA). However, the tertiary signal distribution adds very broad wings to the PSF. The contribution of the tertiary signal over the entire detector area to the total signal amounts to 14%. These wings in the PSF will cause a significant loss of contrast in transmission images as they add up to a significant baseline under the primary signal. Figure 35 shows the dependence of the total neutron induced PSF on incident neutron energy.

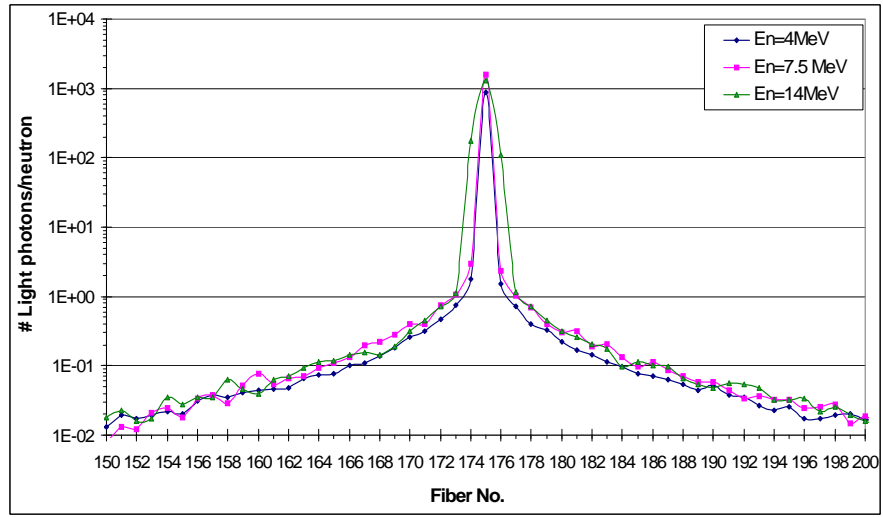

**Fig. 35 The dependence of neutron induced PSF on neutron energy for 4, 7.5 and 14 MeV neutrons**

The central part of the PSF peak broadens with neutron energy due to increased range of the recoil protons.

Figure 36 shows the variation of normalized tertiary PSF with neutron energy.

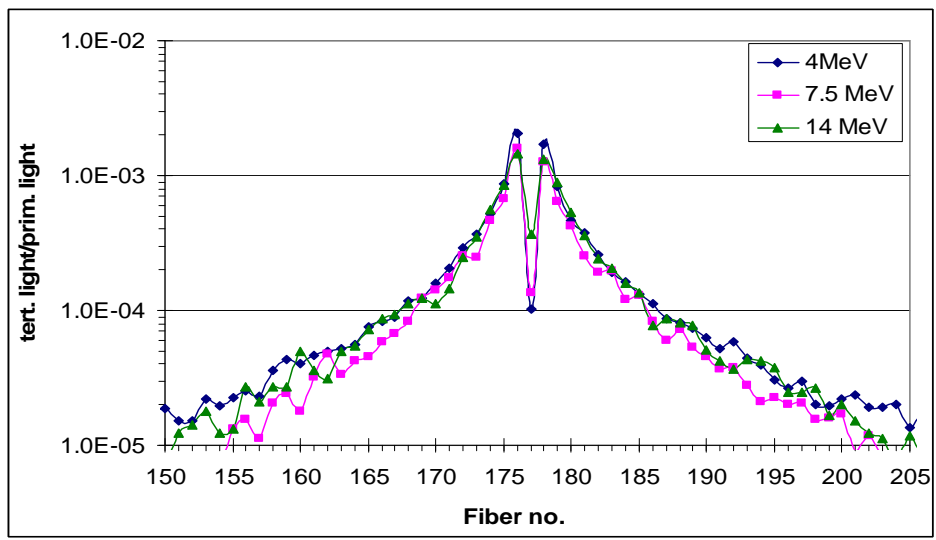

**Fig. 36 Variation of normalized tertiary PSF with neutron energy**

The tertiary light signal was normalized by the corresponding primary light signal for each of the simulated energies. As can be seen, the wings from the tertiary signal are only slightly dependent on neutron energy.

In order to determine how this PSF affects the spatial resolution and image contrast, I have calculated transmission image of 7.5 MeV neutrons through a steel plate 30 cm thick with 9 holes, each 1 cm in diameter. The plate was positioned at a distance of 1 m from the fiber screen. The incident neutron beam was a collimated cone beam emitted from a target spot 3 mm in diameter, at distance of 1200 cm from the screen face. This configuration simulates well our experimental conditions at the PTB cyclotron. Figure 37 shows the mask which was used in this measurement.

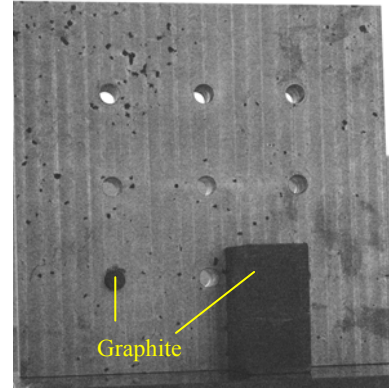

**Fig. 37 Steel mask used in PTB experimental measurement**

The images in Figures 38 a, b are due to primary and tertiary light, respectively. They are displayed at the same grey level range. The secondary light image could not be displayed in this range because the signal difference is too large.

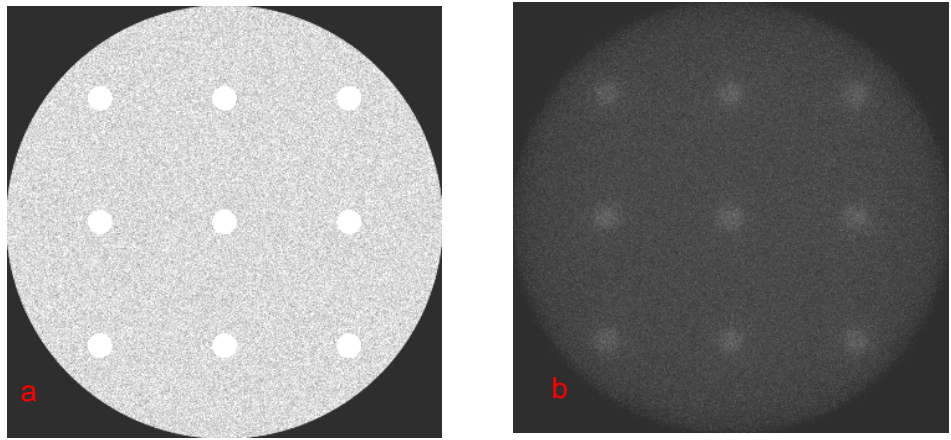

**Fig. 38 GEANT calculated neutron images of a steel mask using 7.5 MeV neutrons using different light components. a) obtained with the primary light only, b) with the tertiary light signal only**

Figure 39 a & b show a horizontal profile taken through the center row of the simulated image with and without (full-transmission, or flat image) the steel plate, respectively. All contributions were normalized to the primary signal of the flat image. As can be observed, only the primary signal shows the expected 7.5 MeV neutron transmission of 0.41 through the 30 mm steel plate. The contribution of the tertiary signal to the total one is quite significant in both images.

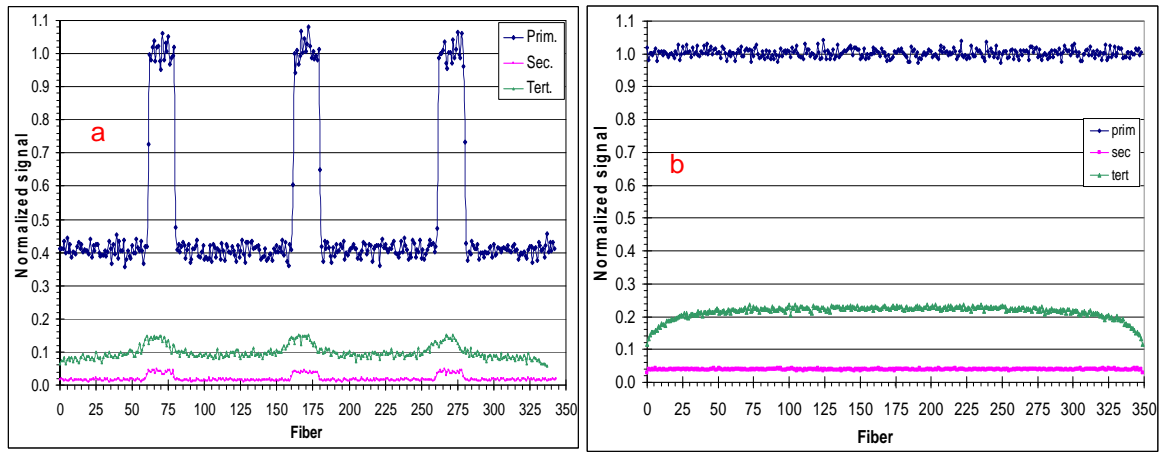

**Fig. 39 Profiles from simulated images of primary, secondary and tertiary light (a) with 30 mm steel plate and (b) without it**

Figure 40 shows a profile of total light (sum of all contributors) of the simulated 30 mm thick steel plate radiograph normalized by the total light of simulated "flat" radiograph (without the object).

As expected, some contrast is lost due to the processes taking place in the scintillating screen, primarily due to the relatively high contribution of the tertiary signal.

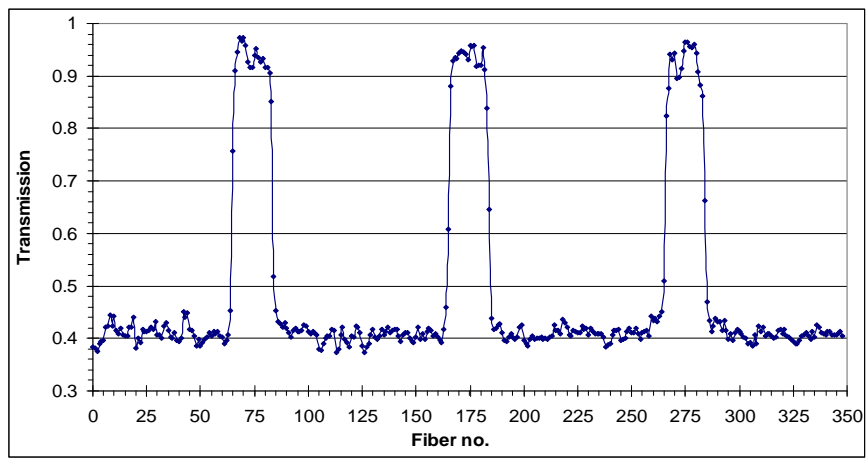

**Fig. 40 Transmission profile due to 30 mm thick steel plate showing loss of contrast due to processes in the scintillating screen**

In summary, the above clearly indicate that the processes which take place within the screen indeed degrade image contrast.

3.2.1.2 Experimental determination of overall PSF in TRION Gen. 2

In TRION Gen. 2, the optical chain is considerably more complex than that of TRION Gen. 1. Nevertheless, as demonstrated in section 3.2.1, it exhibits comparable CTF to TRION Gen. 1, but with an upper limit on resolvable spatial-frequency at 0.5 lp/mm.

In this section, the focus is on the possibility of recovering the lost contrast of the image by determining experimentally the system PSF for deconvolving it from the neutron image. The PSF determined in section 3.2.1.1 was due to neutron interaction in the scintillation screen only. The overall spatial PSF we used was determined by taking a neutron image of a relatively large well-determined aperture allowing the extraction of the PSF function by de-convolution (Grünauer, 2005, 2006).

Figure 41b shows the radiographic transmission image of the steel plate seen on Figure 41a (30 mm thick with 9 holes, each 1 cm in diameter) taken with TRION Gen 2. The energy of neutrons is 7.5 MeV. Since the dimensions of the aperture are known the spatial PSF of the system was determined from this image according to (Grünauer, 2006):

$$I_{blurred} = I_{ideal} \otimes PSF \qquad (10)$$

Where $I_{blurred}$ represents the blurred image and $I_{ideal}$ the ideal image.

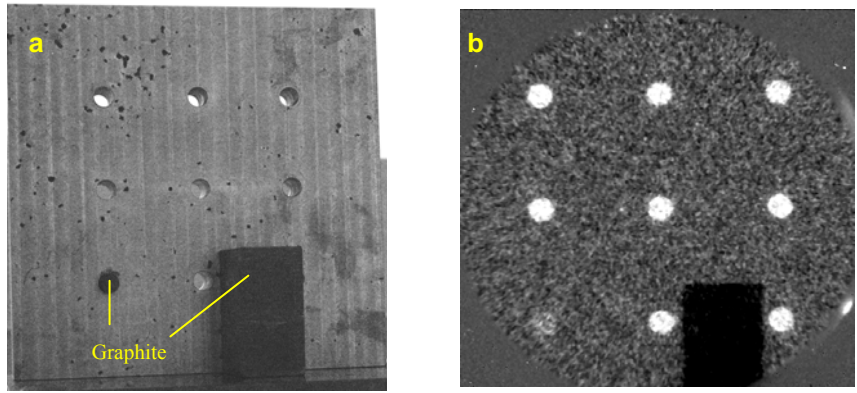

**Fig. 41 a) A visual image of the 30 mm thick steel plate, b) transmission image of 7.5 MeV neutrons through the steel plate**

Figure 42 shows a one dimensional profile of the PSF. The Full-Width Half Maximum (FWHM) of the PSF is 7.16 fibers, which corresponds to 4 mm. This compares unfavorably to the FWHM of neutronic PSF of 1 fiber (see Figure 34a), indicating that the loss of spatial-resolution can be mainly attributed to the optical and electro-optical components in TRION.

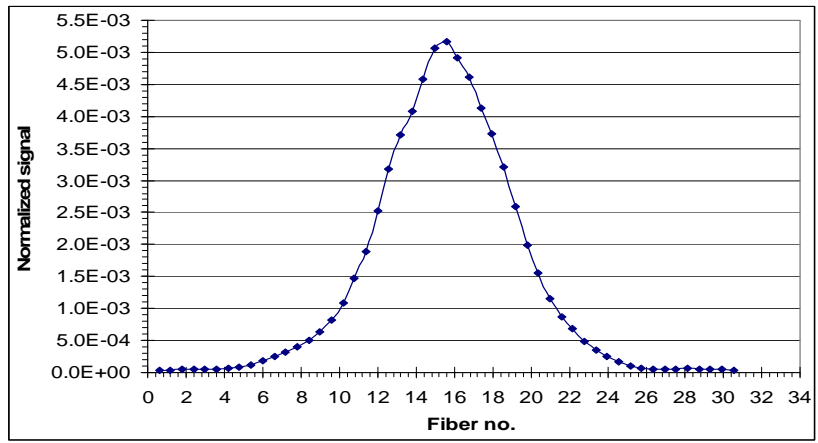

**Fig. 42 Experimentally determined PSF**

Figure 43 (pink squares) shows an experimental CTF obtained using a tungsten CTF mask similar to the one described in section 3.2.1. Figure 43 (blue circles) shows the CTF after the Lucy-Richardson deconvolution procedure (which appears as a standard function in Matlab (MathWorks, 2011)) using the experimentally determined PSF (whose profile is seen in Figure 42). Following deconvolution, the contrast shows significant improvement, especially at low spatial frequencies, however at the expense of increased noise.

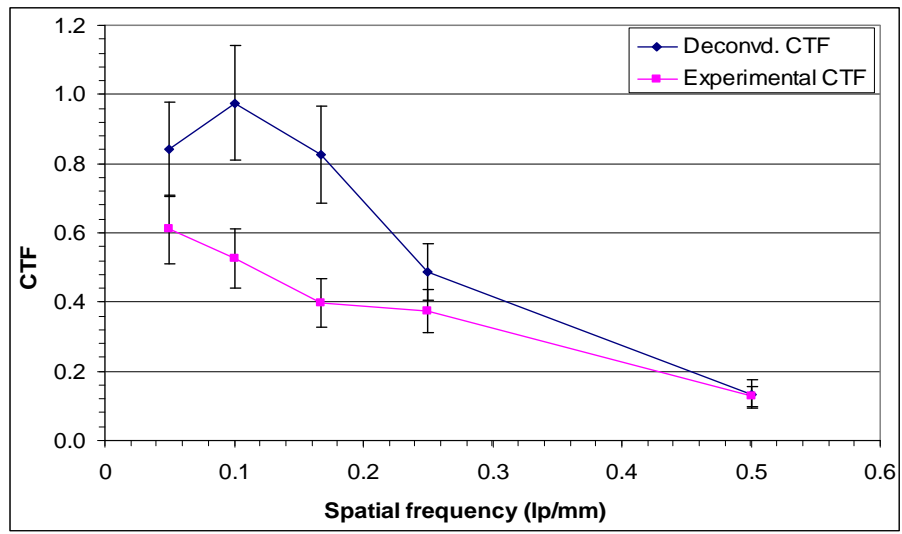

**Fig. 43 Experimental CTF before and after deconvolution**

3.2.2 Imaging results

The imaging results obtained with TRION Gen.2 are presented in the following.

3.2.2.1 Spectroscopic imaging

For the purpose of demonstrating the quality of gamma-ray and neutron images using the TRION Gen.2 detector, a phantom consisting of the following items was radiographed at the PTB facility: 7.65 mm Walther PPK gun, magazine with gas filled bullets, hollow tungsten bar, UO2 powder.

Each of the 4 cameras captured an image for a time window gate-width at a different delay time (energy) corresponding to the following:

- Camera 0 – Gamma-ray spectrum from the $^9$Be(d,n) reaction (Rasmussen et al., 1997)
- Camera 1 - **10.5 MeV** neutrons
- Camera 2 - **7.3 MeV** neutrons
- Camera 3 - **3.1 MeV** neutrons

Figure 44 shows a TOF spectrum resulting from the d-Be reaction, measured at the PTB accelerator facility using TRION Gen.2. The 4 vertical red lines in the figure indicate the locations of the above-mentioned 4 energies on the TOF plot.

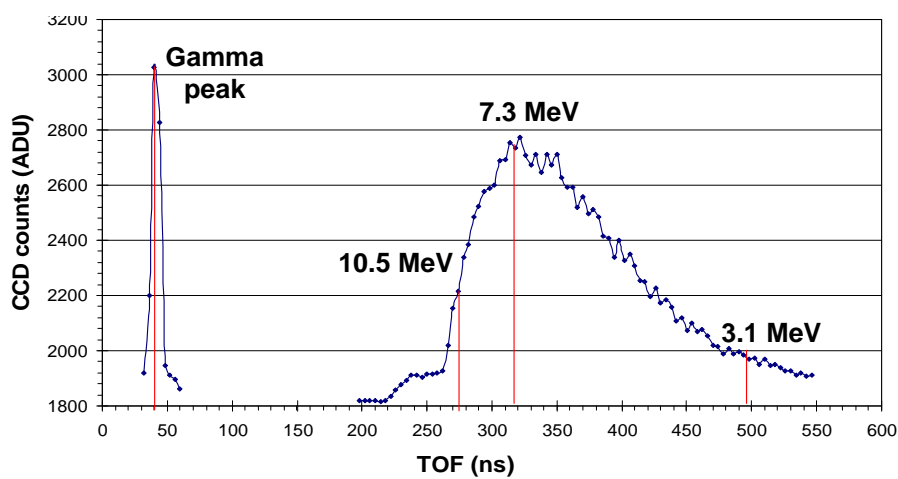

**Fig. 44 TOF Spectra resulting from the d-Be reaction, measured with TRION Gen.2**

Each of the images was normalized by a full-transmission (flat) image taken at the relevant time window.

Figure 45a & b-d show the gamma-ray and neutron images of the phantom. As with TRION Gen.1, the quality of the images is sufficient to permit visual examination of the radiographed objects by an operator. It can be seen that Figures 28b & d suffer from excessive quantum noise due to the low neutron statistics at those energies (Brede et al.,

1989) (see also Figure 44). In addition, 3 MeV neutrons create less scintillation photons per detected event than higher energy neutrons.

These results show that the 4 optical channels can be individually operated, each capturing a different time window in the TOF spectrum with very good spatial and temporal resolution.

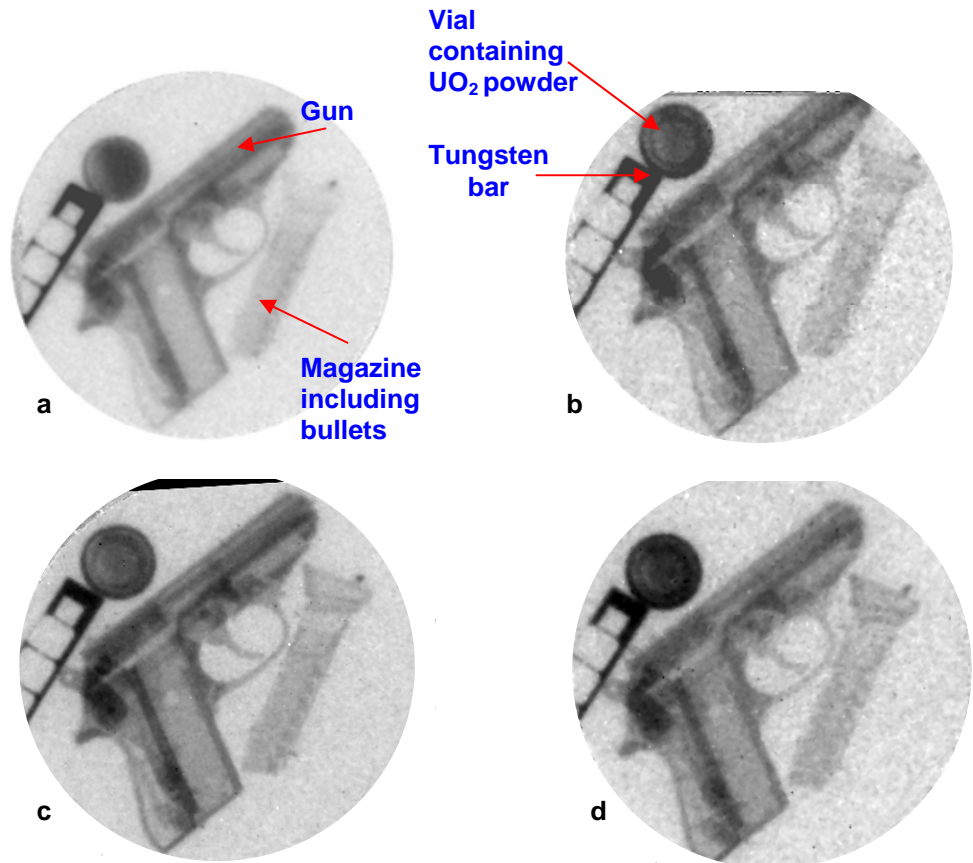

**Fig. 45 Spectroscopic imaging with TRION Gen.2. a) Camera 0: gamma image, b) Camera 1: En = 10.5 MeV, c) Camera 2: En = 7.3 MeV, d) Camera 3: En = 10.5 MeV**

The neutron attenuation of the tungsten bar was calculated from the images and compared to the expected attenuation calculated using tabulated neutron cross-sections (IAEA, 2011). Table 4 shows that there is reasonable agreement between calculated and experimental attenuation values.

**Table 4 Tungsten attenuation: Calculated vs. experimental values**

|  | Expected. attenuation | Calculated attenuation. | % deviation from calc. value |
|---|---|---|---|
| $En_1$ = 10.45 MeV | 0.517 | 0.549 | 5.8 |
| $En_2$ = 7.3 MeV | 0.513 | 0.543 | 5.5 |
| $En_3$ = 3.125 MeV | 0.474 | 0.441 | 7.5 |

The images at 3.1 MeV and 10.5 MeV are noisy since the incident neutron flux at these energies is low. For the purpose of visual examination it may be worthwhile to sum the images taken at different energies in order to get better counting statistics. Figure 46 shows the summed images.

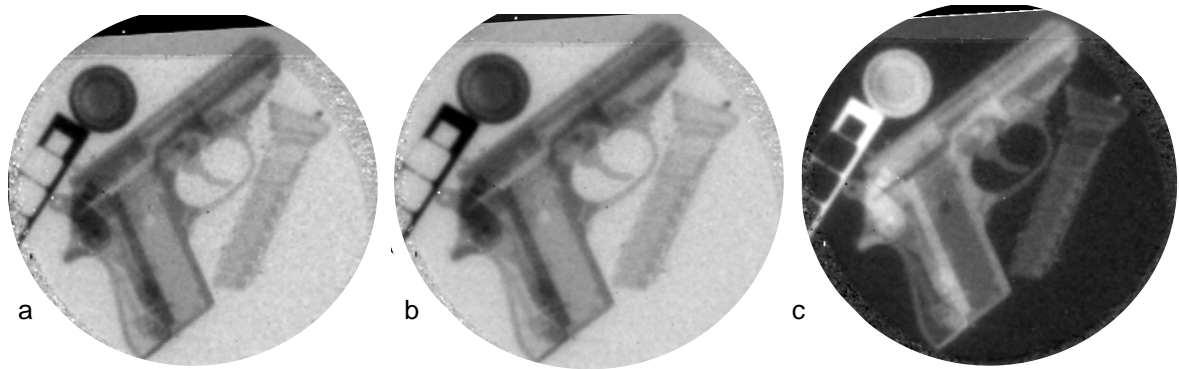

**Fig. 46 a) sum of all neutron images, b) sum of all neutron and gamma-ray images, c) negative of the logarithmic display of (b)**

3.2.2.2 Obscured phantoms

In order to test the influence of massive neutron shielding and scattering on image quality, a 8 cm thick polyethylene block was placed in-front of the above mentioned phantom as shown in Figure 47.

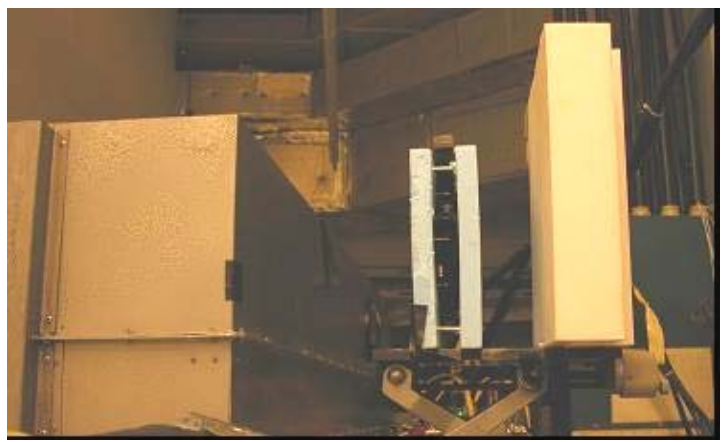

**Fig. 47 An 8 cm thick polyethylene block (right) obscuring the beam path**

Figure 48 shows the gamma-ray and neutron radiographs of the above phantom. Although the images are not as sharp as before, the objects are still discernible.

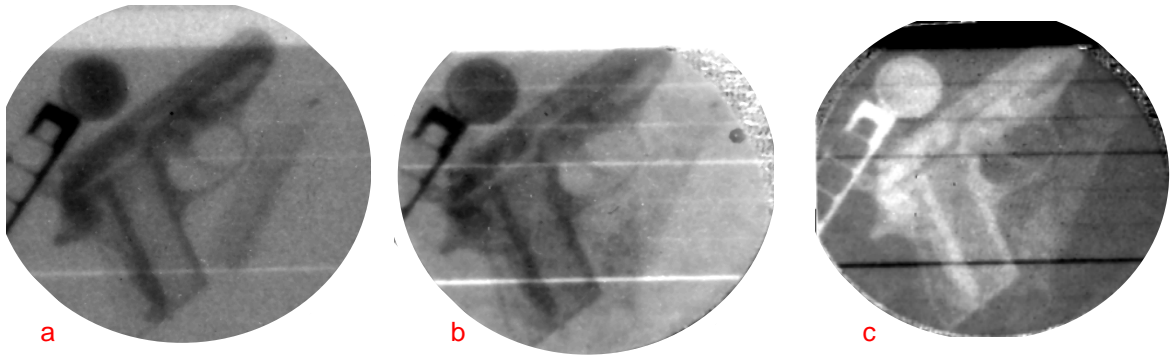

**Fig. 48 Phantom behind 8 cm thick polyethylene block: a) Gamma-ray image, b) Sum of all images (gamma & neutrons), c) negative of the logarithmic display of (b)**

To simulate the scenario in which the above phantom is concealed within normal every-day items, a clutter phantom was added in front of the detector, as seen in Figure 49a. The clutter phantom (Figure 49b) consisted of a plastic bag containing: tooth paste, tooth brush, calculator, kiwi fruits, vaseline jar, trombone lube paste, scissors and tissue paper packet.

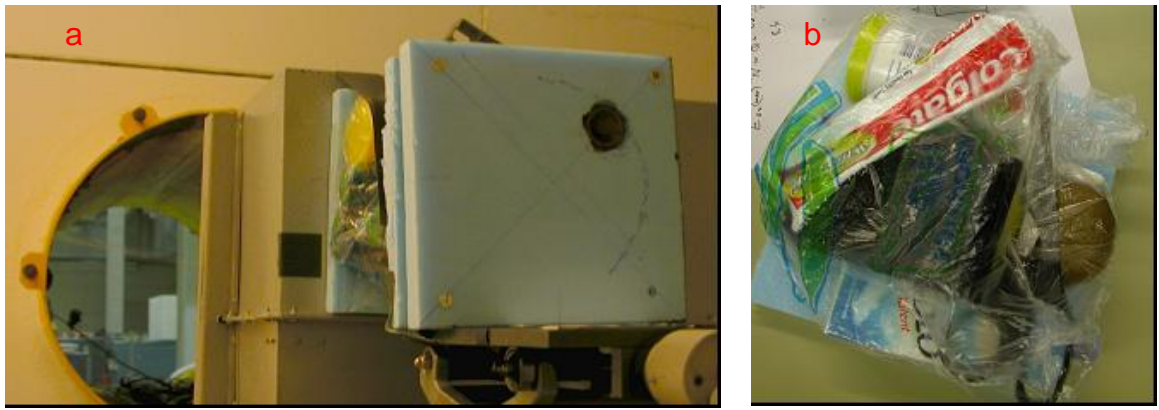

**Fig. 49 a) The clutter phantom (attached to the detector) obscuring the gun phantom, b) The clutter phantom**

Figure 50 shows the gamma-ray (left) and neutron (right) radiographs of this configuration. As expected the two images are complementary and show different details. The hydrogen containing objects (fruits, lubricant) are more visible on the neutron image.

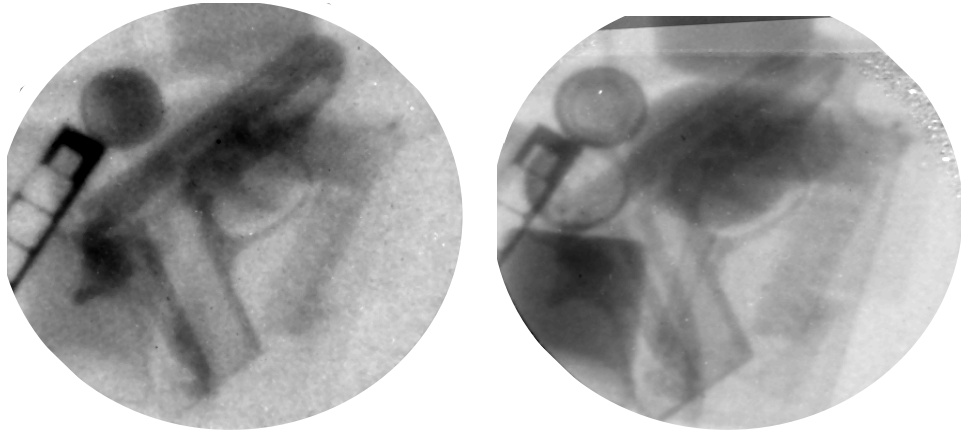

**Fig. 50 Clutter obscuring gun phantom: left) gamma-ray radiograph, right) neutron radiograph**

3.2.2.3 Detection of high Z materials

As originally shown by the Oregon group (Rasmussen et al., 1997), the ratio of natural logarithms of gamma-ray transmission to that of neutron transmission represents the ratio of gamma-ray to neutron attenuation constants. Thus, it is sensitive to the atomic number (Z) of the inspected object, irrespective of its thickness. In this study we have investigated this ratio vs. neutron energy

Figure 51 shows images of the ratio of the natural logarithm of the gamma transmission image (camera 0) to the natural logarithm of the neutron transmission image for the three neutron energies (cameras 1-3).

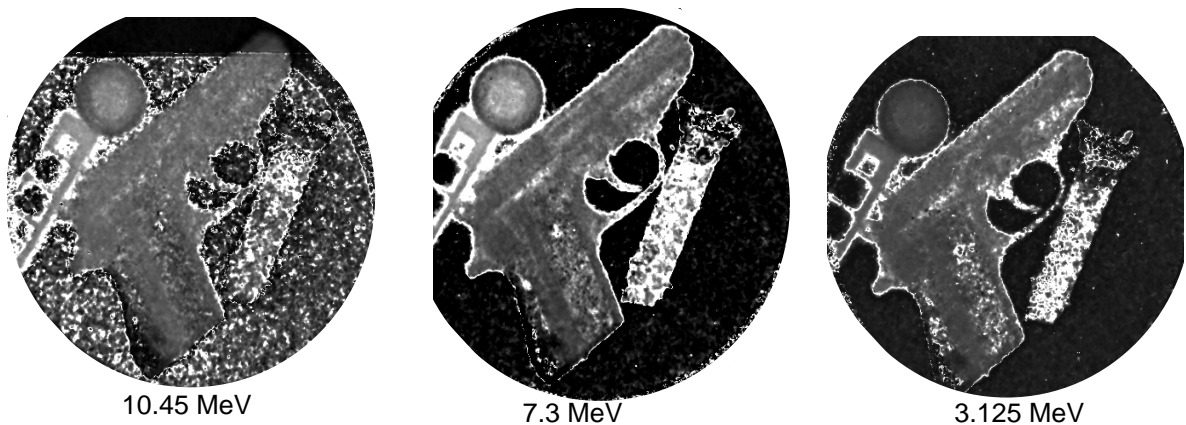

10.45 MeV      7.3 MeV      3.125 MeV

**Fig. 51 Ratio of natural logarithms of gamma-ray to neutron transmission for three neutron energies**

Figure 52 displays the natural logarithm ratio-values for tungsten, $UO_2$, steel (gun) and polyethylene extracted from the above images at each of the neutron energies. As can be observed, at neutron energy of 7.3 MeV the heavy elements (W and $UO_2$) can easily be discriminated from lighter materials. The effect is less pronounced at lower neutron energies.

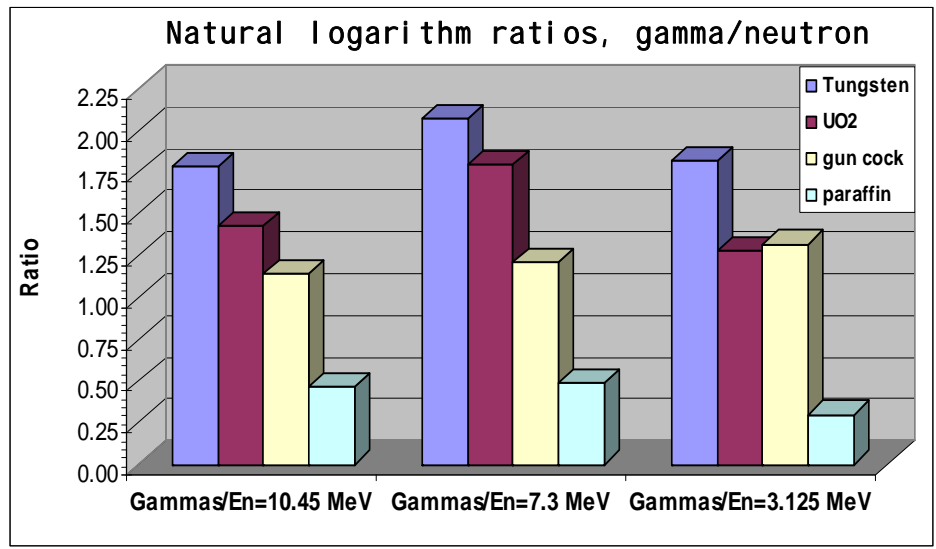

**Fig. 52 Natural logarithm (Ln) ratios, extracted from the ratio images seen in Figure 51**

3.2.3 Temporal resolution

The following will provide an analysis of TRION's temporal performance.

3.2.3.1 Evaluation of the influence of neutron scattering within the scintillating screen on temporal resolution

The temporal resolution of TRION is governed by the following factors:
- Accelerator burst duration: 1.5-2 ns
- Scintillation decay time: 2.5 ns (95%), 12 ns (2.5%), 68 ns (0.2%) (Mor, 2006)
- The minimal achievable I-I gate width: 4-6 ns

As the influence of scintillator decay time and image-intensifier gate-width on temporal resolution had been examined as part of my M.Sc. work (Mor, 2006), I have investigated the influence of an additional factor, namely, the finite TOF required by direct neutrons to traverse the full thickness of the scintillator screen and the temporal behavior of the light signal generated by neutrons scattered within the scintillating screen. If the screen has large dimensions, the probability for multiple scattering within it is significant and the neutron may deposit its energy in the screen over a time interval which is comparable or longer than the characteristic times of the above mentioned processes.

In order to investigate this effect I have calculated the spatial PSF as a function of time, for several incident neutron energies. In this Monte-Carlo simulation, the central pixel of the 20×20×3 cm$^3$ scintillating screen was irradiated with a pencil-beam of neutrons. The time-and-space-history of each neutron were followed since the time it impinged on the screen.

The amount of scintillation light in each pixel due to this beam was determined and displayed, thus creating a scintillation image. A series of images were created vs. time. Figure 53 shows the light images vs. time for a 1 MeV neutron beam (as 1 MeV neutrons will spend the longest time within the screen in comparison to neutrons of higher energies).

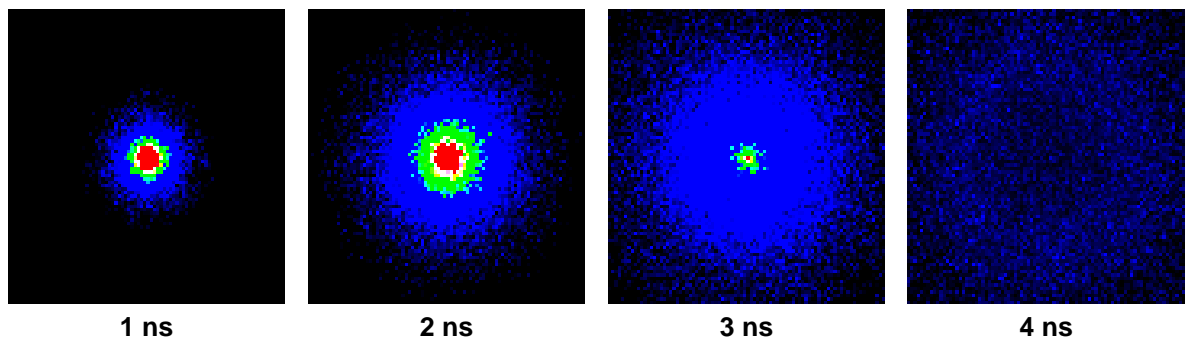

**Fig. 53 Images of light created by 1 MeV neutrons vs. time since their arrival to the central pixel of the screen. Hottest color (red) indicates highest number of light photons**

Figures 54 a, b, c show the profiles of light taken through the central row of fibers in the screen for 1, 2 and 7.5 MeV neutrons respectively.

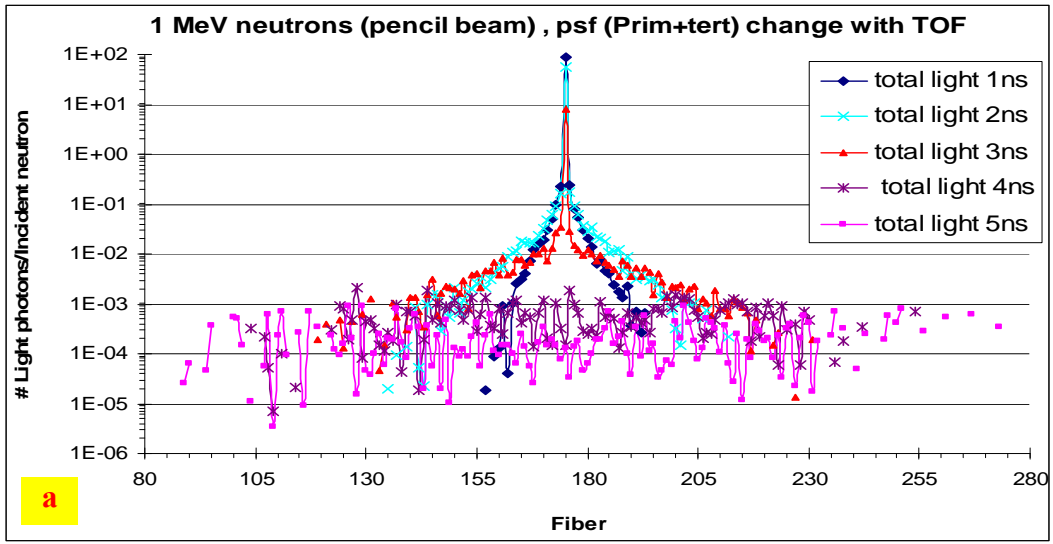

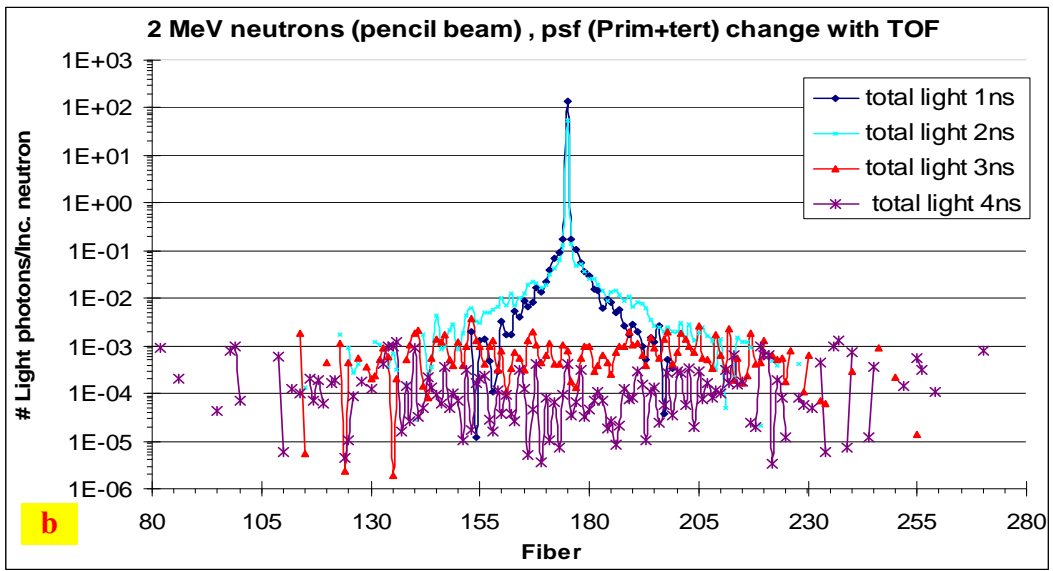

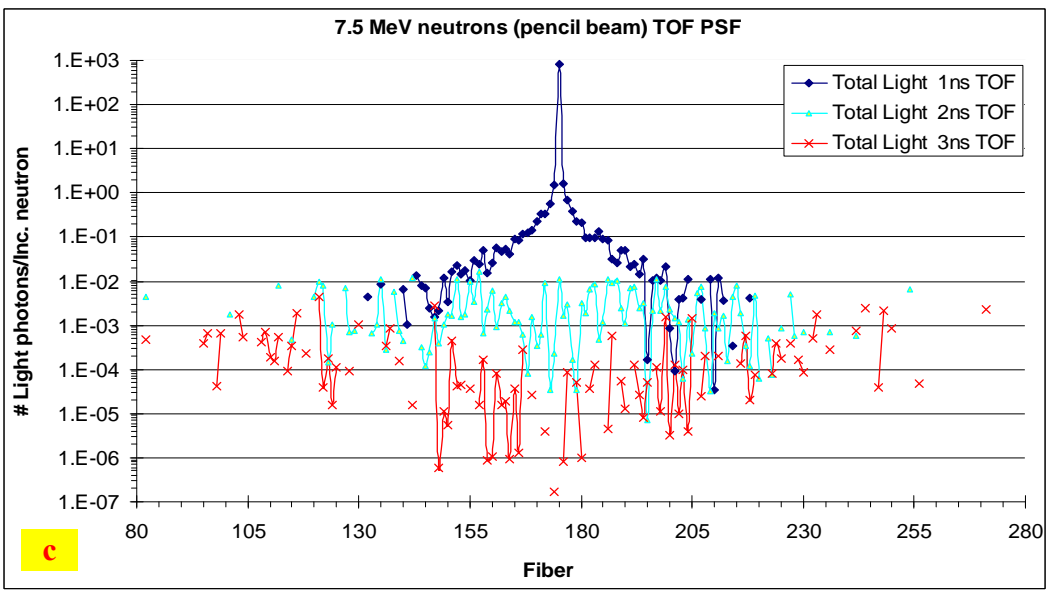

**Fig. 54 a,b,c) PSF vs. time for 1, 2 and 7.5 MeV neutrons, respectively, for 3 cm thick screen**

The PSF broadens with time, after several ns - the time necessary for neutrons to traverse the thickness of the screen, and it becomes a faint circular expanding wave. The time to reach this state decreases with neutron energy, as one would intuitively expect.

For comparison, Figure 55 shows a 7.5 MeV PSF for a 10 cm thick screen. As expected the neutrons spend more time in the larger screen.

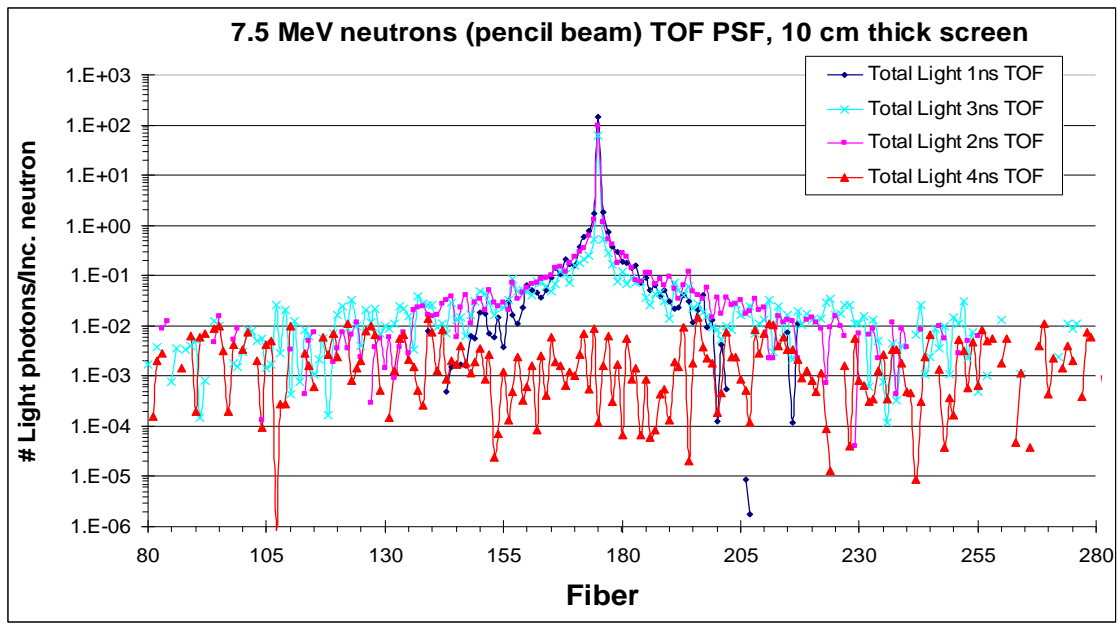

Fig. 55 PSF vs. time for 7.5 MeV neutrons, 10 cm thick screen

The effect of the residual light due to neutron scattering within the screen can be better observed when the entire screen is irradiated with a broad beam of neutrons and the time behavior of the light is observed in the central area of the screen. Figure 56 shows the amount of light per incident neutron vs. time in the central pixel of the 3 cm thick screen. For 2 MeV neutrons the light decays to its 1/10 and 1/100 value after 3 ns and 8 ns respectively, from the time of neutron arrival. For the higher neutron energies the corresponding times are approximately 1.8 ns and 4 ns.

As the time required for light to decay to its 1/100 value is within our gating time (4-6 ns), our conclusion is that for the 3 cm thick screen this effect is not of major importance, however, it should be considered and corrected for in larger screens.

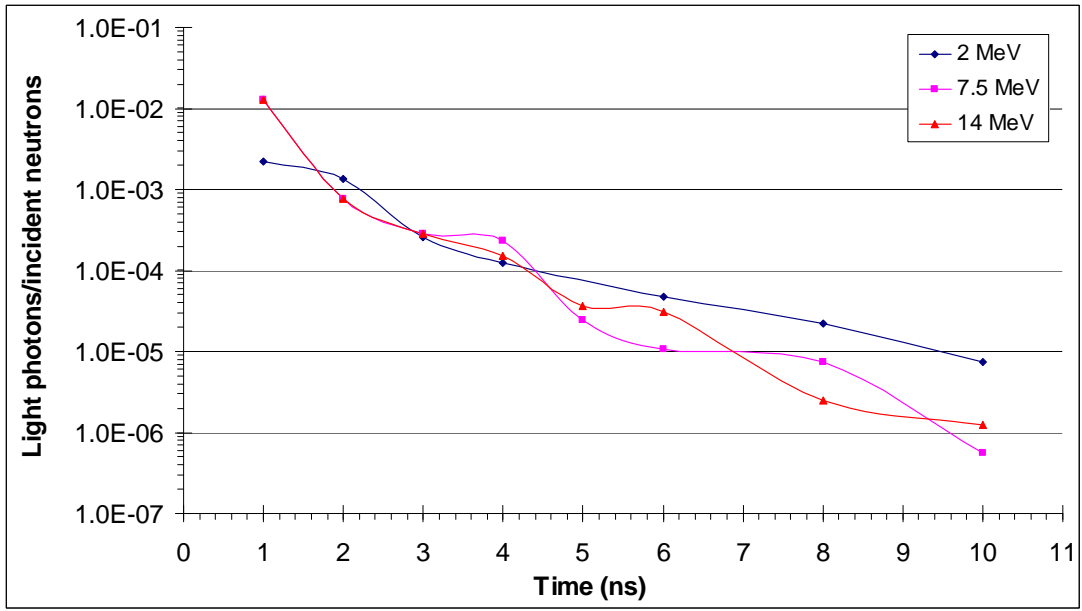

**Fig. 56 Light decay in the central pixel following neutron irradiation of the entire screen. (3 cm thick screen)**

3.2.3.2 Reconstruction of temporal resolution

Figure 57 shows the transmission through 10 cm graphite block vs. time-of-flight (neutron energy) obtained with TRION (blue) and theoretical transmission calculated from known cross-section of carbon (yellow). As can be observed TRION suffers from considerable loss of temporal contrast.

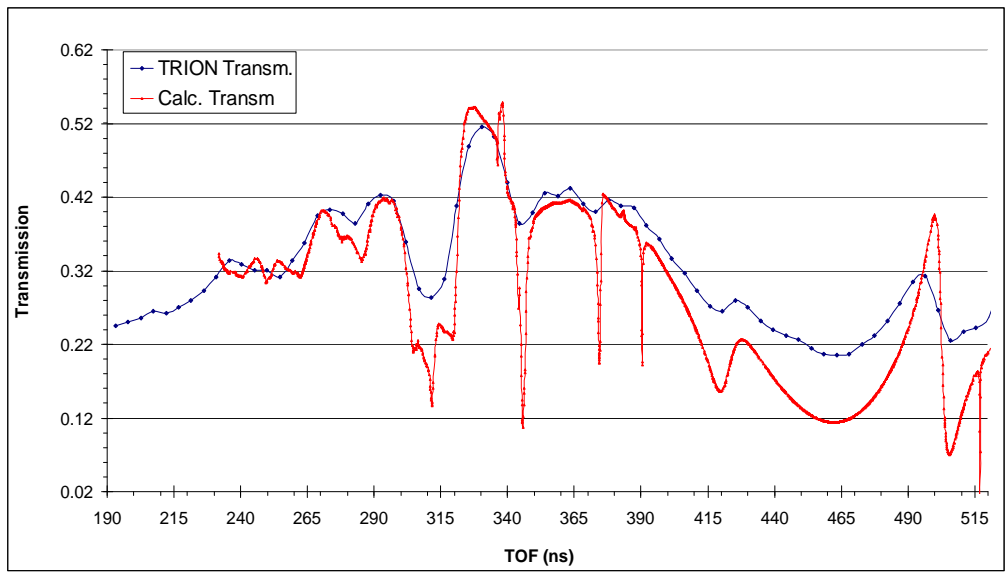

**Fig. 57 Transmission spectra beyond 10 cm graphite obtained with TRION vs. calculated from known cross-section**

As mentioned above, the temporal resolution in TRION is governed by the following factors: finite duration of the deuteron beam burst, scintillation decay time and minimal achievable image-intensifier gate width.

In this section the influence of these factors is deconvolved from the measured TOF spectra. As a kernel or point spread function (PSF) we use the gamma-ray peak, which essentially represents the system response to a sharp (δ-function) time signal. This is a good approximation since, in plastic scintillators, the scintillation decay time-constant for gamma-rays (internally-produced electrons) is not significantly different from that of neutrons (internally-produced protons).

Figure 58 shows the gamma-ray peak as measured by TRION, used as an overall temporal PSF of the system.

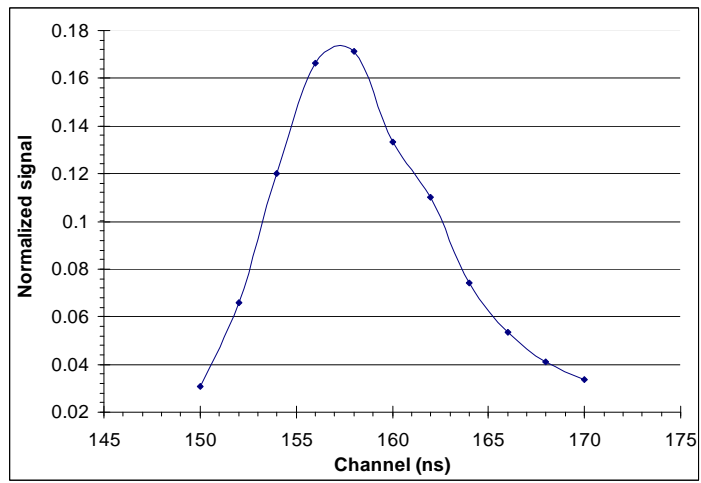

Figure 59 shows the results of deconvolution using the gamma-peak of Figure 58 as a PSF. Clearly, the temporal resolution has been recovered for most peaks of interest. Table 5 shows the peak and dip values before and after deconvolution for the three peaks and dips indicated on Figure 59 by P or D, respectively.

**Fig. 58 The gamma-ray peak as measured by TRION**

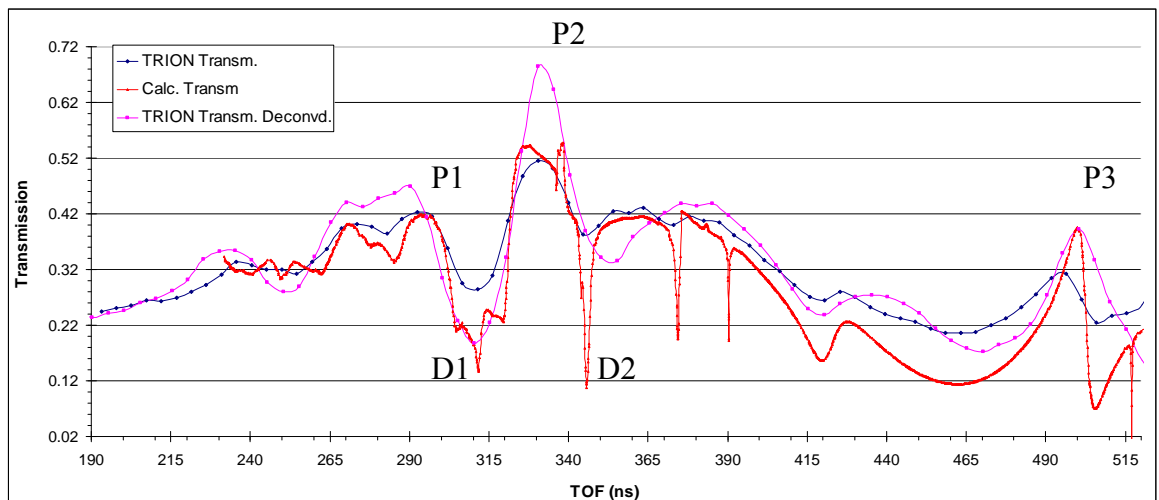

**Fig. 59 Transmission through 10 cm graphite, as measured by TRION Gen.2 before and after temporal deconvolution compared to calculated transmission**

**Table 5 Values of peaks and dips in transmission spectrum through graphite, before and after temporal deconvolution**

|    | Expected Values | TRION - Before Deconvolution | TRION - Following Deconvolution | % Change Following Deconv. |
|----|-----------------|------------------------------|----------------------------------|-----------------------------|
| P1 | 0.413           | 0.41                         | 0.469                            | 12.5                        |
| P2 | 0.531           | 0.52                         | 0.684                            | 24                          |
| P3 | 0.389           | 0.319                        | 0.392                            | 18.6                        |
| D1 | 0.142           | 0.28                         | 0.188                            | 48.9                        |
| D2 | 0.108           | 0.389                        | 0.335                            | 16                          |
| D3 | 0.114           | 0.255                        | 0.172                            | 47.7                        |

3.2.4 Excess noise factor in TRION comparing to event counting mode

In comparison to event-counting detectors where the variance in the signal is dependent only on the number of detected events, in an energy-integrating detector additional factors such as the fluctuations in imparted energy, number of photoelectrons, system gain and other factors will contribute to the noise.

The following provides an analysis of the propagation of the signal produced by the various components of the TRION fast neutron detector. Among others, the physical processes within TRION that influence the signal-to-noise ratio are listed and discussed.

3.2.4.1 Generation of signal in TRION

Figure 60 shows the sequence of signal generation in TRION.

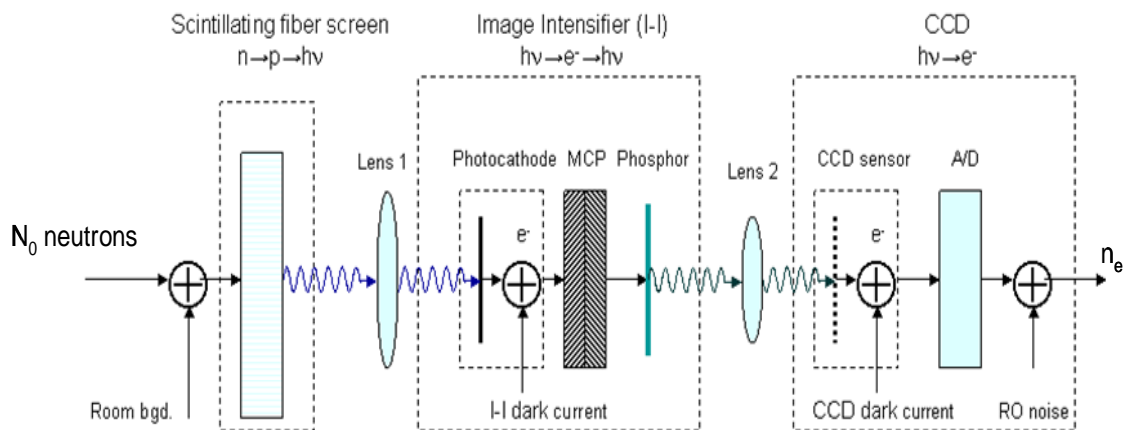

**Fig. 60 Schematic description of the signal sequence developed in TRION**

The physical processes that govern the signal generated on the CCD sensor per incident neutron in TRION are:

1. The interaction of the incident neutron in the scintillating screen, resulting in transfer of part of its energy to one or more protons
2. The conversion of energy dissipated by the proton in the scintillator to light
3. The transport of light photons to the photocathode of the image intensifier through lens-1
4. The absorption of light photons at the photocathode and the emission of photoelectrons
5. The electron multiplication process within the image intensifier and its re-conversion to light at its exit phosphor.
6. The transport of intensified light from the image intensifier to the faceplate of the CCD through lens 2.
7. The absorption of light photons by the CCD sensor and its conversion to electrons.

For **$N_0$ *detected*** neutrons of energy $E_n$, the average number of electrons produced on the CCD sensor can be expressed as (Vartsky et al., 2009):

$$\mathbf{n_e = N_0 \cdot Ph(E_n) \cdot T_1 \cdot p_1 \cdot m_G \cdot T_2 \cdot p_2} \qquad (11)$$

Where,

$N_0$ - number of neutrons detected over the exposure time

$Ph(E_n)$ - average number of light photons emitted from the scintillating fiber per detected neutron of energy $E_n$

$T_1$ - transmission of lens 1

$p_1$ - quantum efficiency of the image-intensifier

$m_G$ - average gain of I-I (photons/electron)

$T_2$ - transmission of lens 2

$p_2$ - quantum efficiency of CCD

In addition to the signal generated by neutrons, additional signals enter the chain at various stages. These are the dark currents of the image intensifier and CCD due to single electrons generated by thermal energy in the intensifier photocathode and in the CCD Si matrix as well as the CCD readout noise (RN). If during the measurement time we detect $N_0$ neutrons and generate $N_1$ and $N_2$ dark current electrons in the image-intensifier (I-I) and CCD respectively, the average overall measured signal will be:

$$n_{eT} = N_0 \cdot Ph(E_n) \cdot T_1 \cdot p_1 \cdot m_G \cdot T_2 \cdot p_2 + N_1 \cdot m_G \cdot T_2 \cdot p_2 + N_2 + RN \qquad (12)$$

In order to obtain the net neutron signal, the dark signals and RN must be determined separately and subtracted from the overall signal. As described in section 3.1.4.1, in TRION we reduce the dark current signals to insignificant levels by cooling the I-I and the CCD.

3.2.4.2 The noise in TRION

The relative noise (reciprocal of signal to noise ratio) in an integrated signal (due to $N_0$ detected neutrons), resulting from a cascade of **m** processes was quantified by (Vartsky et al., 2009) as:

$$\frac{\sigma_{n_{eT}}}{n_{eT}} = \frac{1}{\sqrt{N_0}} \left[ 1 + \sum_m \left\{ \left(\frac{\sigma_m}{m_m}\right)^2 \left(\frac{M_m}{n_{eT}}\right) \right\} \right]^{1/2} \qquad (13)$$

Where $m_m$ and $\sigma_m$ are the mean and standard deviation of a probability distribution function (**PDF**) of the process "**m**" and $M_m$ is the product of the means $N_0 \cdot m_1 \cdot m_2 \cdot m_3 \cdot \ldots m_m$.

Thus, in an integrating system such as TRION, the relative noise will always be larger than the relative quantum noise $1/\sqrt{N_0}$ of the detected incident signal by an excess noise factor (**ENF**) that depends on **weighted** relative errors of all physical processes that contribute to formation of the signal. The inverse square of this factor divided by the detection efficiency is also referred to as **Detective Quantum Efficiency** (**DQE**), it signifies how well the detector reproduces the signal to noise ratio (SNR) of the incident radiation(Dainty and Shaw, 1974).

In TRION, the main contributors to this excess noise are the following processes:
1. Generation of light photons by neutrons in the screen and transport to I-I
2. Creation of photoelectrons on the I-I photocathode (binomial process)
3. Amplification of light in I-I and transport to CCD sensor
4. Creation of photoelectrons in the CCD sensor (binomial process)

The magnitude of the mean value $m_m$ and the relative error $\sigma_m/m_m$ of each process is dependent on its PDF.

The following describes in detail the properties of the scintillating screen, the optical system and the image-intensifier. The PDF of the processes contributing to the ENF is also determined.

3.2.4.3 Determination of PDF and its parameters for each process

The PDF of the above processes were determined either experimentally or by Monte-Carlo simulation.

3.2.4.3.1 Detailed simulation of the response of the scintillating fiber screen

a. Proton energy spectra

In order to understand the contribution of screen response to the signal-to-noise ratio, it is important to determine the distribution of the light output for mono-energetic neutrons.

A neutron entering the screen can transfer its energy to a recoil-proton according to the angle between the scattered proton and the incident neutron, as specified by Eq. 1. The recoil-proton can acquire energies ranging from zero to the full neutron energy. As the neutron-proton reaction is isotropic throughout the energy range of interest here (1-10 MeV), the energy distribution of the protons is flat. Thus, even for mono-energetic neutrons, the distribution of proton energies is very broad. The situation is further modified by the presence of carbon in the screen and by the fact that the screen is composed of fibers of which only the core plays an active role in light creation process. The cladding and EMA paint will absorb part of the proton energy, but will not generate a scintillation signal.

Some of the recoil-protons created in the core may escape it, leaving only part of their energy inside the core. On the other hand, protons created outside the core may enter it and deposit their energy there. Thus the energy distribution of protons capable of creating light in the fiber core is expected to deviate from a pure flat distribution.

A detailed calculation of the distribution of energy deposited in a scintillating fiber core has been performed using the GEANT 3.21 code (CERN, 2003). The simulated setup has been described in section 3.2.1.1.

The simulation calculates the energy deposited by protons in the core of the central fiber (test fiber). As previously described in section 3.2.1.1, for tracking and tallying purposes, protons created by incident neutrons entering the central fiber are termed "***primary*** *protons*". Protons created outside the central fiber that reach its core are referred to as "*secondary protons*" and protons created by neutrons entering the central fiber after being scattered into it from any of the outer regions are defined as *"tertiary protons".*

In the simulation, the same number of neutrons ($6\times10^9$) uniformly incident on the $200\times200$ mm$^2$ area was employed at each of the 5 neutron energies.

Figure 62 shows the distribution of energy deposited by the primary, secondary and tertiary protons in the central fiber at 3 incident neutron energies (2, 7.5 and 14 MeV).

As can be observed, with increasing neutron energy, the distribution of the primary protons deviates from the flat distribution. The reason for this is that, as the proton energy

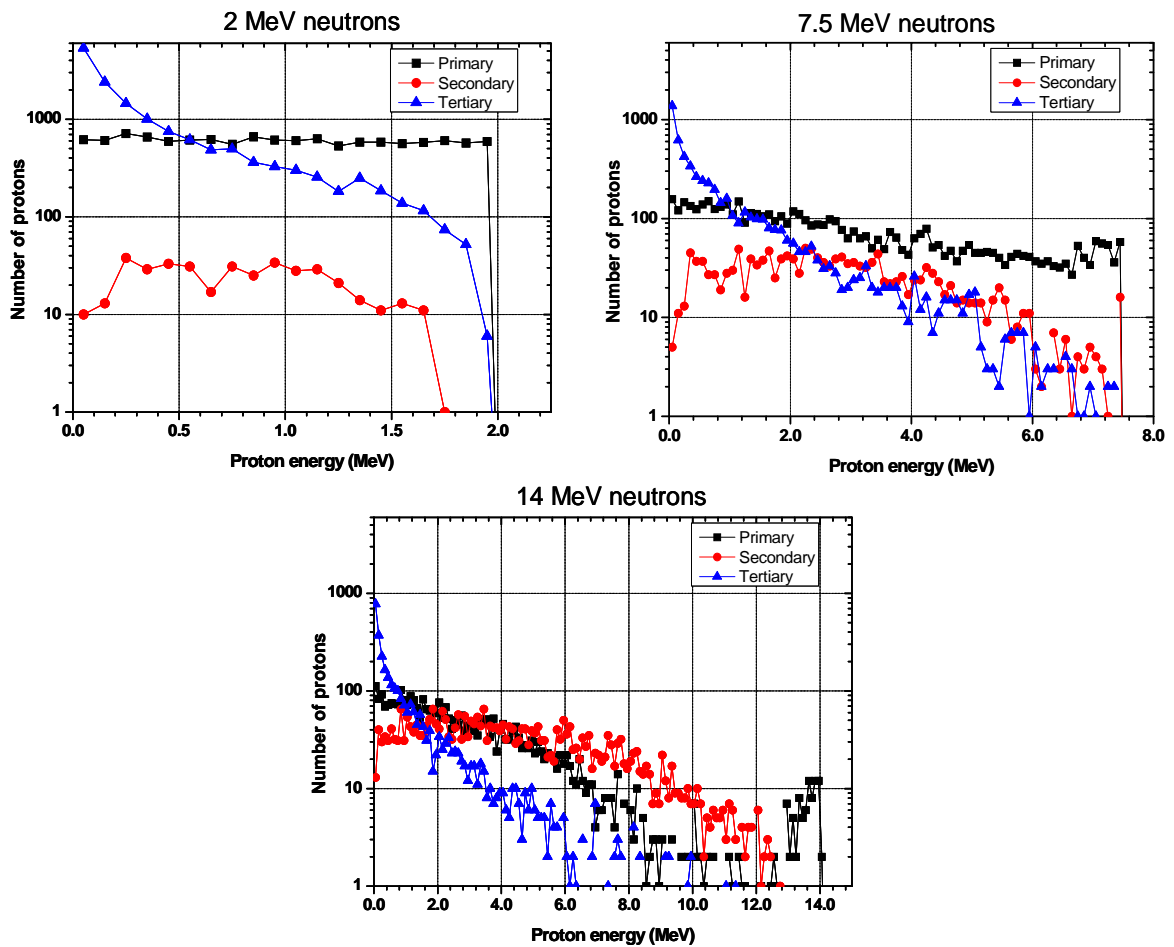

**Fig. 61 Energy deposited in the central fiber by primary (black), secondary (red) and tertiary (blue) protons for 2 MeV (top left), 7.5 MeV (top right) and 14 MeV (bottom) neutrons**

increases, the proton range increases accordingly and if the track crosses a pixel boundary it will deposit only part of its energy in the fiber core in question. The secondary proton energy distribution extends over a wide energy range and that of the tertiary protons is skewed toward lower energies.

Table 6 shows the contributions of primary, secondary and tertiary protons to the total number of protons created in the fiber core.

Table 6 Contribution to total signal and mean energy of primary, secondary and tertiary protons

| Neutron energy (MeV) | Primary protons (%) | | Secondary protons (%) | | Tertiary protons (%) | |
|---|---|---|---|---|---|---|
| | Contribution (%) | Mean energy (MeV) | Contribution (%) | Mean energy (MeV) | Contribution (%) | Mean energy (MeV) |
| 2 | 44 | 0.98 | 1.5 | 0.78 | 54.5 | 0.36 |
| 4 | 45 | 1.82 | 4 | 1.38 | 51 | 0.67 |
| 7.5 | 43 | 2.80 | 13 | 2.73 | 44 | 0.92 |
| 10 | 39 | 3.2 | 25 | 3.6 | 36 | 1.0 |
| 14 | 35 | 2.95 | 34 | 4.3 | 31 | 1.0 |

As can be observed, a large fraction of protons reaching the fiber core originate from neutrons that interacted outside the test fiber. The proportion of secondary protons increases with neutron energy due to the increase in proton range with its energy. Thus there is a larger chance for protons created in the neighbourhood of the test pixel to reach it and deposit part of their energy in its core. The proportion of the tertiary protons is relatively large, however the average energy that they deposit is low.

The total mean proton energy deposited in the fiber is 0.64 MeV, 1.21 MeV, 1.96 MeV, 2.5 MeV and 2.8 MeV for the 2 MeV, 4 MeV, 7.5 MeV 10 MeV and 14 MeV neutrons respectively. This is significantly lower than half the neutron energy, the value usually taken to represent the average proton energy.

b. Light intensity distribution

The scintillation light intensity distribution does not follow exactly the proton energy distribution because of the non-linear behavior of scintillation light generation with proton energy. The response of plastic scintillator to protons has been studied by several investigators (O'Rielly et al., 1996). The response is usually expressed in electron equivalent energy (MeVee), i.e., the electron energy which would produce the same amount of light output as that produced by a proton of energy $E_p$. Figure 63 shows the amount of light $L(E_p)$ vs. proton energy $E_p$ in plastic scintillators obtained using experimental MeVee data of (O'Rielly et al., 1996) together with the specific light yield of 8000 photons/MeVee, see (Saint-Gobain, 2011). A polynomial function fit to this data is also shown on the figure (in red).

As can be observed L(E$_p$) exhibits non-linear behavior with E$_p$ especially below 5 MeV. The amount of light **L** created in the core of a fiber by a proton can be calculated by:

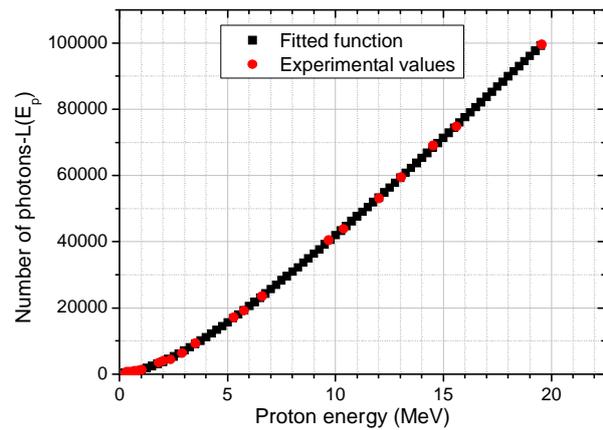

**Fig. 62 Light output L(Ep) of scintillating plastic fiber vs. proton energy Ep (O'Rielly et al., 1996)**

$$L = \int_{E_f}^{E_i} \left(\frac{dL}{dE}\right) dE = L(E_i) - L(E_f) \qquad (14)$$

Where E$_i$ and E$_f$ are the initial and final proton energy, respectively, traveling within the fiber core. The values of E$_i$ and E$_f$ are obtained by the Monte-Carlo calculation for each proton. The above non-linear behavior will cause further skewness in the light distribution toward low light emission.

Only a fraction of the amount of light **L** is transmitted to the end of the fiber and is emitted toward the collecting lens. This fraction (trapping efficiency) is dependent on the refractive indices of core and cladding. According to the manufacturer data (Saint-Gobain, 2011) the trapping efficiency of single-clad square fibers is **4.4%,** i.e., only 4.4% of the total light created in a scintillation travels within the fiber to each end of the fiber. The white EMA coating decreases the amount of light obtained from the fiber, because the coating can interfere with light transmission in the cladding (Saint-Gobain, 2011). The light output of a fiber with white EMA was measured to be about **65%** of that of a bare fiber.

By placing a good quality front face mirror on the front side of the screen face one can increase the above fraction of light exiting the fiber toward the collecting lens by a factor of 1.8. Thus, the amount of light emitted from the fiber is expected to be ~5% of the total light created in a scintillation.

Figure 64 shows the frequency distribution of light emitted from the fiber screen per detected neutron for 3 neutron energies (2, 7.5 and 14 MeV).

The mean number of light photons/neutron emitted from the fiber **Ph(E$_n$)** and the standard deviation of the distribution is **36.6±35.6, 113±112, 271±271 333±332 and 377±376** photons for neutron energies of 2, 4, 7.5, 10 and 14 MeV respectively (Vartsky et al., 2009).

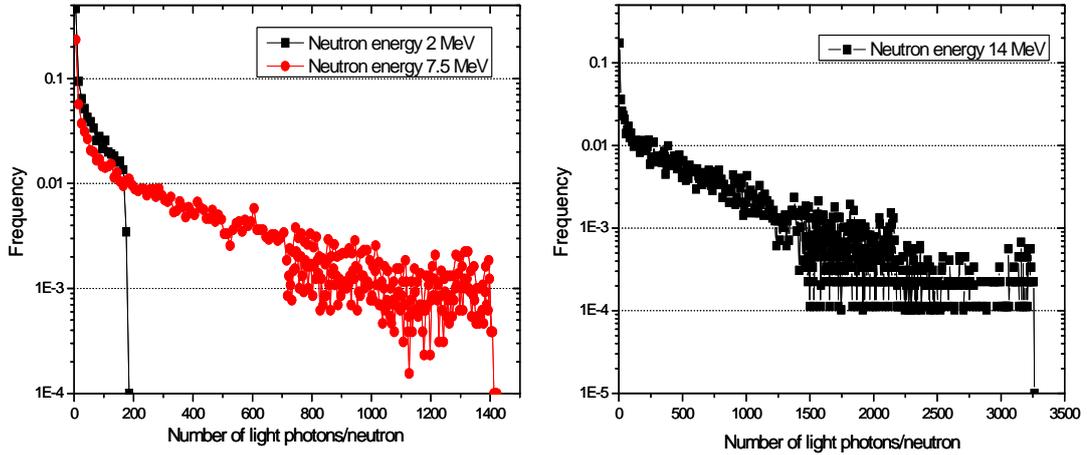

**Fig. 63 Probability distribution of light emitted from the fiber screen**

c. Light collection efficiency

The light emission from the scintillating fiber is limited to a cone, whose apex angle $\theta$ is determined by the refractive indices $n_1$ of the core and $n_2$ of the cladding material. In our fiber screen the maximum emission angle is ~35.7$^o$ (Mor, 2006; Saint-Gobain, 2011).

This light is viewed by a custom made, large aperture lens (120 mm F#0.95), described in detail in section 3.1.3, positioned at a distance of 750 mm from the scintillating screen. The fraction of light collected by the lens is determined by the ratio of the solid-angle subtended by the lens to that defined by the fiber emission cone and was found to be 0.017 (Mor, 2006).

Table 7 presents the mean number of light photons per detected neutron reaching the image-intensifier photocathode for neutron energies of 2, 4, 7.5, 10 and 14 MeV respectively.

**Table 7 Mean number of light photons reaching the intensifier photocathode per detected neutron**

| Neutron energy (MeV) | Mean number of photons/n -$m_{Ph}$ |
|---|---|
| 2 | 0.62 |
| 4 | 1.92 |
| 7.5 | 4.6 |
| 10 | 5.7 |
| 14 | 6.5 |

3.2.4.3.2 Generation of the signal in the image intensifier

a. Conversion to photoelectrons

The image-intensifier described in section 3.1.4 (manufactured by (Photek Ltd., 2011)), is a high-gain, proximity-focus device. The tube is about 25 mm in length and 40 mm in diameter with a rugged metal ceramic construction. The input window is made of fused silica. The photocathode is a bi-alkali one. The manufacturer quotes the quantum efficiency at a wavelength of 420 nm to be 17%. The output window is made of fiber optic and an E36 phosphor screen.

Conversion of photons into photoelectrons (ph-e) is a **binomial** process. Thus, for the mean number of photoelectrons/photon (QE) $p_1$, It follows that the mean number of photoelectrons/photon is $p_1$=**0.17.**

b. Distribution of image-intensifier gain and transfer of light to CCD

Following the creation of photoelectrons in the photocathode they are accelerated (by a photocathode voltage of –200 V) towards the MCP, where they are multiplied and accelerated again (under a voltage of +5400 V) toward the E36 phosphor screen.

Several factors affect the distribution of I-I gain:

- electron transfer from photocathode to MCP-in
- multiplication in MCPs
- generation of light in phosphor

The electronic image-intensifier gain distribution has been determined for a similar intensifier. Both intensifiers were produced by the same manufacturer, only differing in photocathode and phosphor, e.g. different QE (photocathode) and light-yield (phosphor). So, to a very good approximation one can rely on the following measurement (see Figure 64) to describe the gain distribution of the intensifier employed in TRION Gen.2, as both contain the same amplifying MCPs. The image-intensifier gain distribution has been determined by measuring the single electron pulse height distribution on the intensifier phosphor (Dangendorf, 2008). This measurement does not take into account the statistics of the light emission from the phosphor, however, since the average number of photons/electron is large, we do not expect this process to affect the image-intensifier gain distribution to any significant extent. Figure 65 shows the single electron pulse height spectrum.

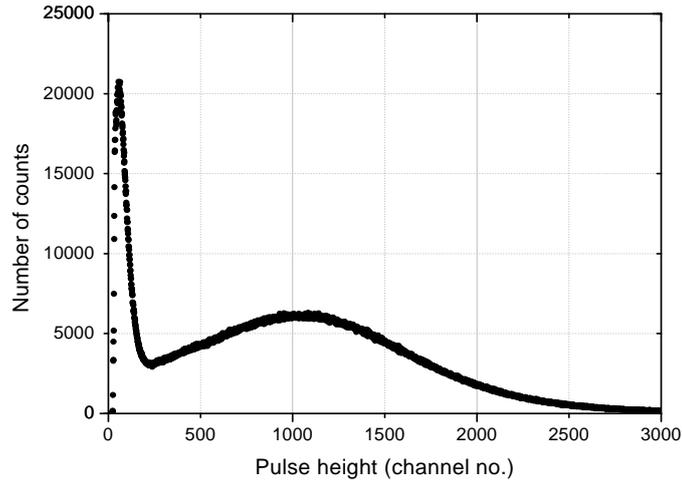

**Fig. 64 Distribution of I-I gain**

3.2.4.3.3 Conversion of light to electrons on CCD sensor

The final process in the cascade is the conversion of photons emitted from the I-I phosphor screen to a CCD camera electronic signal. This conversion occurs at the CCD sensor and is also a binomial process. The quantum efficiency QE of our CCD sensor at 550 nm is 0.55.

3.2.4.4 Excess Noise Factor in TRION

As shown by Eq. 13 (Vartsky et al., 2009), the ENF depends on the weighted squares of the relative standard deviations of all physical processes that contribute to the formation of the signal. Table 8 (Vartsky et al., 2009) shows the value of the above expression for each of the 4 processes previously described for 5 neutron energies.

**Table 8 Value of weighted squares of the relative errors for each process and ENF (Vartsky et al., 2009)**

| Neutron energy (MeV) | $1^{st}$ process Generation of light in fiber screen | $2^{nd}$ process Generation of ph-e on I-I photocathode | $3^{rd}$ process I-I amplification | $4^{th}$ process Generation of ph-e on CCD sensor | ENF |
|---|---|---|---|---|---|
| 2 | 0.97 | 10.8 | 5.7 | $1.4 \times 10^{-4}$ | 4.3 |
| 4 | 0.99 | 3.5 | 1.8 | $4.5 \times 10^{-5}$ | 2.7 |
| 7.5 | 1.0 | 1.4 | 0.8 | $1.9 \times 10^{-5}$ | 2.1 |
| 10 | 0.98 | 1.2 | 0.6 | $1.5 \times 10^{-5}$ | 1.9 |
| 14 | 0.99 | 1.1 | 0.6 | $1.4 \times 10^{-5}$ | 1.9 |

The contribution of the first process is nearly constant with energy and represents the fluctuations in imparted energy and consequently in the amount of light reaching the I-I (see table 7). The contribution of the 2nd and 3rd processes depends inversely on the mean number of photoelectrons produced at the I-I photocathode per neutron and decreases with neutron energy. The contribution of the 4th process is insignificant compared to the other processes due to the large number of photoelectrons generated in the CCD sensor following the light amplification.

The total light created in a scintillating fiber is due to energy deposited by primary, secondary and tertiary protons. Only primary protons carry spatial information of the incident beam. The contribution of the secondary protons will cause minor deterioration of the spatial resolution (especially at high neutron energies) and that of tertiary protons will give rise to a diffused background that may reduce image contrast. However, as is evident from Figure 62 and Table 5, the mean energy of tertiary protons is substantially lower than that of the other contributors, due to the fact that they originate from neutrons scattered within the fiber screen. Thus, although the number of tertiary protons is comparable to the other contributors, they only create small amount of light and will add relatively little to the total number of photoelectrons created at the I-I photocathode.
TRION detector exhibits a relatively large ENF especially at low neutron energies. The main reason for this is the rather low number of photoelectrons produced at the I-I photocathode. In principle, the ENF may be improved by increasing the scintillator light yield, light collection efficiency and quantum efficiency of the image intensifier. It is difficult to achieve a substantial increase in the scintillation yield of a plastic scintillator, however the two other factors can, in principle, be improved.

3.2.5 Summary

A second generation of TRION capable of capturing 4 simultaneous TOF frames within a single accelerator pulse was developed. It exhibits a 4-fold increase in the utilization efficiency of the broad neutron spectrum created following each accelerator pulse. This presents a significant step towards a future real-time operational system.

TRION Gen.2 demonstrates spatial resolution comparable to TRION Gen.1, reduced intensifier thermal noise and improved temporal resolution.

## IV. Spectroscopic neutron detectors developed in this work – part 2

As mentioned in the introduction, two spectrometric detectors were developed as part of this work: an integrative optical detector (TRION) and a fibrous optical detector (or capillary detector).

Chapter 4 will present the capillary detector concept, background, simulations and experimental results obtained with this detector.

## 4. Fibrous optical detector (Capillary detector)

### 4.1 Rationale for the research work

The imaging neutron detector described in the previous chapter performs fast neutron spectrometry by measuring neutron time-of flight. This method requires operating with nanosecond-pulsed neutron source, such as a particle accelerator. The following will describe the development of a detector that permit imaging and spectroscopy of non-pulsed fast neutron sources, such as isotopic neutron sources or reactor beams.

### 4.2 The concept of capillary based detector

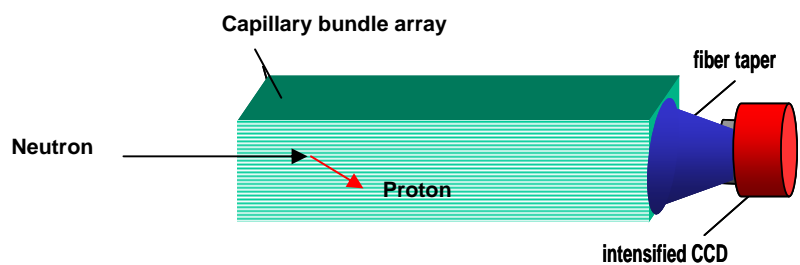

**Fig. 65 Description of the concept of the capillary bundle detector**

Figure 65 shows schematically the concept of the capillary bundle detector. The detector is based on a capillary array filled with liquid scintillator. A fast neutron impinging on the capillary array will undergo an interaction. The dominant fast neutron reactions (0.8 – 14 MeV) within the liquid scintillator are elastic scattering with hydrogen and carbon as both elements have similar atomic density within the scintillator. Scattering by hydrogen is of the highest value since the scattered hydrogen nucleus, also denoted a recoil-proton, is primarily responsible for the excitation of the scintillator molecules that leads to emission of scintillation light. The proton will move within the bundle and create scintillation light within each capillary that it traverses. A fraction of this light will travel to the ends of the capillaries and be recorded by the optical system, thus creating a projection of the proton track.

4.3 Background

In recent years, interest has grown in the methods based on measurement of various parameters associated with recoil proton (track length, scattering direction and amount of light it created). Several groups have been developing such detectors, mainly to determine the direction of the incident neutron (Bravar, 2006; Miller et al., 2003; Ryan et al., 1999; Ryan, 1999), but also for high resolution imaging (Yates and Crandall, 1966).

As mentioned in chapter 1, the energy of the recoil proton after undergoing an elastic scattering with a neutron of energy $E_n$, depends on the scattering angle $\theta_R$ according to:

$$E_R = E_n \cos^2 \theta_R \qquad (15)$$

Where $\theta_R$ is the proton recoil angle relative to the flight direction of the impinging neutron (in the lab reference system). In the past, a hydrogenous radiator and photographic plates were used for measurement of the recoil proton track (Knoll, 2000; Marion and Fowler, 1963). Over the last years there has been increasing use of plastic fiber scintillators or capillary tubes filled with liquid scintillator as neutron detector.

The proton recoiling within a scintillator (liquid or plastic) will lose its energy along a track of tens to hundreds of micro-meters in length (depending on its initial energy) while exciting scintillator molecules, leading to the creation of light. Light intensity within the fiber depends on the amount of energy deposited by the proton within the fiber and its initial energy. Part of the light created will travel along the fiber or capillary according to the total-internal reflection principle. The proton track-length and the intensity of light along its track can be measured by imaging the fibrous array by means of electro-optical techniques (image-intensifiers, CCD cameras).

4.3.1 SONTRAC detector (Bern and New-Hampshire Universities)

This group developed a detector for astrophysical research purposes SOlar Neutron TRACking Telescope (SONTRAC) (Ryan et al., 1999). Their detector is based on a band of plastic fiber scintillators which allows tracking the trajectory of recoil protons scattered twice by non-relativistic neutrons from 14 MeV up to intermediate-energies that come from the sun, or are produced by proton-induced spallation reactions in the atmosphere (Miller et al., 2003; Ryan et al., 1999).

Knowledge of the scattering direction of the recoil proton (in single scatter) and its energy will allow determination of the incident neutron energy (Miller et al., 2003; Ryan et al., 1999). However, when the direction of the incident neutron is not known, as in the case of a neutron arriving from the sun, its energy cannot be determined solely on the basis of a single scatter, but only with two (or more) recoil-protons created in successive scatterings.

The lowest threshold for detectable neutron energy is dictated by the scintillating fiber cross-sectional dimension while the highest threshold is dictated by the size of the fibrous band (Miller et al., 2003). The dimensions of the square fibers which were used were 250 μm and the distance between them was 300 μm.

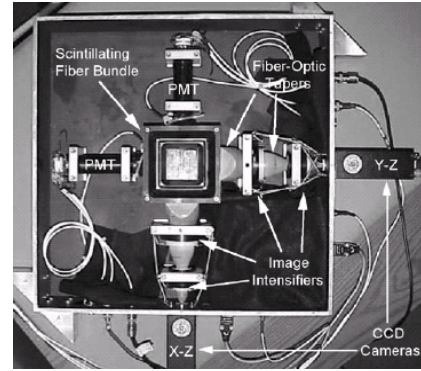

The SONTRAC detector consists of scintillating fiber band of dimensions 5×5×5 cm$^3$, light guides, photomultipliers, fiber-optic tapers and image-intensifiers which are positioned within a light tight enclosure, as seen on Figure 66 (Miller et al., 2003). The two CCD cameras are used as an opto-electronic readout for the scintillating fibers. The fiber band layers are assembled perpendicular to each other such that the proton track can be registered in 3 dimensions. Signals arriving simultaneously at both photomultipliers trigger the image-intensifiers to capture both perpendicular track images and relay them the CCD cameras.

**Fig. 66 The SONTRAC detector (Miller et al., 2003)**

This group has demonstrated that Minimum Ionizing Particles (MIPs) can be distinguished from protons by the shape of their track (Miller et al., 2003). MIPs are characterized by uniform ionization density along their track, while protons exhibit increased ionization density at the end of their track (the Bragg peak), as can be seen in Figure 67. The detector was operated at 20 Hz.

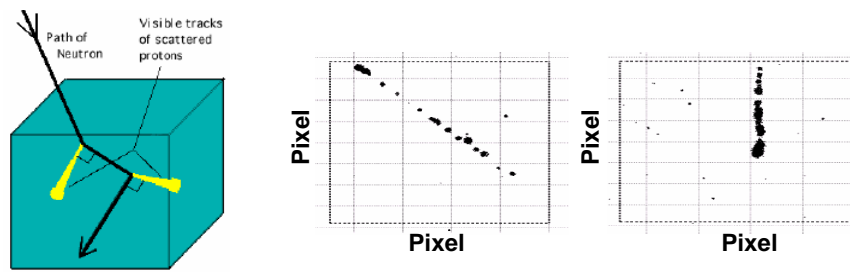

**Fig. 67 The response of the SONTRAC detector to two particle types: left) an example of double neutron scattering, center) a track created by a muon (MIP), right) proton generated track (Miller et al., 2003; Ryan, 1999)**

4.3.2 FNIT detector

In addition to SONTRAC, the above mentioned groups built a Fast Neutron Imaging Telescope – FNIT (designed to cover the energy range of 2-20 MeV) for the purpose of detecting Special Nuclear Material (SNM), such as plutonium or highly-enriched $^{235}$U (Bravar, 2006).

The FNIT detector is based on the following components (Bravar, 2006): scintillating plastic slab BC-404 (12×12×100 cm$^2$ and 1.5 cm thick), 64 round Wave-Length Shifting (WLS) fibers (0.1 cm in diameter) arranged in a perpendicular configuration with spacing of 0.375 cm, a 16-channel Multi-Anode Photo-Multiplier Tube (MAPMT) (4×4 array) and mirrors attached to the free ends of the fibers.

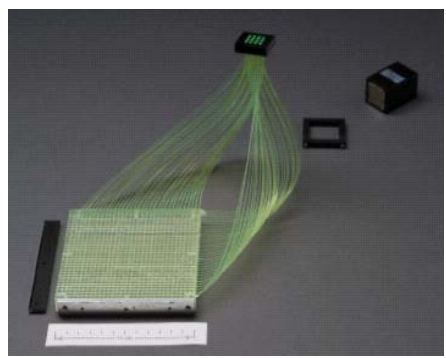

**Fig. 68 The FNIT detector. 64 WLS fibers banded into an array of 4x4. This array is coupled to a MAPMT, seen on the upper right corner (Bravar, 2006)**

Each group of 4 fibers is attached to a single channel of the MAPMT, i.e., 64 fibers feed 16 channels (Bravar, 2005), as shown in Figure 68. The results achieved with this detector and MIP exhibited energy resolution of 16 % at energy of 2.6 MeVee, temporal resolution of 0.6 ns at the same energy, spatial-resolution of 0.85 cm for 2 MeVee and threshold detection energy of 0.125 MeVee (Bravar, 2006). The postscript 'ee' to the MeV labels represents 'electron-equivalent' energy.

4.3.3 Sandia lab

This group has developed a directional detector for neutrons of 14 MeV (Peel et al., 2006). With this detector, the direction of the neutron is reconstructed by finding the direction and energy of the recoil proton from a single elastic scatter.

The detector is composed of an array of 64 square scintillating plastic fibers (BCF12) 0.5× 0.5×100 mm$^3$, such that the spacing between fibers is 2.3 mm (Mengesha et al., 2006). The fiber core is made of polystyrene while the cladding is acrylic, of cross-sections that are 4 % of the core size. Each fiber is attached at its end by optical grease to an anode which is part of an MAPMT array. The PMT is read by suitable electronics. The fibers are covered by a light-tight aluminum envelope. Light cross-talk between fibers was reduced by applying a coat of absorbing black paint. Figure 69 shows a schematic and a photograph of the detector (Mengesha et al., 2006; Sunnarborg et al., 2005).

Although this detector was built to detect the direction of the impinging neutron it can also, in theory, permit performing neutron spectroscopy.

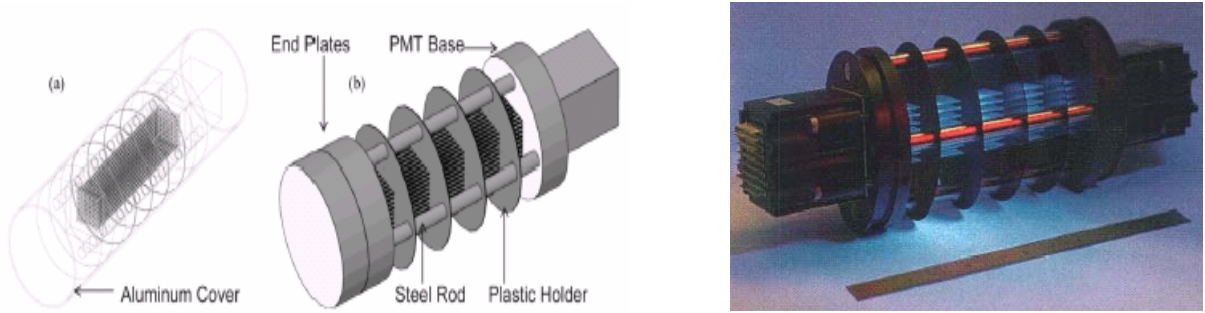

Fig. 69 The Sandia detector. left) The band of fibers consisting the detector as modeled in GEANT simulation code (without the aluminum envelope) (Mengesha et al., 2006), center) An overall view of the detector as modeled in the simulation program, right) The detector in reality (without the aluminum envelope) lit from behind (Sunnarborg et al., 2005)

4.3.4 CAE-DAM group, France

In order to improve image quality of neutrons emitted from the fusion reaction (d-t) as a result of a high-power laser beam, this group (Disdier et al., 2004, 2006) developed a capillary detector composed of 85 μm diameter capillaries filled with liquid scintillator. The size of the capillary array is 100×100 mm² and its length is 50 mm. The capillaries were filled with liquid scintillator with refractive index of 1.58 while the refractive index of the capillary walls was 1.49 (Disdier et al., 2004).

A mirror is attached to one face of the array, while an image-reducer 80 mm in diameter was attached to the other. This reducer demagnified the size of the emitted light-image to the diameter of a time-gated image-intensifier (IIT, 40 mm diameter) coupled to a CCD camera (1000×1000 pixels). Each camera pixel captured 38×38 μm² in the capillary array plane. According to their calculations, for a single neutron, about 330 photons reach the photocathode of the IIT and are translated to ~56 photoelectrons. Due to the limited number of CCD pixels only part of the capillary array was captured (about a third).

Figure 70 (left) provides a schematic illustration of the detector configuration and Figure 70 (right) displays an image of a recoil proton track scattered by a 14 MeV neutron, as captured by the capillary detector.

The spatial resolution for 2.45 MeV neutrons is 200 μm. No neutron spectroscopy was performed with this detector.

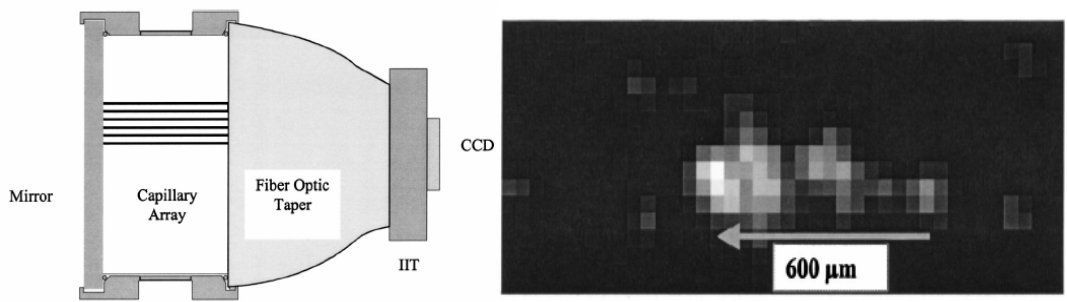

**Fig. 70 The CAE-DAM detector. left) Schematical illustration of the detector configuration, right) a recoil proton track scattered by 14 MeV neutron as captured by the detector (Disdier et al., 2004)**

Figure 71 shows the relation between photon yield and recoil proton track length as measured by (Disdier et al., 2004).

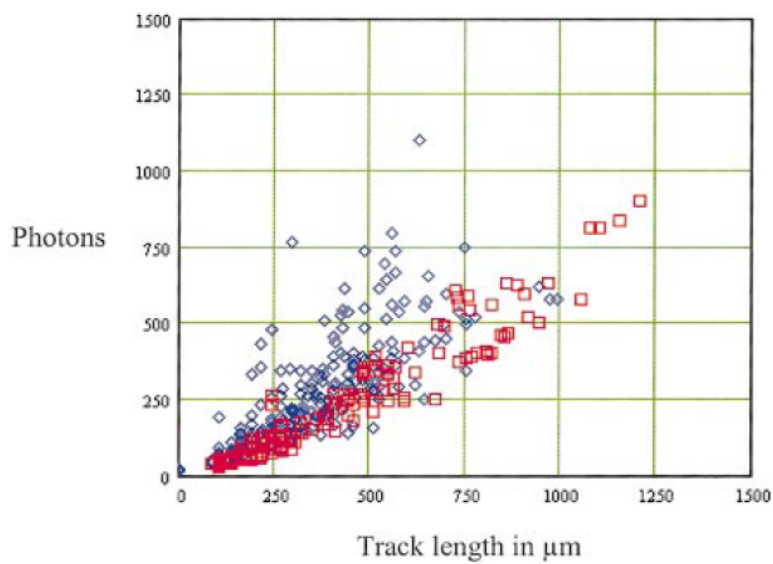

**Fig. 71 Photon production vs. observed recoil proton path length. The number of collected photons per neutron interaction exhibits linear variation with the track length (in μm) when the neutrons are incident normal to the capillary axis (red squares), whereas the relation is not straightforward when the neutron direction is parallel to the capillary axis (blue diamonds) (Disdier et al., 2004)**

When the neutron direction is parallel to the capillary axis, the observed track length is the projected range of the recoil proton on the output face of the detector. As the most energetic recoil protons are scattered along the capillary axis according to Eq. 15, the track projection does not scale with the photon yield. The experimental results shown in Figure 71 demonstrate that there is a linear relation between photon yield and track length when the capillary axis is perpendicular to the incident neutron direction.

4.3.5 Waseda university, Japan

This scintillating fiber camera (Terasawa et al., 2000, 2001) was developed for the purpose of measuring the dose contribution from fast neutrons in the spacecraft environment. As such, there was no need for fast neutron spectral measurement with this detector.

The camera consists of a scintillating fiber stack, an image-intensifier unit and photomultipliers for triggering events. Schematic views of the stack and the camera system are shown in Figure 72 a & b respectively.

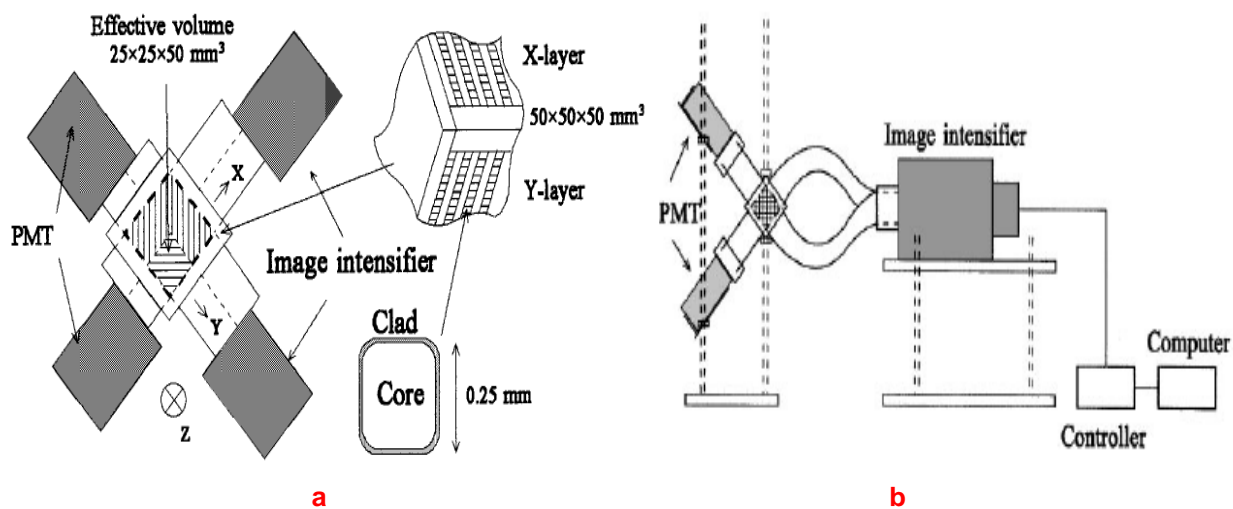

Fig. 72 Structure of scintillating fiber stack and fiber camera. a) Structure of a scintillating fiber stack. Each scintillating fiber has a cross-section of 0.25×0.25mm$^2$. The stack is composed of 200 layers of fiber sheets. Each layer consists of 200 fibers and the layers are alternatively stacked to be perpendicular to each other (Terasawa et al., 2000). b) System of a scintillating fiber camera. The camera consists of a scintillating fiber stack, an image intensifier unit and two photomultipliers (Terasawa et al., 2001)

Each scintillating fiber (SCSF-78, Kuraray Co.) has a cross-section of 0.25×0.25 mm$^2$ and is composed of core and cladding. The core is mainly made of polystyrene doped with two kinds of wavelength shifters which convert UV-light from polystyrene to visible light peaking at around 450 nm (Terasawa et al., 2001).

The scintillating fiber stack is composed of 200 sheets of scintillating fibers. Each sheet consists of 200 fibers and the sheets are alternatively stacked perpendicular to each-other. The effective volume of the camera is limited to 25×25×50 mm$^3$. Two 2D projections of the three-dimensional tracks, which correspond to the Y-Z plane and X-Z plane, respectively, are coupled to the image-intensifier unit, consisting of a two-stage image-intensifier and the CCD camera for the image readout. The Z-direction is perpendicular to both fiber layers and almost parallel to the incident direction of neutrons and charged particles. From the images of two 2D projections on the image-intensifier unit, the three-dimensional image is reconstructed. The opposite ends of the stack are connected to two photomultipliers as seen in Figure 72, and are almost fully viewed by the photomultipliers. The signals from the photomultipliers are used as triggering signals for the image-intensifier and the readout electronics system.

Quasi-monoenergetic neutrons, whose peak energy was 66 MeV, were produced by bombarding a $^7$Li target with 70 MeV protons. Typical images of a primary proton (fired directly into the detector with the Li target removed) and a recoil proton produced by a neutron are shown in Figures 73 a & b, respectively. Figures 73 c & d display a recoil proton track from the D-T reaction and an electron track produced by a γ-ray, respectively (Terasawa et al., 2001).

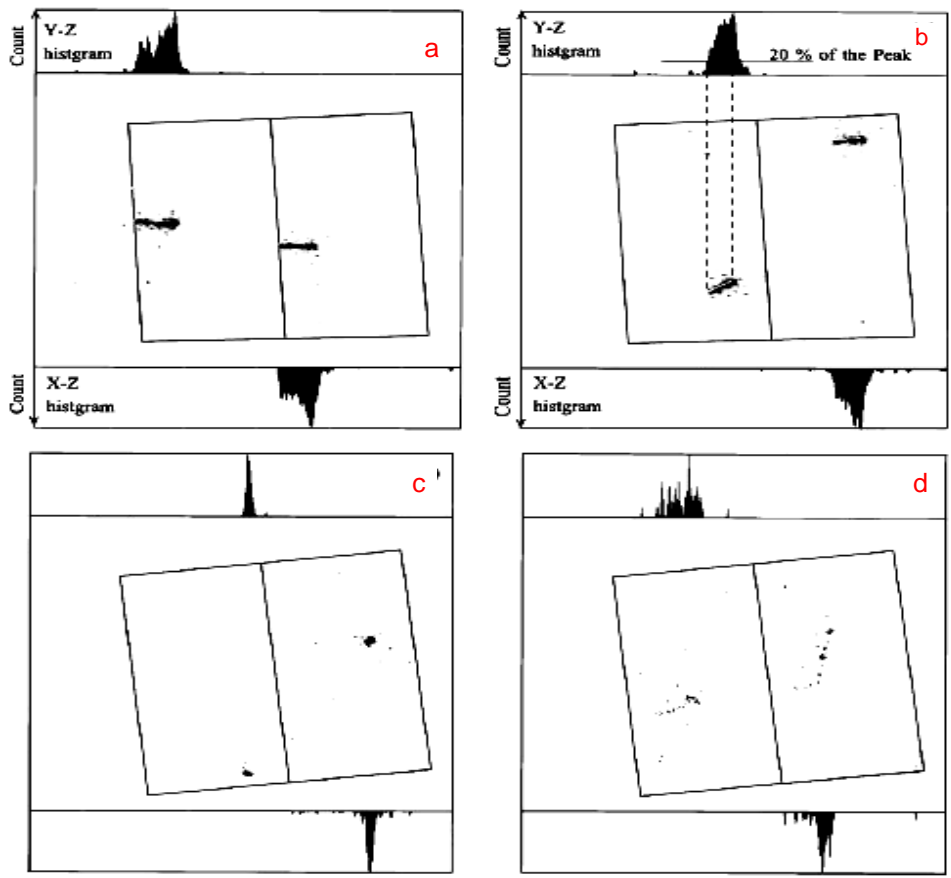

**Fig. 73 Tracking images obtained with the scintillating fiber camera: (a) a primary proton, (b) a neutron (a recoil proton), (c) a D-T neutron (a recoil proton), (d) a γ-ray (an electron). Each region has two sub-regions. The left one corresponds to Y-Z and the right to the X-Z projection (Terasawa et al., 2001)**

In Figures 73(a)-(d), the upper and lower histograms indicate the luminance sum of CCD

pixels, which has the same horizontal axis coordinate on the CCD plane. The upper histogram is for the Y-Z plane and the lower one for the X-Z plane. These histograms are considered as corresponding to that of the light intensity per unit length along the track. Each image is divided to two regions, the left one corresponds to the Y-Z projection and the right one to the X-Z projection.

Figures 74 a & b show the relation between the total number of active CCD pixels (pixels of non-zero luminance) and the sum of the light yield from these pixels (Terasawa et al., 2000). In Figure 74a, recoil protons can be identified from electrons by noting the difference in the scintillation light intensity per unit track length, obtained from the ratio of the light yield to the number of active CCD pixels. However, in Figure 74b, only higher energy protons are identified. The threshold energy for protons to be distinguishable from electrons is 5 MeV. Because of the small dE/dX of electrons, observed tracks of electrons by the camera are not completely continuous, that is, there are non-active pixels along the track. A fiber corresponds to several pixels. If the light yield from fibers is quite large, all pixels corresponding to the fiber can be active. In the case of small light yields, however, some of those pixels are not active. Therefore, the number of active pixels for electrons is evaluated to be smaller. The scintillation light intensity per unit track length for electrons is larger and then the identification capability deteriorates. If the track length of electrons can be precisely measured by a pattern-recognition method, the discrimination capability between protons and electrons at lower energies is expected to improve (Terasawa et al., 2000).

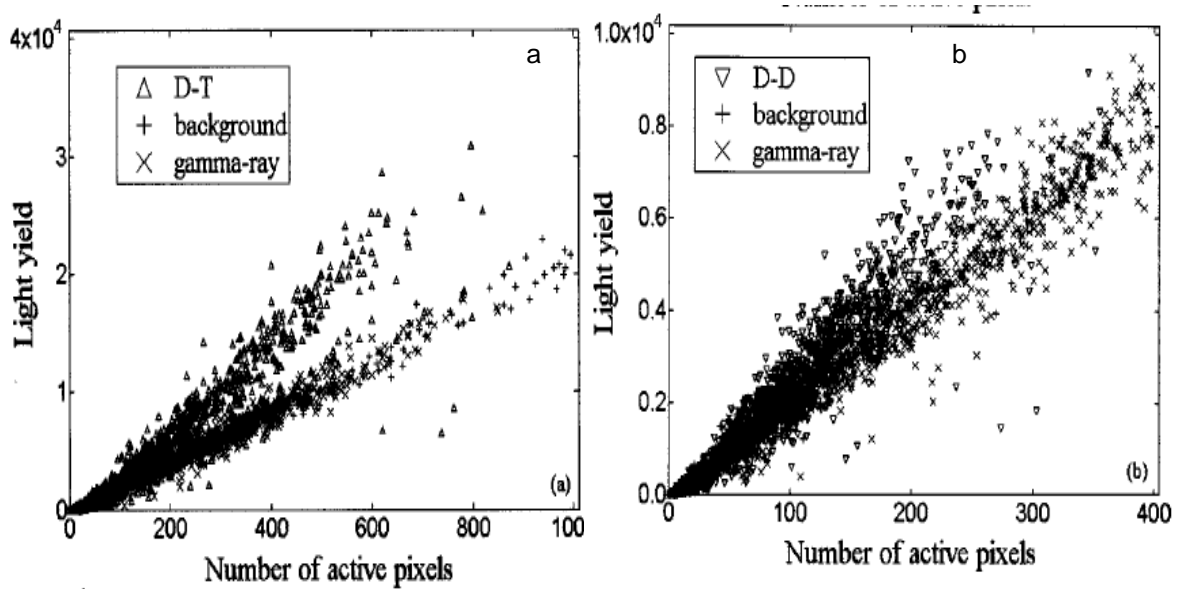

**Fig. 74 The relation between the number of active pixels and the light yield for a) higher energy and b) lower energy particles**

4.4 The novelty of this work

The detectors built for neutron spectroscopy purposes were based on a plastic fiber array (SONTRAC and FNIT). The relatively low energy resolution achieved with these detectors was caused by the relatively large dimensions of the fibers (about 0.5 mm) and the non-negligible spacing between them. For the same reason, the spatial resolution of these detectors is also low and does not allow imaging to better than 1 mm.

The detector built by the CAE-DAM group, based on a capillary array filled with liquid scintillator, presents similar properties to the capillary detector presented in this work. However, the CAE-DAM detector presents lower spatial resolution determined by capillary diameter and the number of available CCD pixels. The lower spatial-resolution affects the energetic resolution.

The detector of the present work is intended for energy resolved high spatial-resolution radiography. If the incident neutron flight direction relative to the capillary bundle axis is known (this is true for radiographic systems) and there are no multiple neutron interaction in the bundle, one can calculate the incident neutron energy using two parameters that characterize the proton track: the amount of light created along the recoil proton track and the recoil proton track-length projection on the capillary array face plane.

Capillaries with a diameter of about 10 microns should allow measurement and reconstruction of the recoil-proton track with sufficient accuracy for neutrons with energy of several MeV. Correspondingly, the expected spatial-resolution defined of this detector will be of the order of tens of microns.

4.5 The capillary detector setup

The capillary detector, seen in Figure 76, consists of the following components: capillary array filled with liquid scintillator, tandem reversed-stacked lens configuration, time-gated image-intensifier, cooled CCD camera.

Scintillation light-image created by an impinging neutron via recoil-proton is captured by a tandem reversed-stacked lens configuration (50 & 200 mm lenses) which transmits it magnified onto time-gated image-intensifier. The gating function of the intensifier plays a role only during initial calibration and detector characterization measurements.

Figure 75 provides a schematic illustration of the detector setup.

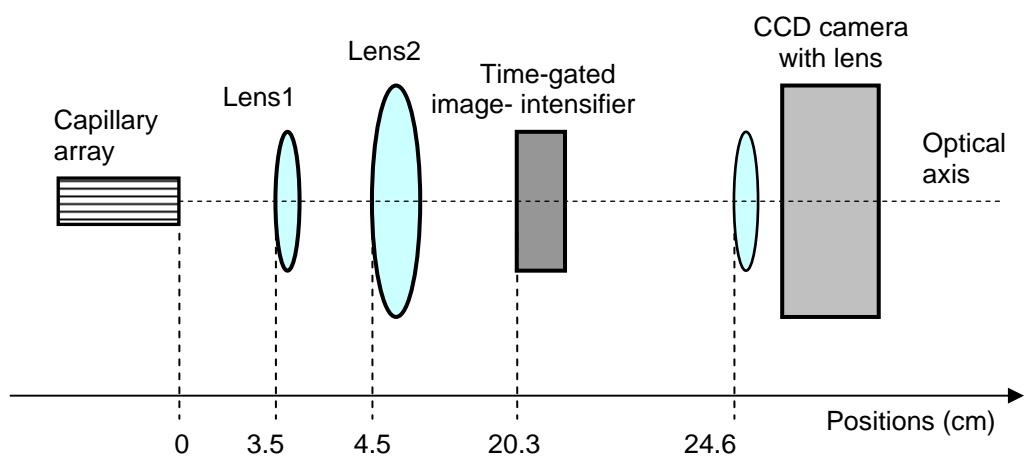

**Fig. 75 Schematic illustration of the capillary detector setup**

Figure 76 shows the actual image of the capillary detector.

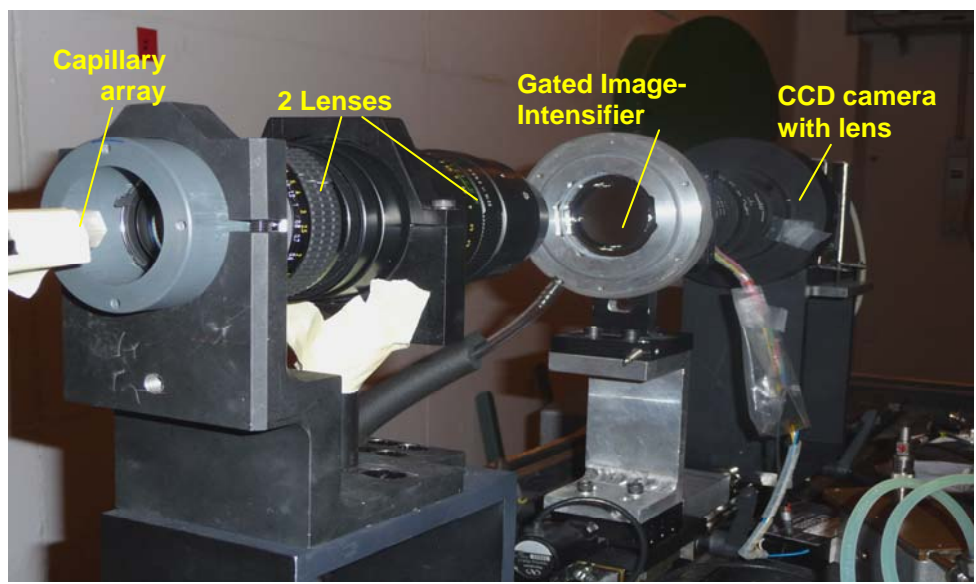

**Fig. 76 The capillary detector**

The following will describe in detail the constituent components of the capillary detector.

4.5.1 Capillary array

In order to achieve high energy-and-spatial resolution, the capillaries cross-sectional dimensions should be as small as possible. Figure 77 shows a small capillary array made from silica glass (1.36 cm in width, 3 cm in length, refractive index n=1.4632) made by the XOS company (XOS, 2011). The dimensions of a single capillary are: inner average diameter 11 μm, wall thickness 1 μm, the available free area (excluding the triangular patterns in Figure 77) is 71.5% (Kamtekar, 2007).

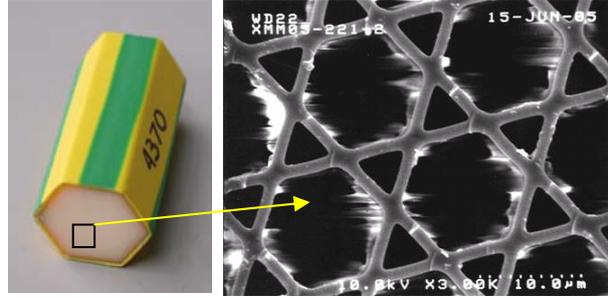

**Fig. 77 left) The capillary array manufactured by XOS (XOS, 2011), right) enlarged section of the capillary array showing individual capillaries imaged by an electronic microscope (Kamtekar, 2007)**

4.5.1.1 Capillary array filling process

The capillaries were filled with EJ309 scintillator manufactured by ELJEN (ELJEN Technology, 2011), presenting the following characteristics: n=1.57, photons per 1 MeV e- = 11,500, density: 0.965 gr/cm$^3$, H/C = 1.25, light % of anthracen: 75%.

This scintillator was chosen as it presents the highest combination of high refractive index, hydrogen content and light output compared to other liquid scintillators (ELJEN Technology, 2011).

The filling process relied on the capillary phenomena. Eq. 16 describes the height attainable by the liquid based on the capillary phenomena (Gunji et al., 2005):

$$h = \frac{4T \cos\theta}{d\rho g} \qquad (16)$$

Where h is the height of the liquid, T is the surface tension, θ is the angle of contact between the liquid and capillary surface, d is the capillary diameter, ρ is the liquid density and g is the gravitational acceleration.

Expected filling height for the above capillaries with EJ309 (assuming T of EJ309 is similar to that of liquid toluene = 27.93×10$^{-3}$ (N/m) and a contact angle of 80$^o$) is of the order of 18 cm, more than sufficient for the above mentioned capillaries.

The filling procedure was based on that of (Gunji et al., 2005). During their filling tests, this group found that when their capillary array was laid upon a cotton cloth dipped in the liquid scintillator and left to fill, more uniform filling was observed, compared to the straightforward approach of dipping the capillary array itself in the liquid scintillator. Both approaches are schematically illustrated in Figure 78.

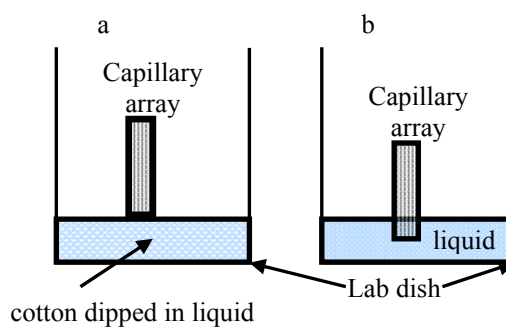

Fig. 78 Capillary array filling: a) The capillary array immersed in the liquid scintillator. b) The capillary array rests upon a cotton sheet immersed in the liquid scintillator

The capillaries were left to fill over the weekend.

The next step involved sealing the exposed top-end of the filled array. Sealing was performed by applying uniformly a thin layer of semi-hardened optical cement BC600 (Saint-Gobain, 2011) and uniformly removing excess material with a thin sharp rim such as a razor-blade, leaving a layer of several microns. As the refractive indices of the scintillator (1.57) and BC600 (1.56) are almost equal, the addition of an extremely thin layer of the optical-cement has only a negligible effect on light emission from the capillary array.

After 24 hours, upon full hardening of the BC600 layer, the above sealing procedure was repeated for the bottom end of the array. Prior to applying the optical cement, any excess liquid was removed.

Upon completion of the sealing process, the uniformity of the filling was tested by transmitting light through the capillaries and examining the captured image. Figure 79 shows a section of the illuminated capillary array. As can be seen, there are some dark regions which do not transmit light and numerous single dark capillaries. Nevertheless, upon inspection of the whole illuminated array image, the conclusion is that the fraction and dispersion of dark regions and unlit capillaries is such that it will not prevent the capillary array from serving its purpose.

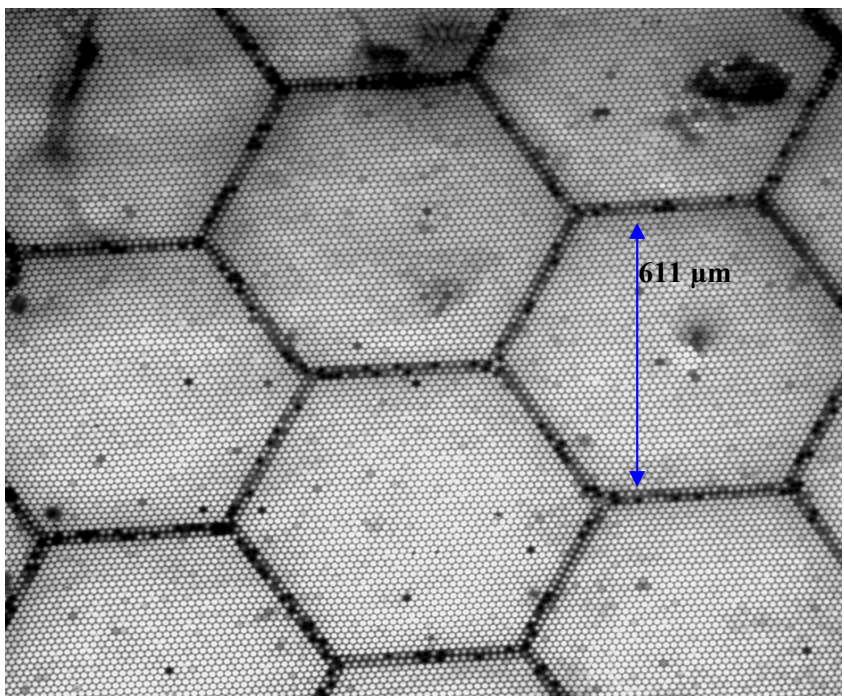

**Fig. 79 A small section of the illuminated capillary array**

4.5.2 Tandem reversed-stacked lens configuration

A lens is designed to accept light in its front end and bend it sharply into a stubby cone converging to the CCD plane. If the lens is turned around, light will follow the same path in reverse -- sharply diverging light from an object very close to the lens' rear element will be focused on the film. This allows one to achieve very high magnifications, especially with wide-angle lenses. The magnification is determined by the prime lens magnification divided by that of the reversed lens. The detector configuration utilized 200 and 50 mm lenses, as seen in Figure 80, resulting in a magnification ratio of 4.

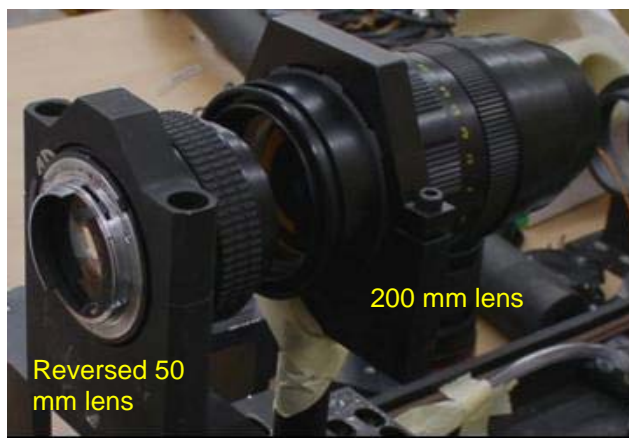

**Fig. 80 Tandem reversed-stacked lens configuration, used in the capillary detector**

4.5.3 Image intensifier

The light emitted from the capillary array and relayed via the tandem reversed-stacked lens configuration is captured by an image-intensifier (25 mm in diameter), seen in Figure 81, consisting of two Multi-Channel Plates (MCP) (Proxitronic, 2011). The entrance and exit windows are composed of optical fibers (6 μm in diameter).

The image-intensifier was able to operate in pulsed mode, allowing the verification of reconstructed neutron energy against expected values based on neutron time-of-flight measurements.

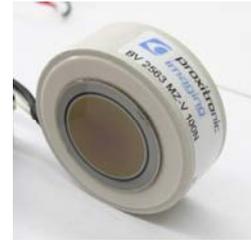

**Fig. 81 The image-intensifier manufactured by Proxitronic**

### 4.5.4 CCD camera

The cooled CCD camera employed in the capillary detector was ML16083, manufactured by Finger Lakes Instrumentation (Finger Lakes Instrumentation, 2011). CCD size is 4096×4096 pixels, pixel size 9×9 um.

### 4.6 Irradiation geometries

Two irradiation geometries, schematically described in Figures 83 & 84, were considered:

Irradiation **parallel** to the capillary array axis and irradiation **normal** to the capillary array axis. Figure 82 illustrates the two configurations.

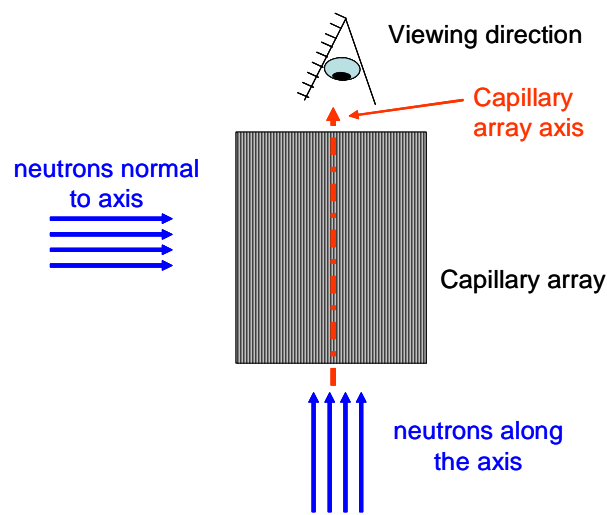

**Fig. 82 Schematic illustration of irradiations parallel and normal to capillary array axis**

4.6.1 Irradiation parallel to the capillary array axis x

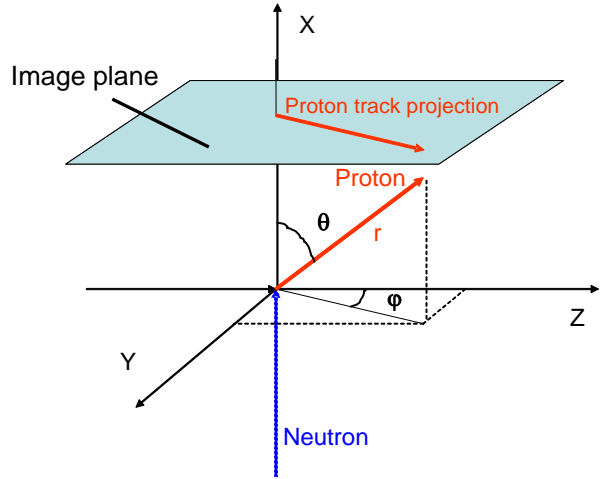

**Fig. 83 Irradiation along capillary array central axis**

In a configuration such as seen in Figure 83, axial symmetry exists and the length of track projection is dependent only on θ and proton energy. The energy of the proton is $E_p = E_n \cos^2 \theta$. The length of the track **r** is a function of proton energy. In such configuration the track projection **Proj** is related to track length **r** by:

$$\textbf{Proj = r sin}\theta \qquad (17)$$

The projections are usually short, since forward going protons (small θ) will have high energies (thereby generating more light) but short track projections. As θ increases the proton energy decreases, resulting in shorter tracks and lower light intensities.

This configuration permits neutron imaging with very high position resolution of tens of microns.

4.6.2 Irradiation normal to the capillary array axis x

Here the projection **Proj** is related to track length **r** by:

$$\text{Proj} = (r^2 \sin^2 \theta' \cos^2 \varphi' + r^2 \cos^2 \theta')^{1/2} \qquad (18)$$

There is no axial symmetry along the capillary axis but, contrary to the previous configuration the projection length always increases with the length of the proton track.

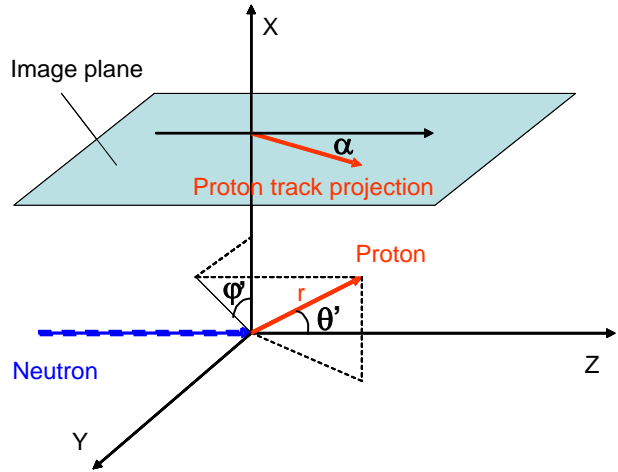

**Fig. 84 Irradiation normal to capillary array central axis x**

The projection angle **α** relative to the direction of the incident neutron is:

$$tg\alpha = tg\theta'\cos\varphi' \qquad (19)$$

After substituting the above expression with Eq. 18 we obtain:

$$\text{Proj} = r\cos\theta'(1+tg^2\alpha)^{1/2} \qquad (20)$$

For **φ'=90°** and **α=0°** the projection **Proj** is reduced to:

$$\text{Proj} = r\cdot\cos\theta' \qquad (21)$$

Thus for small **α**'s (or forward going protons) one can use the above expression.

In this configuration neutron imaging is not possible and the detector can only be used as a spectrometer.

4.7 Determination of neutron energy

When using the capillary detector, the only two measurable quantities available to us are:
- Total amount of light created along the recoil proton track- **L**
- Track projection length- **Proj**.

Ignoring for the time being the energy loss in capillary walls, the amount of light **L** is proportional to the proton energy $E_p$. Knowledge of $E_p$ should, in principle, yield information on the real track length **r** in 3D. Knowing the relation between measured

**Proj** and **r** provides us an estimate of **θ**. Using $E_p$ and the relationship $E_p = E_n \cos^2\theta$ we can now determine $E_n$.

Clearly in a physical system of liquid capillaries where capillary walls cannot be neglected, the situation is more complicated. In order to study this issue we simulated the capillary system via Monte-Carlo simulations.

4.7.1 Simulations of capillary detector

As experimental results provide only limited information about the recoil-proton, namely, total light per track (L) and track projection length (Proj), detector simulations were performed in order to gain better understanding of its operation and to find missing relations among the measurable parameters that are relevant to the recoil-proton behavior within the capillary array. The missing relations should facilitate reconstruction of the impinging neutron energy. These relations are:

- Recoil proton real track-length (**r**) in 3D to its energy (Ep)
- Scintillation light (L) to recoil-proton real track-length (**r**) in 3D or its energy (Ep)

4.7.1.1 Simulation description

An array of 500×500 round silica capillaries, of which a section is shown in Figure 85, was simulated using Geant 3.21 (CERN, 2003). Capillary dimensions were: 11 μm in diameter, 1 μm wall thickness, 3 cm long. The scintillator within the capillaries was EJ309: density: 0.964 gr/cm$^3$, H/C 1.249 (ELJEN Technology, 2011). The neutron source was a pencil beam source uniformly distributed over an area of 0.1×0.1 cm$^2$.

As described above, the simulation work examined irradiation of the capillary array both <u>parallel</u> to the capillary array axis and <u>perpendicular</u> to it.

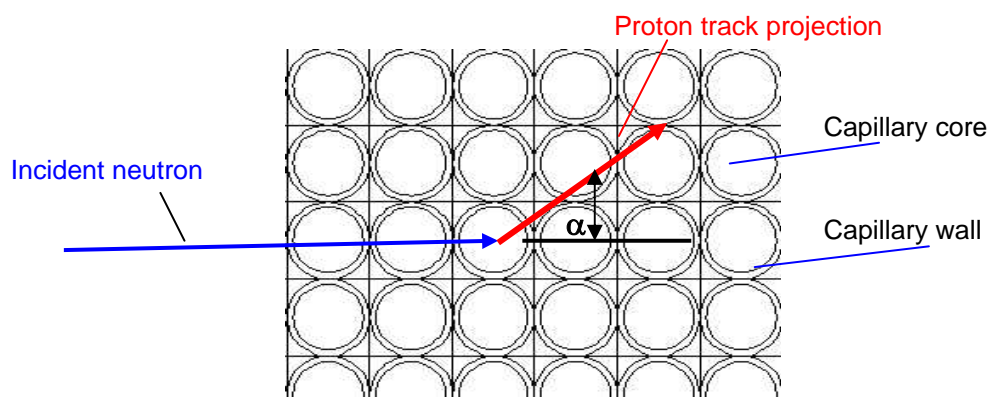

**Fig. 85 A section of the simulated capillary array (perpendicular irradiation)**

4.7.1.1 Neutron energy reconstruction from simulations

The following parameters were recorded for each recoil proton:

- Number of capillaries traversed, e.g. track projection length (**Proj**)
- X,Y positions of the capillaries along the track (for image creation)
- Full track length in 3D (**r**)
- Proton creation energy
- Proton energy upon entering and exiting each capillary (for light creation calculation)
- Deposited energy per capillary
- Number of photons created in each capillary along the track (see section 3.2.4.3.1-b for detailed explanation on conversion from proton deposited energy to electron equivalent energy, and then to number of scintillation photons)
- Proton origin – indicates whether the proton was created in capillary core or in the capillary wall
- Determination of $\alpha$ angle (for perpendicular irradiation)

4.7.1.1.1 Irradiation normal to capillary array

The advantages presented by perpendicular irradiation in comparison to parallel one are:

a) lower radiation induced background events directly in the electronic and optical detector components
b) minimal radiation damage to detector components
c) simpler correspondence between proton energy and track projection-length. Specifically, this means that for a given neutron energy, the highest energy protons will continue in the direction of the impinging neutron, resulting in traversal of a maximal number of capillaries, while in parallel irradiation, the highest energy protons will create the shortest track-projections as they move parallel to the capillary central axis, depositing all their energy within.

The principal drawbacks of the perpendicular irradiation approach are:

i) inability to perform imaging
ii) lack of axial symmetry along the capillary central axis, such that track projection is dependent on both polar angle $\theta$ and azimuthal angle $\varphi$.

The advantages presented by parallel irradiation geometry are:

- ability to perform imaging
- existence of axial symmetry along the capillary central axis.

The main goal of this preliminary study was to examine the spectroscopic capability of this detector rather than its imaging capability. Additionally, since the advantages presented by the perpendicular irradiation outweigh its drawbacks as well as the advantages presented by the parallel one, the perpendicular geometry was selected for further investigation.

Figures 86-87 show plots of two relations necessary for the reconstruction procedure. These were derived from simulation of 20, 10 and 4 MeV neutrons impinging perpendicularly on the capillary array as described above. The following protons were excluded from the following plots: protons which escaped the capillary array before depositing their full energy within, those originating from the capillary walls and those exhibiting track-projections shorter than 4 capillaries.

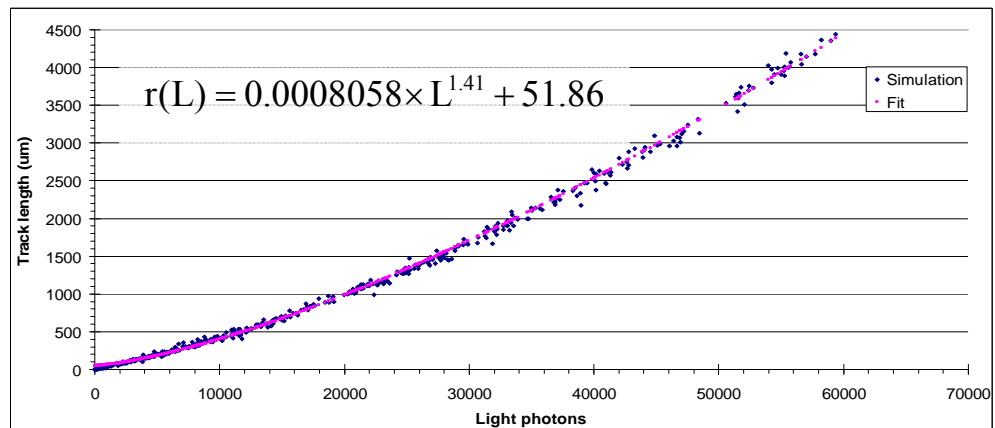

**Fig. 86 Track length (r) vs. total number of light photons (L) created per track**

$$r(L) = 0.0008058 \times L^{1.41} + 51.86$$

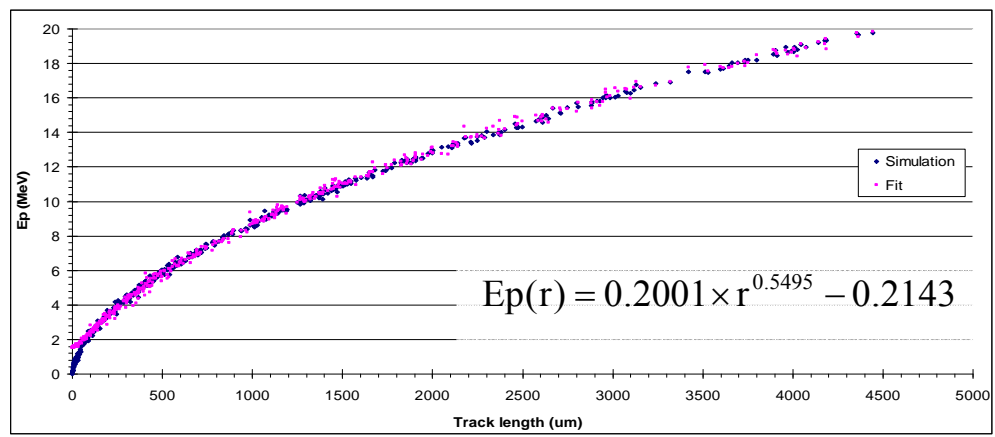

**Fig. 87 Proton energy (Ep) vs. track length (r)**

$$Ep(r) = 0.2001 \times r^{0.5495} - 0.2143$$

A mathematical expression was fitted for each of the above relations (see Figures 86 - 87)

using Matlab (MathWorks, 2011):

a. Track length (**r**) vs. total number of light photons (L) created per track (with 95% confidence bounds) -

$$r(L) = (0.0008058 \pm 3.9 \times 10^{-5}) \times L^{(1.41 \pm 0.005)} + (51.86 \pm 2) \quad (22)$$

b. Proton energy (Ep) vs. track length (**r**) (with 95% confidence bounds) -

$$Ep(r) = (0.2001 \pm 3.8 \times 10^{-3}) \times r^{(0.5495 \pm 0.0023)} - (0.2143 \pm 0.0256) \quad (23)$$

The above relations provided the basis for reconstruction of the impinging neutron energy per each of the recoil-protons which were recorded in the simulation.

Figure 88 a & b show two recoil proton track-projections (different magnification scale) obtained from the previously-described simulations (20 MeV incident neutrons), 132 and 6 capillaries long (1716 and 78 μm), respectively. As can be seen, each track exhibits increased scintillation light emission at one end indicating a Bragg peak.

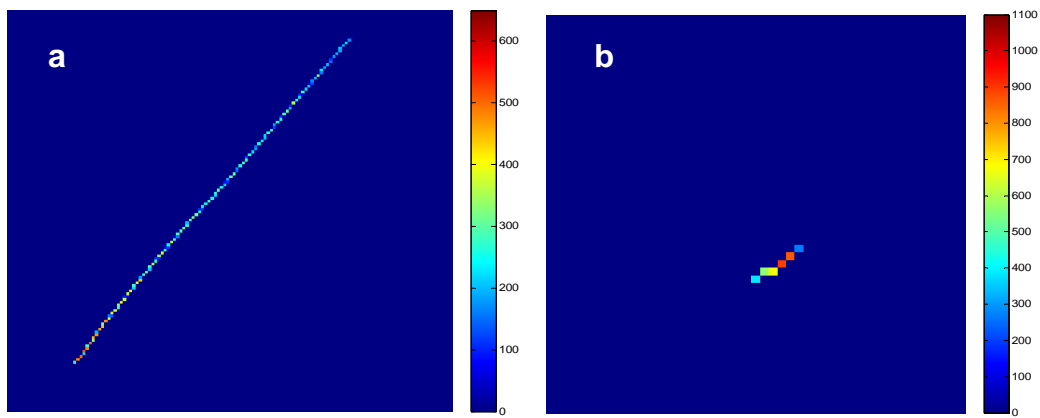

**Fig. 88 Images of simulated recoil proton track-projections at different magnification scales: a) 132 capillaries long (1716 μm), b) 6 capillaries long (78 μm)**

4.7.2 Neutron energy reconstruction from simulations

The simulations were performed with 4 neutron energies: 4, 10, 13 and 20 MeV.

The reconstruction procedure was performed in a manner similar to the experimental one where only track projection length (Proj) and amount of light created along the track (L) are known:

1. Based on Eq. 22, real track length in 3D (**r**) (as function of light **L**) was obtained

2. Based on Eq. 23, proton energy (**Ep**) (as function of real track length in 3D) was determined

3. Cosine of the proton scattering angle (θ') was calculated in two ways:

a. Based on Eqs. 20 and 21 (Proj/r combined with α angle-based correction)

b. Based on Proj/r with α angle-based filtration, selection of α angles from within a predefined range (no α angle-based correction). This was performed in order to demonstrate that for relatively small alpha angles, one can achieve correct reconstruction with no correction of the proton scattering angle (θ') Cosine. To obtain acceptable energy resolution (~2 MeV) in this approach, only recoil protons of $-20°<α<20°$ were selected for the energy reconstruction procedure. As the α angle selection range is narrowed, energy resolution improves at the expense of counting efficiency.

4. Neutron energy found based on the above: $E_n = E_p / \cos^2(\theta')$

Figure 89 a-d show two series of reconstructed neutron energy spectra for incident neutrons of 4, 10, 13 and 20 MeV. The left-hand column of spectra ending with '1' (a1-d1) show spectra reconstructed following α angle-based correction while the theright-hand column of spectra ending with '2' (a2-d2) show α angle-based filtration $(-20°<α<20°)$.

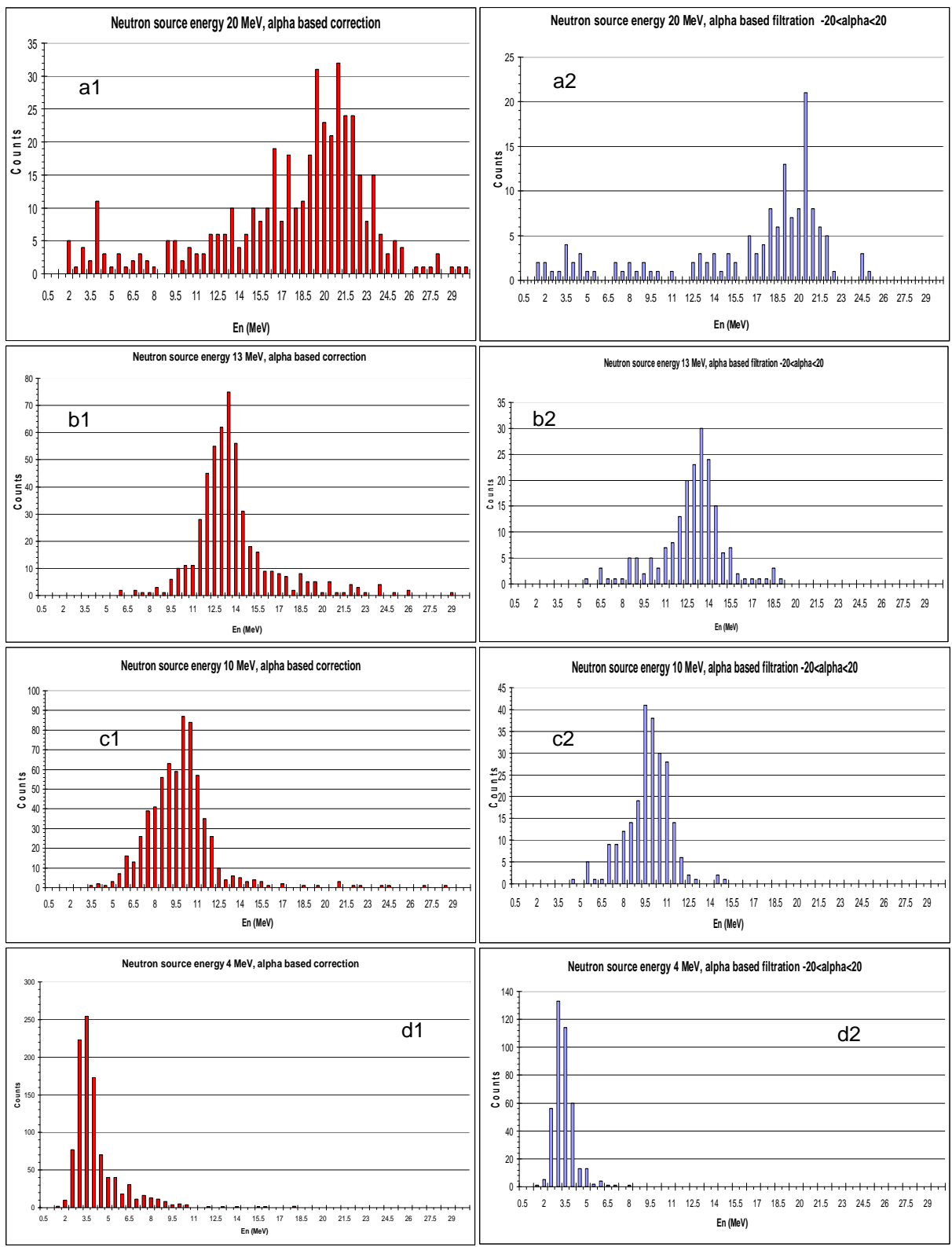

**Fig. 89** Two series of reconstructed neutron energy spectra from simulation for incident neutrons of 20, 13, 10 and 4 MeV. The series ending with '1' (a1-d1) show spectra reconstructed following <u>*α* angle-based correction</u> while the series ending with '2' (a2-d2) show <u>*α* angle-based filtration</u> (-20°<*α*<20°)

As rejection of short tracks improves the accuracy of the reconstructed distribution, both series of reconstructed spectra excluded track projections shorter than 4 capillaries. The Full Width at Half Maximum (FWHM) resolution of the above spectra ranges between 1-3 MeV. Noticeably, the peaks are quite broad at the base of the distribution. The main reason for this is the inaccuracy in counting the number of capillaries and converting them into projection length (in μm), especially for short projections. Therefore, keeping the diameter of the capillaries and the thickness of the walls as small as possible is of utmost importance.

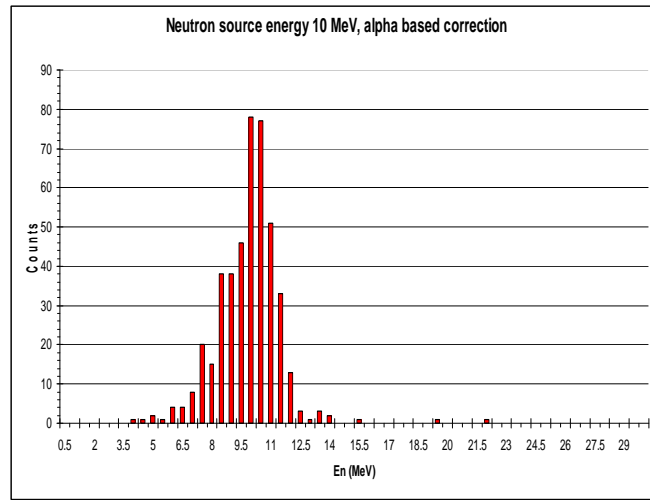

The above distributions can be improved by extending the rejection criteria. Figure 90 shows a reconstructed 10 MeV neutron spectrum for tracks with projection length extending beyond 15 capillaries. As can be observed, the spectrum is narrower (FWHM improved to 2 MeV from 3 MeV), with significantly reduced tails. However, rejecting track projections shorter than 4 and 15 capillaries reduces the counting efficiency at this neutron energy by 20 and 47 %, respectively.

**Fig. 90 Reconstructed neutron spectrum from 10 MeV incident neutrons, analysing only track projection lengths>15 capillaries**

4.8 Experiment description and results

During the year 2010, the capillary detector (described in detail in section 4.5) was tested in a neutron beam using the PTB cyclotron. The capillary array was irradiated from the side, i.e., perpendicular to its central axis, by neutrons created with the $^9$Be(d,n) reaction using a 12 MeV pulsed deuteron beam impinging on a thick Be target. This reaction yields an intense, broad-energy neutron spectrum with sufficient neutron intensity up to about 16 MeV. The deuteron beam was pulsed in ~1.5 ns bursts, at 2 MHz repetition rate. Neutron energy was selected using the Time-of-Flight (TOF) technique. This was accomplished by time-gating the image-intensifier such that it acts as a very fast shutter (on ns time scale), collecting light after a pre-determined time relative to the time of the neutron pulse creation. The image intensifier opening time was about 10 ns.

The detector-target distance was 3.54 m. Figure 91 shows the detector positioned in the neutron beam-line during final adjustments. Upon completion of optical adjustments, the detector was enveloped in by a light-tight enclosure.

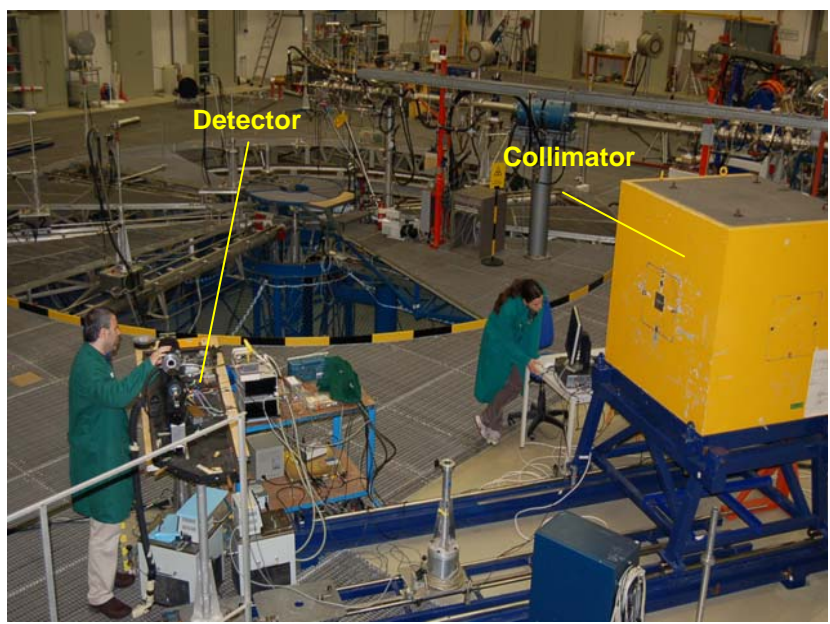

**Fig. 91 The capillary detector positioned at the PTB beam-line during final adjustments**

4.8.1 Analysis of experimental data

Track images were collected by exposing the CCD camera for durations ranging from 0.1 – 4 seconds depending on the neutron flux for the imaged energy. As the CCD chip was relatively large, readout time of each frame was 7-8 seconds.

Figure 92a shows an entire image (all CCD pixels in reduced scaling) obtained for neutron energy 15 MeV. This is a net image, with the readout and dark noises having been removed. In addition, a threshold was applied such that all values below 16 ADC units (ADU) were set to zero.

One can observe a very large number of point-like low-light-intensity events and a single continuous track of a proton (indicated by the arrow). The projection length of this track is 298 μm. Figure 92b shows the proton track in high magnification. Figures 92c & d show the same image region around the track with light intensity threshold at 16 ADU and increased to 180 ADU, respectively. The separation between gamma-ray and neutron events is quite evident from the latter figures. Protons generated by neutrons in the scintillating liquid exhibit bright continuous tracks with a Bragg peak at their end. Gamma-ray-induced electrons generate small, faint blobs of light that appear in Figures 92a & c as a multitude of specks. The separation of electron from proton events, can in principle, be performed automatically by using light and track length thresholding, pixel connectivity analysis and the existence of a Bragg peak. As is evident from Figure 92d, a simple thresholding procedure removes most of the gamma-ray induced events.

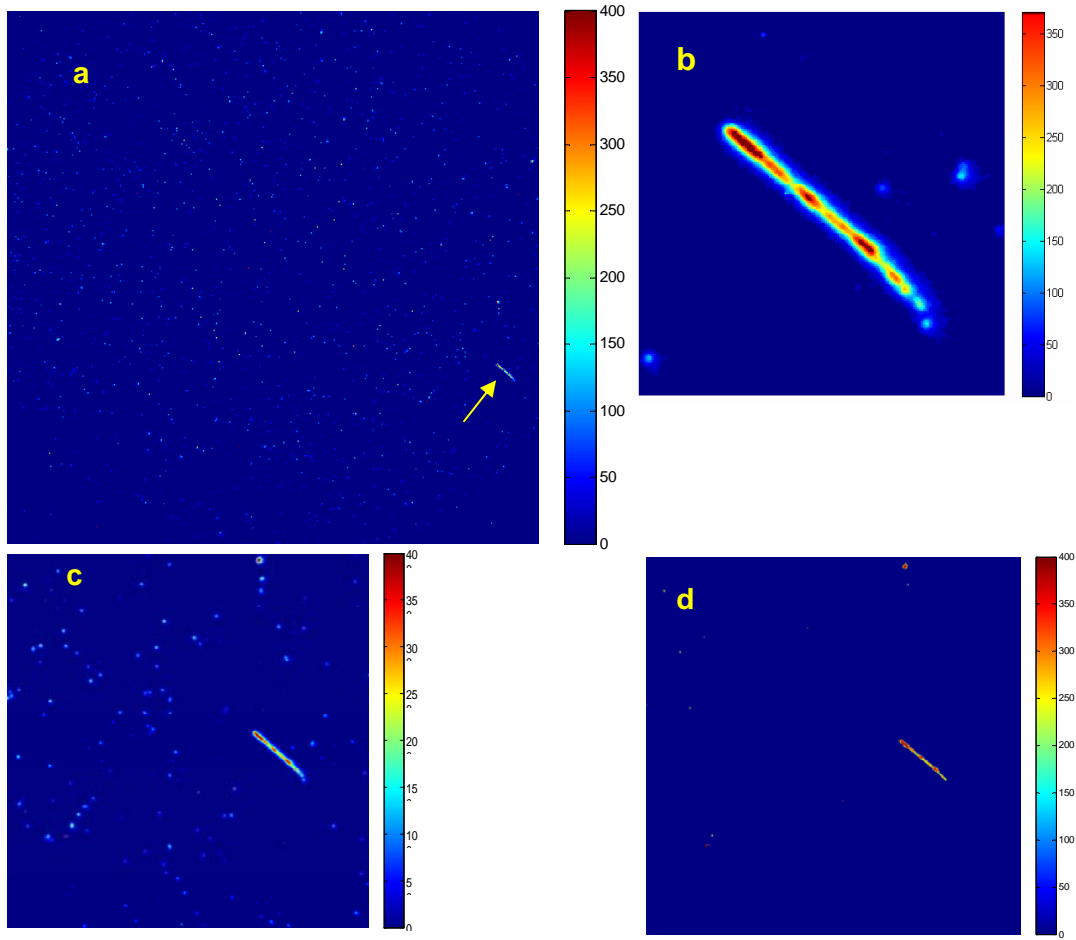

**Fig. 92 Experimental track projection image from 15 MeV incident neutron: a) entire CCD frame, b) A highly magnified region including the proton track, c) Area around the proton track, image threshold 16 ADU, d) Same as (c) but with image threshold raised to 180 ADU**

Figure 93 a-c shows three additional experimental images of recoil proton track projections, 110, 320 and 82 μm long (from protons of 3.3, 6 and 2.69 MeV), respectively, produced by 15 MeV neutrons incident perpendicularly on the capillary detector. As expected, each track exhibits increased scintillation light emission at one end, indicative of the Bragg peak.

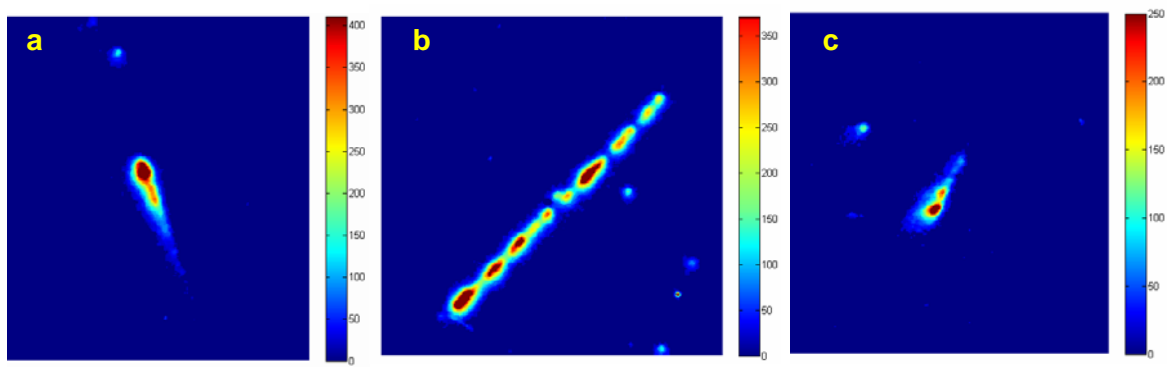

**Fig. 93 Experimental track-projection images from 15 MeV neutrons: a) 110 μm, b) 320 μm, c) 82 μm**

Images were obtained for 4 neutron energies 7.3, 8.6, 13, 12.5 and 15 MeV. In order to avoid excessive background and image complexity at neutron energies with higher flux, exposure times of the CCD camera were kept very short (0.1 – 0.5 s). Upon completion of analysis, the apparent conclusion was that this exposure regime was too short, resulting in only two exploitable neutron energies (15 and 8.6 MeV) with a very low number of usable events (based on acceptable range of alpha angles -90°<α<90°). As evident from Figure 92a, proton tracks are easily discernible from the background, reaffirming the above statement regarding the extension of CCD exposure time.

The neutron energy reconstruction was performed in the manner described in section 4.7.2 (reconstruction from simulation results) with limited success.
This may be attributed to the detectors' light collection efficiency. The simulation fits (see Eqs. 22 & 23) assume complete collection of all light photons created within each of the capillaries along the track. However, this is not the case in reality, as only 1.5% of the created light emitted from each capillary does arrive at the image-intensifier photocathode and only 13 % of it will be converted to photo-electrons. Thus, un-negligible number of capillaries along the proton track will appear either unlit or will show statistics of a single electron, reducing the expected total light or, losing the proportionality between the signal and the emitted light.

To test this hypothesis the simulation results were revisited, excluding track-projections shorter than 12 capillaries (majority of the experimental tracks range between 4 - 12 capillaries) thus modifying Eqs. 22-23 such that the new fits compensate for the above described light loss, as seen in Figure 94 (from 8000 photons and downward).

The modified fits are:
a. Track length (**r**) vs. total number of light photons (L) created per track-

$$r(L) = (0.0003638 \pm 9.7 \times 10^{-5}) \times L^{(1.482 \pm 0.022)} + (110.5 \pm 20) \qquad (24)$$

b. Proton energy (Ep) vs. track length (**r**)-

$$Ep(r) = (0.214 \pm 0.0211) \times r^{(0.5418 \pm 0.0108)} - (0.3349 \pm 0.21) \qquad (25)$$

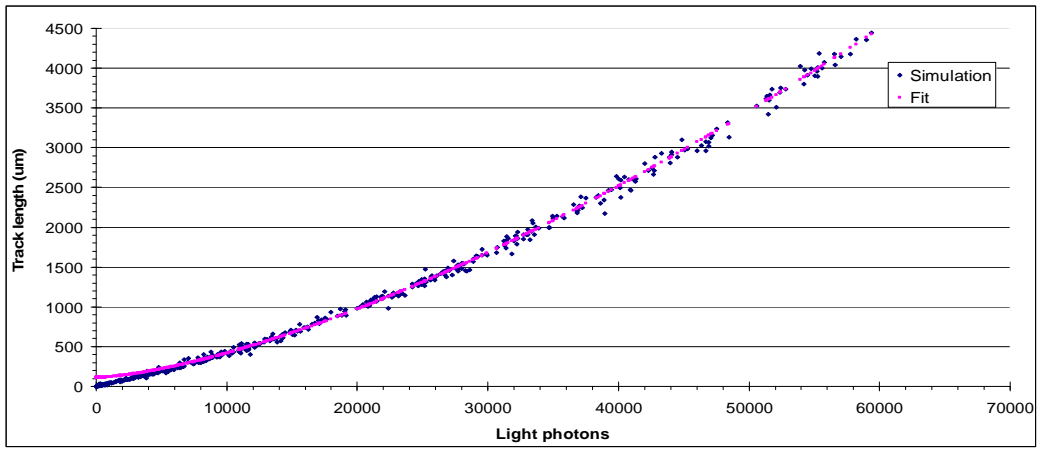

**Fig. 94 New fit for simulation results of track length (r) vs. total number of light photons (L) created per track following exclusion of all tracks below 12 capillaries**

Reiteration of the neutron energy reconstruction described in section 4.7.2 (reconstruction from simulation results) with the modified fits, applied separately to each of the neutron energies, yielded improved results, as seen in Figures 95a & b.

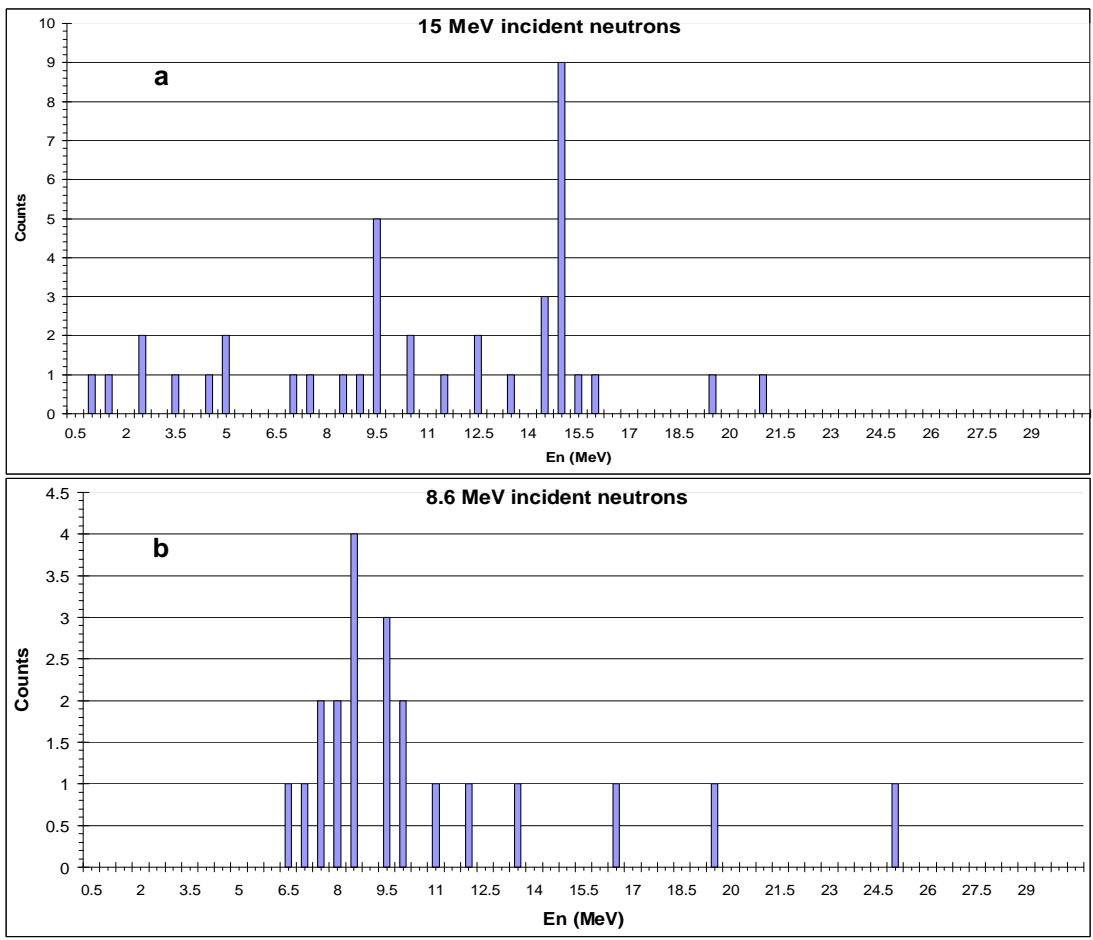

**Fig. 95 Reconstructed experimental result for incident neutrons of a) 15 MeV and b) 8.6 MeV**

As mentioned above, experimental results suffer from very low counting statistics which does not permit clear-cut statements. However, the above experimental reconstruction results may be taken as an indication that the above described modification of the original simulation fits were indeed called for.

4.9 Conclusions and recommendations for further development

A novel imaging neutron detector consisting of micro-capillaries filled with a high refractive index liquid scintillator has been developed. The key properties of the detector are high position resolution (tens of microns), good rejection of gamma-ray events and the ability to construct efficient, large-area detectors.

The energy reconstruction method presented here, based on the determination of light yield and the projection length of each proton track resulted in 10-15 % energy resolution in the range 4 - 20 MeV. Such energy resolution is poorer by a factor of 2-4 than that obtainable with <u>non-imaging</u> organic spectrometers, which use unfolding algorithms (Brooks and Klein, 2002), and is also inferior to that of TOF spectrometry. Nevertheless, its energy resolution is much better than that of the Bonner multi-sphere method (Brooks and Klein, 2002) and that of capture-gated detectors (Hyeonseo, 2010). In addition, it can be operated at significantly higher count-rates compared to recoil-telescopes.

Finally, identification of the proton tracks has been performed by visual inspection of the CCD images. This is a rather labor-consuming task and a computerized track recognition procedure needs to be developed for rapid automatic identification of a large number of proton tracks.

5. Summary and conclusions

The present work describes efficient, large-area fast-neutron detectors for combined sub-mm spatial imaging and energy spectrometry, capable of operation in mixed high-intensity neutron-gamma fields.

The combination described here of high-spatial-resolution fast-neutron imaging with energy spectrometry is <u>novel and unique</u>.

Within the framework of this work two detection systems based on optical readout were developed:

a. Integrative optical detector

b. Fibrous optical detector operating in event-counting mode

a. Integrative optical detector

The main part of this Ph.D. work dealt with the development of a $2^{nd}$ generation Time-Resolved Integrative Optical Neutron (TRION) detector (TRION, Gen. 2). It is based on an integrative (as opposed to event-counting) optical technique, which permits energy-resolved imaging of fast-neutrons, via time-gated optical readout. This mode of operation permits loss-free operation at very high neutron-flux intensities.

The TRION neutron imaging system may be regarded as a stroboscopic photography of neutrons arriving at the detector on a few-ns time scale. As this spectroscopic capability is based on the Time-of-Flight (TOF) technique, it has to be operated in combination with a pulsed neutron source, such as an ion accelerator producing 1-2 ns wide beam pulses.

As part of this development work, various properties related to this detector's performance were investigated: Spatial resolution, temporal resolution and excess noise.

A detailed Monte-Carlo calculation of the Point-Spread Function (PSF) in the scintillating screen was performed using the GEANT 3.21 code (CERN, 2003). It has shown that the FWHM of the total PSF is only 1 fiber which corresponds to 0.572 mm. However, the tertiary signal distribution (see text) adds very broad wings to the PSF, causing a considerable loss of contrast in transmission images, because they add up to a significant baseline under the primary signal. Nevertheless, following deconvolution using an experimentally-determined PSF, the contrast shows significant improvement, especially at low spatial frequencies; this is, however, at the expense of increased noise.

With respect to the temporal behavior, both the finite TOF required by direct neutrons to traverse the full thickness of the scintillator screen and the temporal behavior of the light signal generated by neutrons scattered within the scintillating screen were investigated. If the screen has large dimensions, the probability for multiple scattering within it is significant and the neutron may deposit its energy in the screen over a time period which is comparable to (or longer than) the characteristic time-scales of the above-mentioned processes. In order to investigate this effect, a Monte-Carlo calculation was performed to simulate the spatial PSF as a function of time, for several incident neutron energies. The outcome has shown that, for 2 MeV neutrons, the light decays to 1/10 and 1/100 of its value after 3 ns and 8 ns, respectively, relative to the time of neutron arrival. For higher neutron energies, the corresponding times are approximately 1.8 ns and 4 ns.

As the time required for light to decay to 1/100 of its value is within our gating time (4-6 ns), our conclusion is that this effect is not of major importance for the 3 cm thick screen. However, it should be taken into account and corrected for in thicker screens.

The temporal resolution in TRION is governed by the following factors: finite duration of the deuteron beam burst, scintillation decay time and minimal achievable image-intensifier gate width. The influence of these factors was deconvolved from the measured TOF spectra, resulting in recovery of the temporal resolution for most neutron energy groups of interest.

In comparison to event-counting detectors, where the variance in the signal depends only on the number of detected events, in an energy-integrating detector additional factors such as the fluctuations in imparted energy, number of photoelectrons, system gain and other factors will also contribute to the noise. The propagation of the signal produced by the various components of the TRION fast neutron detector was analyzed and the physical processes within them that influence the signal-to-noise ratio were presented in detail. It has been shown that TRION exhibits a relatively large excess noise factor (ENF) especially at low neutron energies. The main reason for this is the rather low number of photoelectrons produced at the image-intensifier (I-I) photocathode. In principle, the ENF may be improved by increasing scintillator light yield, light collection efficiency and quantum efficiency of the image intensifier. While it is difficult to achieve a substantial increase in scintillation yield of a plastic scintillator, the two other factors can in principle be improved.

In summary, TRION Gen. 2 is capable of capturing 4 simultaneous TOF frames within a single accelerator pulse. Relative to TRION Gen. 1, it thus exhibits a 4-fold increase in the utilization efficiency of the broad neutron spectrum created. In addition, it has demonstrated spatial resolution comparable to TRION Gen.1, reduced intensifier thermal noise and also improved temporal resolution.

The characterized, optimized and fully-functioning TRION Gen. 2 detection system constitutes an important R&D step towards the next-generation inspection systems for cargo and passenger vehicles, comprising <u>automatic</u> detection of small quantities of concealed standard and improvised explosives as well as special nuclear materials.

b. Fibrous optical detector (capillary detector)
An imaging neutron detector consisting of micro-capillaries filled with a high refractive index liquid scintillator has been developed. The important properties of this detector are high position resolution (tens of microns) and good rejection of gamma-ray induced events. The detection efficiency is dependent on the size of the capillary array and in principle can be in the range of 10-20%.
This detector works in event-by-event detection mode. It is based on the extraction of various response parameters related to detection of an impinging fast neutron, that combine to permit neutron spectrometry. These parameters are: length of the recoil proton track-projection within the detector, as well as the amount of energy deposited by the recoil proton that is recorded as scintillation light.
The detector developed is based on a micrometric glass capillary array loaded with high refractive index liquid scintillator. Capillaries with diameters of about 11 microns should allow reconstruction of the recoil-proton track with sufficient accuracy.

Two irradiation geometries were considered: Parallel and perpendicular to the capillary array central axis. The advantages presented by perpendicular irradiation geometry are:
   a) lower radiation induced background events directly in the electronic and optical detector components
   b) minimal radiation damage to detector components
   c) simpler correspondence between proton energy and track projection-length. Specifically, this means that for a given neutron energy, the highest energy protons will continue in the direction of the impinging neutron, resulting in traversal of a maximal number of capillaries, while in parallel irradiation, highest energy protons will create the shortest track-projections as they move parallel to the capillary central axis depositing all their energy within.

The drawbacks of the perpendicular irradiation geometry are:

i) inability to perform imaging

ii) lack of axial symmetry along the capillary central axis. The most severe consequence of this is that track-projection depends on both polar angle θ and azimuth angle φ.

The advantages presented by parallel irradiation geometry are:

- ability to perform imaging
- existence of axial symmetry along the capillary central axis.

The main goal of this preliminary study was to examine the spectroscopic capabilities of this detector rather than its imaging ones. In addition, since the advantages, as far as determining particle energy, are found to be far better for neutrons that hit the detector perpendicularly, we have decided on a perpendicular geometry for the feasibility study presented in this work, though this geometry is not suitable for imaging. We note however, that imaging capabilities can be rendered rather easily by mounting the detector on a mechanical turn-table and rotating it by 90º to allow for parallel geometry as well.

As experimental results provide only limited information about the recoil-proton, namely, total light per track (L) and track projection length (Proj), detector Monte-Carlo simulations were performed in order to have better understanding of its operation and to find hitherto unknown relations between parameters that are relevant to recoil-proton behavior within the capillary array. The calculated relations facilitated reconstruction of the impinging neutron energy. The reconstruction procedure was performed in a manner similar to the experimental one where only track projection length (Proj) and amount of light created along the track (L) are known. This resulted in 10-15 % energy resolution in the range 4 - 20 MeV. Such energy resolution is poorer by a factor of 2-4 than that obtainable with <u>non-imaging</u> organic spectrometers, which use unfolding algorithms (Brooks and Klein, 2002) and is also inferior to that of TOF spectrometry. Nevertheless, its energy resolution is much better than that of the Bonner multi-sphere method (Brooks and Klein, 2002) and than that of capture-gated detectors (Hyeonseo, 2010). In addition, it can be operated at significantly higher count-rates compared to recoil-telescopes. It may prove advantageous to investigate incorporating the standard unfolding reconstruction procedures within the capillary detector approach, to achieve better energy resolution while maintaining the high position resolution.

The capillary detector was tested with a broad-spectrum neutron beam. A limited number of recoil proton track images were produced for two energies: 8.6 and 15 MeV. Due to very low counting statistics, the reconstruction procedure was successful only to a limited extent, since it did not permit drawing clear-cut conclusions.

It should be emphasized that the capillary detector possesses an intrinsic capability for discriminating between neutrons and gamma-rays. Experimental images show that protons generated by neutrons in the scintillating liquid exhibit bright continuous tracks with a Bragg peak at their end. In contrast, gamma-ray-induced electrons generate small, faint blobs of light that appear as a multitude of specks. The separation of electron from proton events can, in principle, be performed automatically, by performing light and track length thresholding, invoking pixel connectivity criteria and finally, checking whether the track signature typical of a Bragg peak is apparent.

The expected spatial-resolution defined by this detector is of the order of tens of microns. The technique is <u>not</u> based on TOF spectroscopy and is designed to work with un-pulsed neutron sources. This fact should render fibrous optical detector suitable for operation with any fast-neutron source.

Future work on this detector needs, among others, to be devoted to automatic track recognition and a detailed investigation of the gamma-ray rejection factor.

תקציר

מטרת עבודה זו היתה פיתוח גלאים יעילים, בעלי שטח גדול, אשר יכולים לעבוד בשדות מעורבים של קרני-גאמא וניוטרונים בעוצמה גבוהה, וכן מסוגלים לבצע ספקטרומטריה אנרגטית של ניוטרונים מהירים בשילוב עם דימות ברזולוציה מרחבית גבוהה. השילוב של דימות בעזרת ניוטרונים מהירים ברזולוציה מרחבית גבוהה (תת-מילימטרית) עם יכולת ספקטרומטריה אנרגטית הוא חדשני וייחודי.

במסגרת העבודה פותחו שתי שיטות גילוי המבוססות על שיטת קריאה אופטית:

**א. גלאי אופטי - אינטגרטיבי**

זהו הדור השני של גלאי Time-Resolved Integrative Optical Neutron (TRION) המאפשר לכידת תמונת ניוטרונים (בתחום אנרגיה מוגדר מתוך ספקטרום רחב) הנוצרת על מסך נצנץ לאחר פרק זמן המתאים לזמן המעוף של ניוטרונים אלו, תוך שימוש בקריאה אופטית אינטגרטיבית (בניגוד ל״מניית מאורע בודד״) ממותגת. אופן מניה זה מאפשר עבודה אמינה בשטפי ניוטרונים בלתי מוגבלים בעצמתם ללא-איבודים. ניתן להתייחס אל מערכת TRION כאל מערכת המבצעת צילום סטרובוסקופי של ניוטרונים המגיעים אל הגלאי בתחום זמנים של ננו-שניות בודדות. מאחר ויכולת ספקטרומטרית זו מתבססת על טכניקת זמן המעוף, עליה להיות משולבת עם מקור ניוטרונים פועם, כגון מאיץ יונים המייצר פולסים של קרן ברוחב טיפוסי של כ-1 ננו-שניות.

גלאי דור שני זה מסוגל ללכוד תמונות ניוטרונים בכל אחד מ-4 חלונות זמן-המעוף (חלונות אנרגטיים) בכל אחד מפולסי הניוטרונים הנוצרים במאיץ, ולסכמם על פני זמני חשיפה שונים. הוא מציג יעילות ניצול ספקטרום הניוטרונים הרחב הנוצר בכל אחד מפולסי הקרן שמקורם במאיץ הגבוהה פי 4 מזו של גלאי הדור הקודם. זהו צעד משמעותי לקראת הקמת מערכת מבצעית עתידית. בנוסף לכך, מאופיין גלאי הדור השני ע״י כושר-הפרדה מרחבי המשתווה לזה של הדור הראשון, הפחתה משמעותית ברעש התרמי של מגבר-האור הממותג וכן שיפור בכושר-ההפרדה הזמני. עקרון פעולתו של גלאי הדור השני, וכן סימולציות ותוצאות של ניסויים לאיפיונו יתוארו בעבודה זו.

**ב. גלאי אופטי - סיבי (Fiber)**

גלאי זה מבוסס על אגדים של צינוריות זכוכית נימיות בקוטר מיקרומטרי, מלאות בנצנץ נוזלי בעל מקדם שבירת אור גבוה. ספקטרומטריית ניוטרונים מתבצעת ע״י גילוי מאורעות בדידים ושחזור אנרגיית הניוטרון הפוגע בהתבסס על מאפייני מסלולו של פרוטון הרתיעה: אורך היטל מסלול הפרוטון בתוך הגלאי וכמות האור אשר נמדדה לאורך מסלולו, ממנה ניתן להקיש על כמות האנרגיה אותה השאיר. בנוסף לכך, לגלאי זה יכולת לאפשר דימות ניוטרונים מהירים עם כושר-הפרדה מרחבי בקנה-מידה של עשרות מיקרומטרים.

עקרון פעולתו של הגלאי הנימי, סימולציות ותוצאות ניסיוניות אשר התקבלו בעזרת אב-טיפוס קטן יתוארו בעבודה זו. בנוסף, יוצגו תמונות מסלולי פרוטוני רתיעה אשר נוצרו מניוטרונים בעלי אנרגיה של 4-20 MeV וכן תוצאות ראשוניות המציגות את יכולתו הספקטרומטרית של הגלאי.

I

העבודה נעשתה בהדרכת

<u>ד״ר דוד ורצקי</u>

<u>המכון למחקר גרעיני נחל-שורק</u>

<u>פרופ׳ אמנון מועלם</u>

במחלקה <u>לפיסיקה</u>

בפקולטה <u>למדעי הטבע</u>

<u>ד״ר יצחק אוריון</u>

במחלקה <u>להנדסה גרעינית</u>

בפקולטה <u>להנדסה</u>

# גלאי נויטרונים מהירים בעלי כושר-הפרדה מרחבי גבוה למטרות דימות וספקטרומטריה

מחקר לשם מילוי חלקי של הדרישות לקבלת תואר "דוקטור לפילוסופיה"

מאת

אילן    מור

הוגש לסינאט אוניברסיטת בן גוריון בנגב

אישור המנחה _______________________________

אישור דיקן בית הספר ללימודי מחקר מתקדמים ע"ש קרייטמן

_______________________

כה טבת, תשע"ג                                                07/01/2013

באר שבע

# גלאי נויטרונים מהירים בעלי כושר-הפרדה מרחבי גבוה למטרות דימות וספקטרומטריה

מחקר לשם מילוי חלקי של הדרישות לקבלת תואר "דוקטור לפילוסופיה"

מאת

אילן    מור

הוגש לסינאט אוניברסיטת בן גוריון בנגב

כה טבת, תשע״ג                                                                 07/01/2013

באר שבע